\documentclass[fleqn,usenatbib]{mnras}

\usepackage{newtxtext,newtxmath}

\usepackage[T1]{fontenc}

\DeclareRobustCommand{\VAN}[3]{#2}
\let\VANthebibliography\thebibliography
\def\thebibliography{\DeclareRobustCommand{\VAN}[3]{##3}\VANthebibliography}

\usepackage{graphicx}	
\usepackage{amsmath}	
\usepackage{pdflscape}
\usepackage{longtable}
\usepackage{microtype}

\newcommand{\sectionref}[1]{{Section~\ref{#1}}}
\newcommand{\figureref}[1]{Fig.~\ref{#1}}
\newcommand{\tableref}[1]{Table~\ref{#1}}

\title[Reclassification of type IIn supernovae]{A systematic reclassification of type IIn supernovae}

\author[C. L. Ransome et al.]{
C. L. Ransome,$^{1}$\thanks{E-mail: C.Ransome@2018.ljmu.ac.uk}
S. M. Habergham-Mawson,$^{1}$
M. J. Darnley,$^{1}$
P. A. James,$^{1}$\newauthor
A. V. Filippenko,$^{2,3}$
and E. M. Schlegel$^{4}$
\\
$^{1}$Astrophysics Research Institute, Liverpool John Moores University, Liverpool Science Park iC2, 146 Brownlow Hill, Liverpool, Merseyside, L3 5RF, UK\\
$^{2}$Department of Astronomy, University of California, Berkeley, CA 94720-3411, USA\\
$^{3}$Miller Institute for Basic Research in Science, University of Califoria, Berkeley, CA 94720, USA\\
$^{4}$Department of Physics and Astronomy, University of Texas at San Antonio, One UTSA Circle, San Antonio, TX 78249, USA\\
}

\date{Accepted for publication into MNRAS}

\pubyear{2021}

\begin{document}
\label{firstpage}
\pagerange{\pageref{firstpage}--\pageref{lastpage}}
\maketitle

\begin{abstract}
\noindent
Type IIn supernovae (SNe\,IIn) are a relatively infrequently observed subclass of SNe whose photometric and spectroscopic properties are varied. A common thread among SNe\,IIn are the complex multiple-component hydrogen Balmer lines. Owing to the heterogeneity of SNe\,IIn, online databases contain some outdated, erroneous, or even contradictory classifications. SN\,IIn classification is further complicated by SN ``impostors'' and contamination from underlying \ion{H}{ii} regions. We have compiled a catalogue of systematically classified nearby (redshift $z<0.02$) SNe\,IIn using the Open Supernova Catalogue (OSC). We present spectral classifications for 115 objects previously classified as SNe\,IIn. Our classification is based upon results obtained by fitting multiple Gaussians to the H$\alpha$ profiles. We compare classifications reported by the OSC and Transient Name Server (TNS) along with the best matched templates from \texttt{SNID}. We find that 28 objects have been misclassified as SNe\,IIn. TNS and OSC can be unreliable; they disagree on the classifications of 51 of the objects and contain a number of erroneous classifications. Furthermore, OSC and TNS hold misclassifications for 34 and twelve (respectively) of the transients we classify as SNe\,IIn. In total, we classify 87 SNe\,IIn. We highlight the importance of ensuring that online databases remain up to date when new or even contemporaneous data become available. Our work shows the great range of spectral properties and features that SNe\,IIn exhibit, which may be linked to multiple progenitor channels and environment diversity. We set out a classification sche me for SNe\,IIn based on the H$\alpha$ profile which is not greatly affected by the inhomogeneity of SNe\,IIn.
\end{abstract}

\begin{keywords}
transients: supernovae --- stars: circumstellar material --- catalogues --- stars: winds/outflows --- technique: spectroscopic --- line: profiles
\end{keywords}



\section{Introduction} 
\label{sec:intro}

Type IIn supernovae (SNe\,IIn) are a somewhat uncommon subclass of SNe accounting for around 7\% of detected core-collapse SNe (CCSNe) \citep{Li_2011}. \citet{Schlegel_1990} analysed optical spectra of Type II SNe and noticed that three of the spectra (SN\,1987F, SN\,1988Z, and SN\,1989C) exhibited relatively narrow features superimposed on the broader hydrogen Balmer emission lines. Those three objects had a very blue continuum and lacked the P\,Cygni profiles one would associate with a normal SN\,II. These so-called SNe\,IIn (with the ``n'' denoting the narrow features) are characterised by multicomponent Balmer line profiles with the most obvious example being the H$\alpha$ line. One may observe a relatively narrow feature with a full width at half-maximum intensity (FWHM) on the order of hundreds of km\,s$^{-1}$ (and sometimes also an even narrower, unresolved feature), an intermediate feature a few thousand km\,s$^{-1}$ wide, and a broad component up to tens of thousands of km\,s$^{-1}$ wide \citep{Filippenko_1997}. SN\,1988Z is often used as a prototype SN\,IIn and may be used as a subtype of SNe\,IIn (other notable SNe\,IIn, such as the well-studied SN\,1998S, are also used as subtypes).

\begin{figure} 
\centering
  \includegraphics[width=\columnwidth]{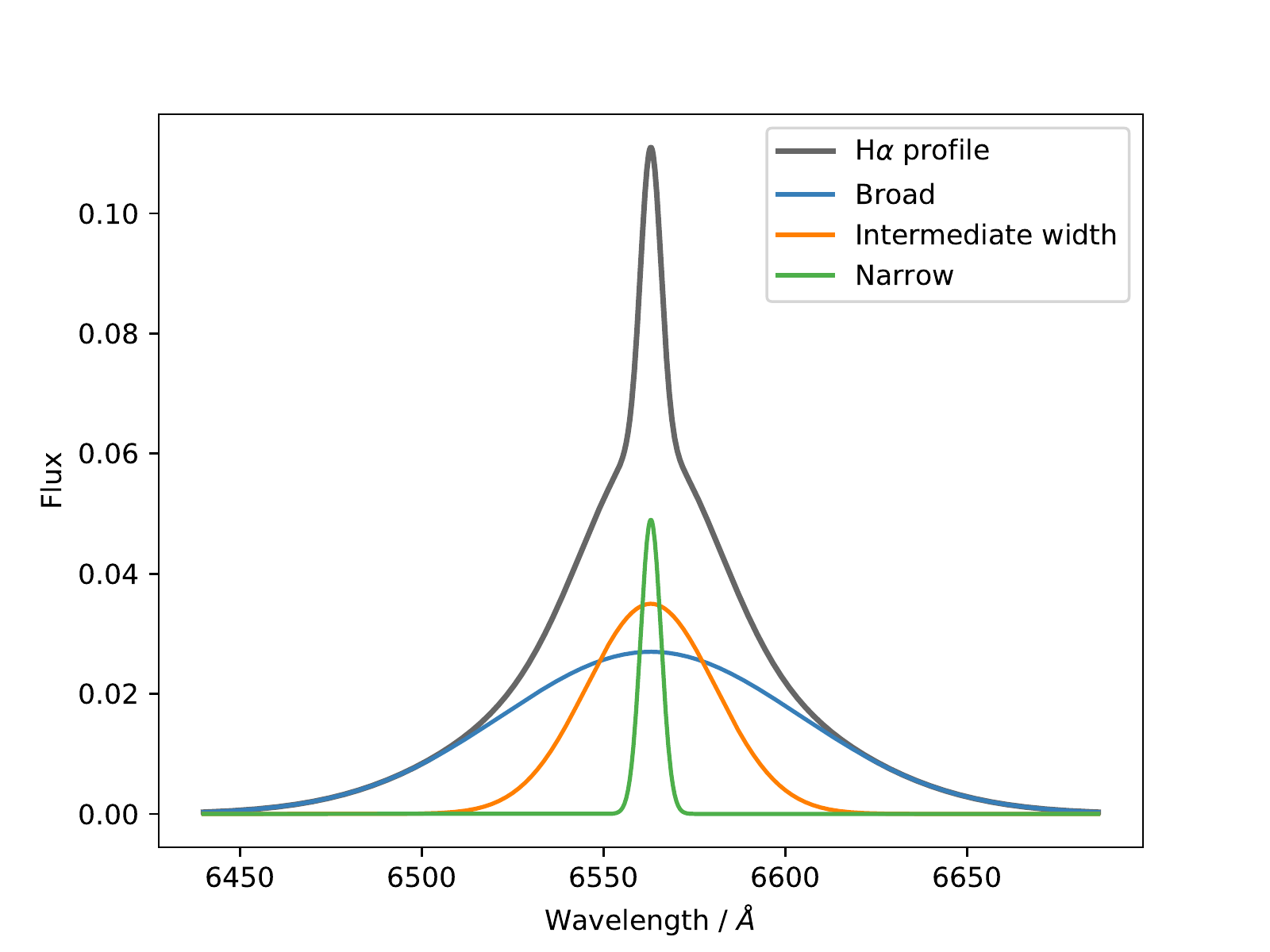} 
  \caption{An example SN\,IIn H$\alpha$ profile with the narrow, intermediate-width, and broad components shown. This is an example of the classic SN\,IIn ``Eiffel Tower'' shape which is the product of three model Gaussian components.}
  \label{fig:halpha}
\end{figure}

The mechanism which drives the formation of the narrow H features is the SN ejecta shocking a pre-existing, cold, slow, and dense circumstellar medium (CSM). Emission from the initial interaction then ionises the surrounding unshocked CSM resulting in an H$\alpha$ excess \citep{Chugai_1991, Chugai_1994}. The broad components of the lines originate from the fast-moving SN ejecta. The intermediate components may form from the SN ejecta interacting with a dense, clumpy wind \citep{Chugai_1994}, broadened by the scattering of photons from thermal electrons \citep{Chugai_2001, Dessart_2009,Humphreys_2012,Huang_2018}, or from the dense, shocked shell \citep{Kiewe_2012}. These features can constitute the classic SN\,IIn H$\alpha$ profile shape, reminiscent of the Eiffel Tower, as can be seen in  \figureref{fig:halpha} which shows an example model H$\alpha$ profile constructed from three Gaussian components (with FWHM $\approx 200$, 1100, and 2700\,km\,s$^{-1}$).

There are several possible origins for this pre-existing CSM. It has been proposed that the progenitor systems of SNe\,IIn undergo mass-loss episodes preceding their terminal SN explosion \citep{Smith_2014}. Using light-curve modelling analysis on the bolometric light curves of a sample of SNe\,IIn, \citet{Moriya_2014} found that the mass-loss rate of many SN\,IIn progenitors exceeds $10^{-3}$\,M$_\odot$\,yr$^{-1}$. Some SNe\,IIn undergo significant mass-loss episodes in the years preceding the SN explosion \citep{Ofek_2014}. SN\,2009ip was initially recognised as an ``SN impostor'' (see \sectionref{imps}) at its 2009 eruption, and then in 2012 it evolved into a SN\,IIn following its terminal explosion or a period of eruptive mass loss from its luminous blue variable (LBV) progenitor (see \sectionref{lbv}; \citealp{Mauerhan13a} \citealp{Ofek_2014}, \citealp{Pastorello_2017}). This latter mode of mass loss was also observed in SN\,2015bh \citep{Boian_2018}, SN\,2016bdu \citep{Pastorello_2017}, and SN\,2018cnf \citep{Pastorello_2019}. 

\begin{figure} 
\centering
  \includegraphics[width = 8cm]{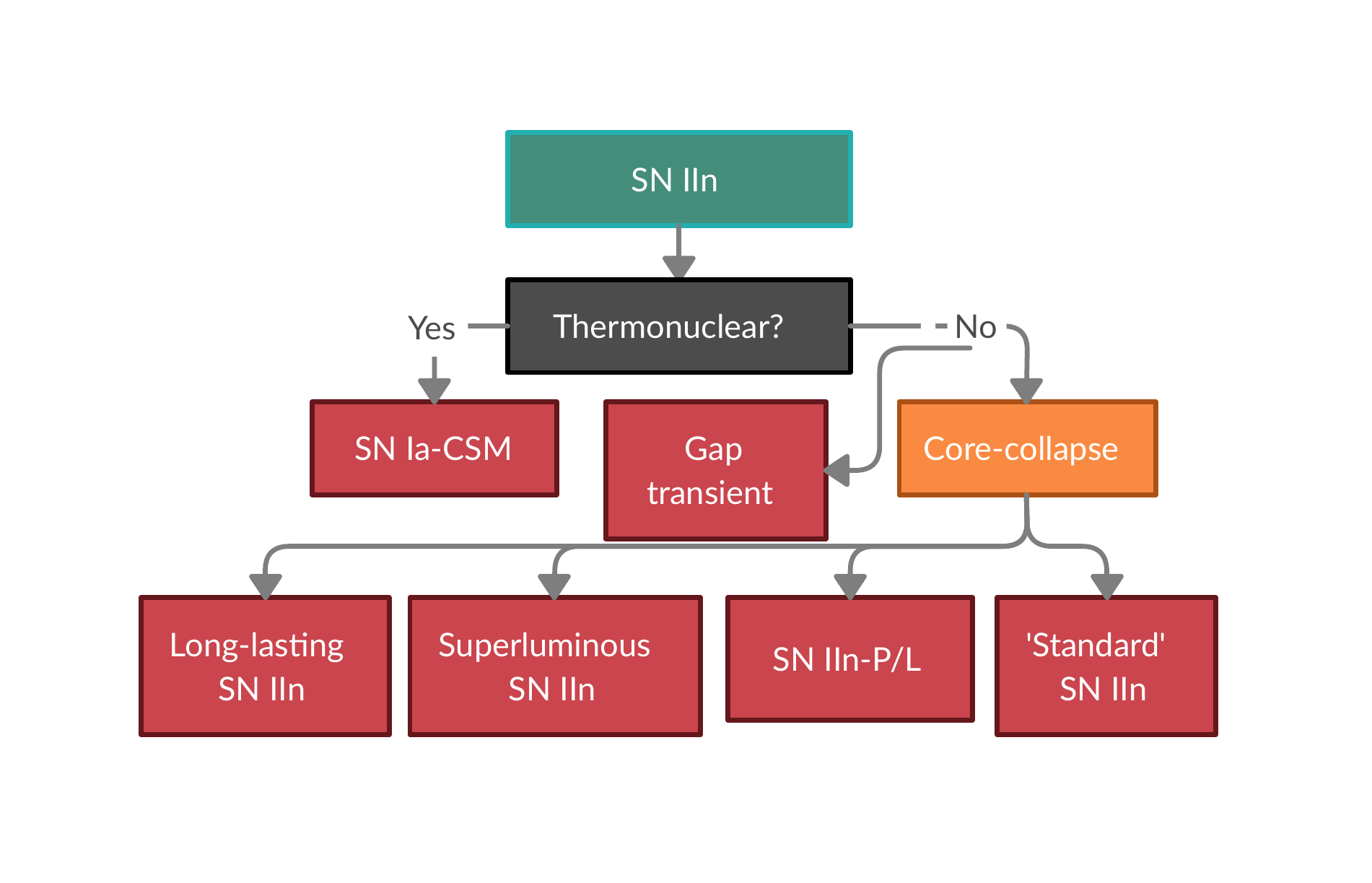}
  \caption{A breakdown of the routes a SN\,IIn can take based on the nature of the explosion and other observed spectroscopic and photometric properties.}
  \label{routes}
\end{figure}

SNe\,IIn are observationally very diverse. There is a wide luminosity range, with some SNe\,IIn inhabiting the standard CCSN space in the luminosity--timescale phase-space diagram \citep{kasphd} at peak absolute magnitudes around $-17$ to $-19$ \citep{Li_2011, Kiewe_2012, tadd13}. Some SNe\,IIn, such as SN\,2006gy, reach $-22$ mag \citep[the most luminous SN observed at the time;][]{Smith_2007}. This may be considered a superluminous SN\,IIn (SLSN-IIn). Using light-curve modelling, \citet{moriya13} found that the CSM shell contained around 15\,M$_\odot$ of H-rich material and the SN ejecta contained similar mass. The light curve of SN\,2006gy is also very long lived, lasting $\sim 3000$\,days \citep{ori15}. A further example of long-lasting SN\,IIn emission is SN\,2005ip, which remained the brightest H$\alpha$ source in its host galaxy (NGC\,2906) for several years \citep{Stritzinger_2012, Smith_2016, Smith17} with a decline only recently \citep{Fox_2020}. \citet{Smith_2016} find that such long-lived emission from SNe\,IIn requires mass-loss episodes from the SN progenitor to occur for 1000\,yr preceding the explosion. Some SNe\,IIn also show signs of pre-existing dust in the CSM shell revealed through infrared (IR) observations (e.g., SN\,2010jl, \citealp{Bevan20}; SN\,2014ab, \citealp{Moriya_2020}; SN\,2017hcc, \citealp{Smith20}). Other transients show CSM interaction for a brief period, perhaps indicating a more confined CSM (e.g., SN\,2008fq, \citealp{tadd13}; SN\,2018zd, \citealp{Zhang_2020}). Some SNe\,IIn follow the standard light-curve decay timescale for CCSNe of a few hundred days. In other cases the light-curve shape is that of a Type IIP SN with a plateau phase that may be due to the inward recombination front and the outward expansion of the SN ejecta. These two processes occur at a similar rate, resulting in the brightness of the object being almost constant. The light-curve shape may resemble that of a Type IIL SN, where the light curve drops off linearly (in magnitudes) with time, owing to the progenitor star having a smaller H-rich envelope following greater mass loss  \citep[e.g., SN\,2011ht, SN\,IIP-like][SN\,2013fs, SN\,IIP-like, and SN\,2013fr, SN\,IIL-like]{Mauerhan13a, Bullivant18}. However, as more SNe are observed, the clear distinction between light-curve morphologies is blurred, and there may be a continuum of light-curve shapes rather than a larger number of discrete categories. \citet{Anderson_2014} investigated the diversity in the $V$-band light curves of 116 SNe\,II and found a continuum between lower-luminosity events with a plataeu in the light curve and brighter events with a linear, more rapid decline \citep[see also][]{Hamuy_2003b, Sanders_2015, Valenti_2016}. The possible subtypes of SNe\,IIn are outlined in \figureref{routes}.

Along with this photometric diversity we note that there are accompanying differences in the spectra of SNe\,IIn. Generally speaking, one may expect to see lines of \ion{O}{i}, \ion{He}{i}, \ion{Ca}{ii}, and \ion{Fe}{ii} along with \ion{H}{i} emission. However, the common defining traits are the multicomponent H$\alpha$ lines with the signature narrow features (see \figureref{fig:halpha}) and a blue continuum. However, even these H$\alpha$ profiles can be diverse in strength, duration, and shape. Some SNe\,IIn exhibit their narrow features and multicomponent H$\alpha$ (and other Balmer line) profiles throughout their evolution. Other SNe\,IIn seem to show narrow features only in the early stages of their evolution, with these features lasting just a few days or even hours. \citet{Ofek_2014} terms such transients with short-lived SN\,IIn phases as ``weak'' SNe\,IIn, with an example being PTF11iqb \citep{Smith_2015}. These ``weak'' SNe\,IIn may suggest that lower-mass SN progenitors experience the mass loss required for the CSM interaction as seen in some SNe\,IIP and SNe\,IIL (for example, ASASSN-15oz, \citealp{Bostroem_2019}; SN\,2013fs and SN\,2013fr, \citealp{Bullivant18}). While SNe\,IIn do not typically show the broad P\,Cygni features one observes in other SNe\,II, we do sometimes see narrow P\,Cygni absorption associated with the narrow H$\alpha$ features. Moreover, asymmetry in both the SN ejecta and the surrounding CSM can profoundly affect the shape of the H$\alpha$ profile. In some cases we observe SNe\,IIn exploding within an aspherical and/or clumpy CSM where one may see asymmetric redshifted or blueshifted components in the H$\alpha$ profiles (for example, there may be blueshifted wings in the later epochs of the spectra of SN\,2008J; see \figureref{2008jTS}). In the case of a SN\,IIn having an LBV progenitor (see \sectionref{lbv}), we can easily infer how this asymmetry and complex structure (e.g., SN\,1998S \citealp{Mauerhan_2012, Shivvers_2015}) can arise by observing the Homunculus nebula around the LBV $\eta$\,Carinae with its asymmetric lobes \citep{Smith_2003, Smith_2006, Akashi_2020, Millour_2020, Kurfurst_2020}. When one considers the viewing angle to the observer, it is clear that CSM geometry is important to our observations as the results of this asymmetry are a deviation from the classic SN\,IIn H$\alpha$ profile shape.

\subsection{Possible progenitor systems of SNe\,IIn} \label{progen}

\subsubsection{Luminous blue variables} \label{lbv}

LBVs arise from massive stars undergoing a transition phase from an O-type star to a Wolf-Rayet (WR) star where the outer H-rich layer has been expelled \citep{HD_1994, Weis_2020}. LBVs are characterised by mass loss through winds at a rate of up to $\sim 10^{-4}$\,M$_\odot$\,yr$^{-1}$ \citep{Vink_2018} and also episodic eruptions. The mass loss from these eruptive episodes (for example, the great eruption of $\eta$\,Car) can reach up to 10\,M$_\odot$ with $\dot M \approx 1$\,M$_\odot$\,yr$^{-1}$ \citep{Smith_2003, Smith_2010}. The H-rich CSM that surrounds SNe\,IIn may originate from the material ejected in these dramatic mass-shedding events \citep[see][and references therein]{Smith_2014}. Notable examples of the LBV class include P\,Cygni \citep{Smith_Hartigan_2006} and $\eta$\,Car \citep{Smith_2003} in the Milky Way and S Doradus in the Small Magellanic Cloud \citep{snt15}. These mass-loss periods of LBVs correspond to increases in the luminosity of the star. \citet{Ofek_2014} found that around half of SN\,IIn progenitor systems experience such brightening at least 120\,days before the final terminal SN explosion. According to \citet{Gal-Yam_2007}, archival data show that SN\,2005gl in NGC\,266 had an LBV progenitor. \citet{GalYam_2009} found that the LBV progenitor had a mass that exceeded 50\,M$_\odot$. This was the first confirmed LBV progenitor system for an SN\,IIn.

As LBVs are high-mass stars, they are expected to reside in young environments close to O-type stars. However, \citet{Tombleson_2015} found that many LBVs are isolated from any cluster of O-stars. The authors propose a scenario where a WR star is a mass donor in an interacting binary system with an LBV mass gainer. Through these interactions the LBV may be kicked out of its place of birth, leading to the observed isolation. This is consistent with the findings of \cite{hab14} described in \sectionref{envs}: many SNe\,IIn reside in environments devoid of star formation. Another route for LBV formation is through binary mergers which may account for their apparent isolation \citep[][]{Justham_2014, Aghakhanloo_2017}.

\subsubsection{Yellow hypergiants} \label{yellow}

Yellow hypergiants (YHGs) are evolved post-red-supergiant (RSG) stars with $M_{\rm ZAMS} \approx 20$--60\,M$_{\odot}$ which undergo enhanced mass loss that may form the required CSM for the SN\,IIn phenomenon \citep[for a review, see][]{deJager_1998}. \citet{BrennanA_2021, BrennanB_2021} present a detailed analysis of the SN\,IIn candidate AT\,2016jbu, which was found to be an SN\,2009ip-like object with a double-peaked light curve and a dusty circumstellar shell. Using archival {\it Hubble Space Telescope (HST)} imaging, the authors found that the progenitor of AT\,2016jbu was consistent with a YHG of $M \approx 20$\,M$_{\odot}$. With age estimates of the possible progenitor of AT\,2016jbu and the age of the environment of the transient, a CCSN scenario was favoured over a nonterminal explosion while an ``impostor'' event could not be ruled out.

\subsubsection{Gap transients and supernova impostors} \label{imps}

Some transients are initially classified as SNe\,IIn because they exhibit the multicomponent and narrow H$\alpha$ emission features. However, it later becomes apparent, when the progenitor star is reobserved after the brightening phase, that some events originate from nonterminal explosions within dense CSM. These transients typically lie within gaps in the timescale-luminosity phase space of exploding transients \citep[][]{kasphd}. With absolute magnitudes ranging from $-10$ to $-14$ (between novae and SNe), they may be generally termed ``gap transients.'' A subset of them are sometimes named SN ``impostors''; they have spectra similar to those of SNe\,IIn, but they are subluminous, typically $M_V \approx -11$ to $-14$ \citep{Kochanek_2012}. Such events may be due to mass-loss episodes from LBVs, similar to the great eruption of $\eta$\,Car \citep{Smith_2011}. After this initial eruption, the progenitor may be obscured by dust that forms after the initial eruption \citep{Kochanek_2012} however some objects such as SN\,2002kg do not much dust \citet{Kochanek_2012,Humphreys_2014, Humphreys_2017}. Other examples that appear in this work include; SN\,1997bs, SN\,1999bw, SN\,2000ch, SN\,2001ac and SN\,2002kg \citep{Kochanek_2012}. These outbursts can sometimes precede a possible full SN explosion --- for example, SN\,2006jc \citep[an SN\,Ibn rather than an SN\,IIn;][]{Pastorello_2007, Smith_2013}, SN\,2009ip  \citep{Foley_2011, Mauerhan13a, Pastorello_2013}, and SN\,2015bh \citep{Boian_2018, Thone_2017}. 

In some cases, such as SN\,2009ip, the nature of the ``final'' eruption is debated. \citet{Pastorello_2013} and \citet{Smith_2014_2009ip} noted that the high luminosity of the ``final'' eruption ($\sim -18$\,mag) of SN\,2009ip and its broad, persistent emission features suggested a large ejecta mass and explosion energy (around $10^{51}$\,erg), pointing toward a terminal CCSN explosion. However, the CCSN scenario is disputed; \citet{Fraser_2013} found that the spectrum of SN\,2009ip in Dec. 2012 resembled the one after the initial 2009 eruption. Furthermore, they found no evidence of nucleosynthesised elements, and the peak luminosity could be achieved by considering efficient kinetic-energy transfer from the ejecta colliding with CSM from eruptive LBV mass-loss episodes, suggesting a nonterminal scenario for SN\,2009ip. \citet{Fraser_2015} found that by Dec. 2014, SN\,2009ip had not yet dimmed below the level of the pre-discovery images and still had not entered the nebular phase expected in CCSNe. This may suggest a nonterminal explosion mechanism, but the authors do not rule out continuing CSM interaction after core collapse being responsible for these features. \citet{EliasRosa_2018} found that SN\,Hunt151 was remarkably similar to the eruption of SN\,2009ip; they suggested that SN\,Hunt151 may be an $\eta$\,Car-like LBV eruption within a dense CSM cocoon.

Other notable ``impostors'' include SN\,2008S \citep[the namesake for a group of similar objects;][]{Thompson_2009} in NGC\,6946 \citep{Arbour_2008}, where a progenitor was unable to be recovered. \citet{Prieto_2008} reported that the culprit may have been a lower-mass ($\sim 10$\,M$_\odot$) star in a dusty environment, hence this object could be an enshrouded ``impostor.'' \citet{Kochanek_2011} argued that SN\,2008S and the similar transient NGC\,300-OT are consistent with an explosion taking place in an environment with very dusty winds. \citet{Berger_2009b} found that the progenitor of NGC\,300-OT may be a compact object such as a WR star or blue supergiant (BSG), contrary to the RSG interpretation of \citet{Kochanek_2011}. \citet{Andrews_2020} suggested that the progenitor of the SN\,2008S-like transient AT\,2019krl is a BSG/LBV, based on archival {\it HST}, \textit{Spitzer Space Telescope}, and Large Binocular Telescope data. They also discussed the possibility that the transient was caused by binary interaction in the form of a merger that resembled an LBV eruption \citep[briefly discussed in \sectionref{lbv}, but see also][]{Smith_2016b, Smith_2018}. An alternative scenario to the LBV/BSG and nonterminal path \citep{Berger_2009, Bond_2009, Kochanek_2011, Smith_2009, Andrews_2020} for SN\,2008S-like transients may be RSG/super-AGB (asymptotic giant branch) progenitors which suffer terminal electron-capture supernova (ecSN) explosions, discussed in \sectionref{lowmass}. \citet{Botticella_2009} found that the bolometric light curve of SN\,2008S was consistent with the decay of $^{56}$Co and suggested that the progenitor was a super-AGB star with $M \approx -6$ to $-8$\,M$_\odot$ that ended its life as an ecSNe. \citet{Adams_2016} reported that SN\,2008S had become fainter than its progenitor, arguing that the explosion was terminal. The dusty environments of SN\,2008S-like transients complicates attempts to detect a surviving progenitor; as such, the nature of these objects remains elusive.

While SN ``impostors'' are subluminous, there may still be luminosity overlap with some genuine SNe such as SNe\,IIn-P and SNe that are obscured by dust --- as in the aforementioned terminal scenario for SN\,2008S-like transients. The outbursts of SN\,2009ip reached absolute magnitudes of $-15$ and $-18$ \citep{Smith_2013}; if the final 2012 event was not terminal, then SN impostors driven by CSM interaction may overlap most of the luminosity space of SNe.  Therefore, distinguishing between ``impostor'' events and genuine SNe can be difficult without late-time imaging showing the surviving progenitor (or the lack thereof).

\subsubsection{Lower-mass progenitors and their explosion mechanisms} \label{lowmass}

At the lower end of the progenitor mass range for CCSNe, the SN\,IIn phenomena may also arise from ecSNe rather than the very high-mass stars which become LBVs. In the case of an ecSN, the deflagration of the degenerate core of an 8--10\,M$_\odot$ star is triggered by electron capture in $^{24}$Mg and $^{20}$Ne. The resulting explosion is less energetic than regular SNe\,II, leaving behind a neutron star \citep{Miyaji_1980, Nomoto_1984, Nomoto_1987}. A recent example of an ecSN is SN\,2018zd in NGC\,2146 \citep{Zhang_2020,Hiramatsu_2020}. 

\citet{Smith_Crab} and \citet{Moriya_2014} explore the possible link between these lower-mass ecSN progenitors and SNe\,IIn by analysis of the Crab Nebula (M\,1), the remnant of the Galactic SN\,1054 \citep{Mayall_1942, Duyvendak_1942}. Historical accounts of SN\,1054 seem to contradict predictions that an ecSN would be less luminous than a regular SN\,II, and \citet{Moriya_2014} explore CSM interaction as a possible solution to this. In the 8--10\,M$_\odot$ progenitor mass range, we may find super-AGB stars which undergo mass loss via massive winds, in turn forming the CSM required for the SN\,IIn phenomenon. As discussed in \sectionref{imps}, a possible progenitor path for SN\,2008S-like transients is dust-enshrouded super-AGBs that explode as ecSNe; however, this cannot be confirmed without late-time imaging showing that the progenitor has disappeared. \citet{Adams_2016} found that SN\,2008S had dimmed to levels below its progenitor, but the dimming of a surviving progenitor can also be explained by extreme dust (dusty environments are typical in SN\,2008S-like transients).

SN\,IIn-like features may be produced by a Type Ia SN detonating within a dense CSM \citep{Deng_2004, Dilday_2012, Silverman_2013}. The origin of the CSM is poorly understood. \citet{Dilday_2012} investigated the SN\,Ia-CSM PTF\,11kx and found that the progenitor system was a white dwarf (WD) with a red giant or supergiant companion in a symbiotic system. In this particular case, the CSM may have consisted of the material ejected by the giant's wind and perhaps material ejected from previous nova eruptions \citep[see also][]{Wang_2018, Darnley_2019}.

SN\,2002ic was the first unambiguous detection of an SN\,Ia-CSM which displayed prominent and persistent multicomponent H$\alpha$ and H$\beta$ emission lines \citep{Hamuy_2003}. A number of SNe\,IIn have been reclassified as SNe\,Ia-CSM once their thermonuclear origin became apparent. Their spectrum evolves to show the characteristic broad SN\,Ia absorption (e.g., Si\,II) features, but the Balmer emission endures, as in SN\,1997cy \citep{Germany_2000,Prieto_2005} and  SN\,2005gj \citep{Prieto_2005, Aldering_2006}. The nature of some SNe\,Ia-CSM is more ambiguous. \citet{Inserra_2016} presented photometric and spectroscopic data spanning 525\,days, finding that the spectrum is CSM-interaction dominated throughout all epochs. The authors modelled the ejecta and energetics of the explosion and argued that the CSM mass and energies required to power the light curve are inconsistent with a thermonuclear explosion, instead requiring a CCSN. Out of a sample of six SNe\,Ia-CSM (SN\,1997cy, SN\,1999E, SN\,2002ic, SN\,2005gj, PTF11kx, and SN\,2012ca) that the authors compared, only PTF11kx was consistent with the thermonuclear detonation of a WD in dense CSM.

\subsection{Environments of SNe\,IIn} \label{envs}

Reflecting the potential diversity in SN\,IIn progenitors, we see a range of properties in their local environments, particularly in terms of association with star formation (SF). As discussed in  \sectionref{lbv}, LBVs are the progenitors of some SNe\,IIn, but as LBVs are very massive and short lived one would expect them to be strongly associated with ongoing SF as traced by H$\alpha$ emission. 

In order to gauge the association of the location of an SN with H$\alpha$ emission, the pixel statistics technique, normalised cumulative ranking \citep[NCR;][]{ja06, and12}, is used on continuum-subtracted H$\alpha$ images of SN host galaxies. \citet{and08} found that the SNe\,IIn in their sample of twelve objects did not trace the H$\alpha$ emission compared to SNe with massive progenitors such as Type Ic SNe. They also found that the SNe\,IIn more closely followed near-ultraviolet (NUV) emission, particularly as observed by the {\it Galaxy Evolution Explorer (GALEX)} space telescope. The isolation of some of these SNe\,IIn from SF, such as HSC16aayt, may suggest that there could be a small amount of SF being enshrouded by dust \citep{Moriya_2019}.

\citet{hab14} used the H$\alpha$ NCR technique on a sample of 17 SNe\,IIn, finding that the SNe did not follow ongoing SF as traced by H$\alpha$ emission. They concluded that the diversity in the SNe\,IIn and the diversity in their environments points toward multiple progenitor paths to SNe\,IIn. 

\subsection{Motivation for a systematically reclassified target list}

There is great variety in the spectral and photometric features of SNe\,IIn. With this diversity in mind, it is worth noting that spectral classification software packages such as SuperNova IDentification \citep[\texttt{SNID};][]{SNID} use templates for SN spectral classification. The SN\,IIn template selection of \texttt{SNID} is limited to three objects: SN\,1996L, SN\,1997cy, and SN\,1998S. Furthermore, \texttt{SNID} has templates over a number of epochs; but as mentioned above, some objects can be classified as a SN\,IIn at a particular epoch when it may be a gap transient or an SN\,Ia-CSM, or vice versa. This issue may be exacerbated when the target only has a single spectrum available.

Contamination from the interstellar medium (ISM) or the underlying \ion{H}{ii} region may influence some classifications if the spectrum has not had an ISM model for that galaxy subtracted. As there is a strong and narrow H$\alpha$ emission feature in the ISM, it can often appear that the broad SN ejecta components have a narrow component superimposed and may seem to represent the H$\alpha$ profile of an SN\,IIn. This becomes challenging to untangle from the SN data if the spectrum has low resolution and the [\ion{N}{ii}] lines that lie on either side of the H$\alpha$ line are blended together with the H$\alpha$ emission. Online databases may retain the SN\,IIn classification for some of these objects that have no strong evidence of CSM interaction.

The literature describes numerous ways to categorise SNe\,IIn. \citet{Ofek_2014} call SNe\,IIn that exhibit CSM interaction features for a short period of time (such as PTF\,11iqb) ``weak'' SNe\,IIn. SNe\,IIn typically have spectra that exhibit enduring CSM interaction. Some subcategories are also used such as SN\,IIn-P and IIn-L, but also SN\,IId where the narrow H$\alpha$ line has a P\,Cygni profile, such as in SN\,2013gc \citep{Benetti_2000, Reguitti_2019}. Furthermore, at later epochs an object may be reclassified (for example, as a standard SN\,II) and online resources mark the object with the new classification despite the previous evidence of CSM interaction.  

The SN\,IIn subclass was coined by \citet{Schlegel_1990}. There are a number of pre-1990 SNe that have subsequently been recognised as potential SNe\,IIn (for example, SN\,1978K, SN\,1987C, and SN\,1989R). However, public databases may retain the original classification, so when one searches for SNe\,IIn, the pre-1990 SNe may be missed.

The aim of this study is to provide a systematically reclassified/reconfirmed sample of nearby (within redshift $z<0.02$) SNe\,IIn based on simple criteria -- the presence of a detectable narrow feature in the H$\alpha$ profile. \sectionref{methods} sets out the methodology employed to classify our SNe\,IIn spectra. In \sectionref{results} we present our results and spectral analysis, examine the spectral category breakdown, and show exemplar spectra and H$\alpha$ profile of each category. Our findings are discussed in \sectionref{discuss}.

\section{Methodology} \label{methods}

\subsection{Collating the sample}

An initial catalogue of local SNe\,IIn (within $z<0.02$) was constructed from the Open Supernova Catalogue (OSC; \citealp{ocs}). As of early 2019 there are 144 SNe that have at some point held a SN\,IIn classification\footnote{The object has a cited SN\,IIn classification, but this may not currently be the primary classification. The object appears in searches for SNe\,IIn.}. All associated spectral data were downloaded from the OSC. There were no data for a number of our objects in the OSC. We also checked the Transient Name Server\footnote{\url{https://www.wis-tns.org/}}. However, for some of the SNe\,IIn in the catalogue there were no publicly available data. Where the OSC entries provide references for the classifications, we contacted the original observers to request spectral data.

\subsection{Classifying the spectra}

We present the spectra collected from our archival sources in \sectionref{methods}. Where available, these are ordered in a time-series plot using \texttt{pyplot} \citep[v3.3.3;][]{Hunter_2007} and placed into their rest frame using redshifts from the OSC entries.

As the main defining feature of SNe\,IIn spectra are the multicomponent Balmer emission lines, we focus on the strongest line, H$\alpha$. We use the Python package \texttt{pyspeckit} \citep[v0.1.22][]{pyspeckit} to fit Gaussians to a multicomponent H$\alpha$ profile, in velocity space.

We begin constructing our H$\alpha$ profile models by fitting a baseline to the continuum. Fitting is restricted to the region around H$\alpha$ (6400\,{\AA}--6800\,{\AA}). We use the \texttt{specfit} routine in \texttt{pyspeckit} to fit Gaussians to the spectra. Fits for relatively narrow (\textless 1000\,km\,s$^{-1}$), intermediate ($\sim1000$--2000\,km\,s$^{-1}$), and broad (\textgreater 5000\,km\,s$^{-1}$) components are trialled. A reduced $\chi^{2}$ (calculated, along with uncertainties in the parameters of the fit, by \texttt{specfit}) test is used to determine the goodness of fit, which allows us to decide if a two- or three-component fit is most appropriate by favouring the number of components with the lowest reduced $\chi^{2}$ value. We propagate the errors in the Gaussian component parameters to calculate the uncertainty of the total fit. False ``IIn'' classifications are present in the literature which seem to arise from the background \ion{H}{ii} region of the host galaxy contaminating the spectrum, giving an appearance of narrow lines upon a broad feature which originates from the SN itself. To tackle this, we also fit for the nebular [\ion{N}{ii}] 6584\,{\AA} and 6548\,{\AA}) doublet to determine whether any narrow features seen can simply be attributed to \ion{H}{ii} region pollution rather than to interaction with CSM.

We also include SNe which may show flash-ionisation features in our analysis. Flash-spectroscopy SNe exhibit very narrow (usually unresolved) lines that are indicative of CSM ionisation for a brief period post explosion (typically \textless 10\,days) before the SN ejecta sweep up the CSM and the typical SN broad lines dominate the Balmer line profiles thereafter \citep{Khazov_2016}. A number of SNe\,IIn show this brief CSM ionisation, such as SN\,1998S \citep{Leonard_2000,Shivvers_2015} and PTF\,11iqb \citep{Smith_2015}. In these cases CSM interaction was observed at later epochs, so these flash-ionisation events cannot be eliminated from our analysis. However, if there are later spectral epochs that show the transient is a standard SN\,II, then we do not classify these objects as an SN\,IIn. We include the SN\,Ia-CSM subtype as part of the SN\,IIn class, as the SN\,IIn phenomenon is an environmental effect rather than being due to the intrinsic properties of the progenitor and explosion.

 The FWHM of the Gaussian line profiles allow us to estimate velocities of both the narrow component from the CSM interaction and the broader components from electron scattering and the SN ejecta. For an object to be classified as a SN\,IIn, the spectrum must have at least a narrow component and either an intermediate-width or broad component. Some of the objects will have narrow, intermediate, and broad components.  

\begin{figure*} 
\centering
  \includegraphics[width = 12cm]{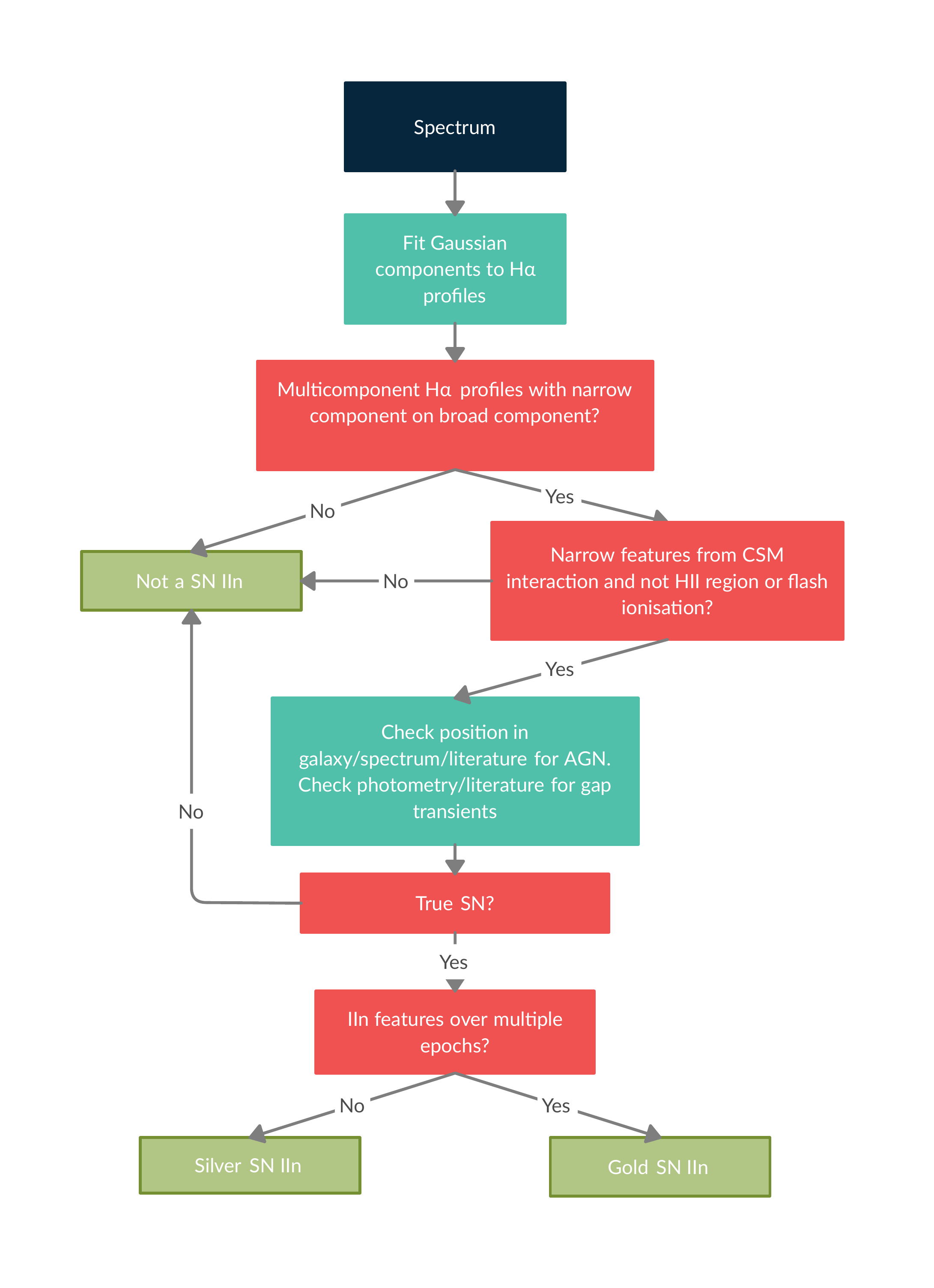}
  \caption{A decision tree describing our classification scheme for the SN\,IIn spectra.}
  \label{decision}
\end{figure*}

Gap-transient (see \sectionref{imps}) spectra can be very similar to those of SNe\,IIn. We check for a peak magnitude (SN ``impostors'' typically have a peak $\gtrsim -14$\,mag) and the literature to see if a surviving progenitor has been discovered or the object is otherwise a gap transient rather than a full-fledged SN. Similarly, the H$\alpha$ profiles of active galactic nuclei (AGNs) are similar to those of SNe\,IIn with broad lines from the AGN and narrow lines from the galaxy \citep{Filippenko_1989,Osterbrock_AGN}. Potential AGNs can be ruled out of our sample by looking at whether the SNe are located in the nucleus of the host galaxy; \citet{Ward_2020} find that very few AGNs are offset significantly from the nucleus. If a SN is central, we can look for the nebular [\ion{O}{iii}] $\lambda\lambda4959$, 5007 lines which are found in AGNs; if absent, the object is likely an SN rather than an AGN. We also check the literature to see if any central transients are known to either be an AGN or have been studied as SNe. A decision tree for our classification scheme is shown in \figureref{decision}.

The now systematically reclassified spectra can be distributed into three categories as follows.

\textbf{Gold sample}: These are SNe\,IIn that maintain their relatively narrow components throughout their evolution. We require a gold SN\,IIn to have multiple spectral epochs that span more than $\sim 10$\,days, which is longer than the typical CSM flash-ionisation timescale. Gold SNe\,IIn may have the classic ``Eiffel Tower'' H$\alpha$ profile or perhaps a more complex structure. These may be further assigned a more specific classification such as SN\,1988Z-like or SN\,1998S-like. Objects in this gold sample must be surrounded by dense CSM for there to be sufficient ongoing interaction to maintain the narrow features throughout their spectral evolution. 

\textbf{Silver sample}: The silver sample contains SNe with spectra that have some unambiguous narrow features that are consistent with CSM interaction. However, the interaction may not be particularly strong compared to gold-sample objects, and these could be younger transients that have not been monitored spectroscopically. Furthermore, silver-sample objects may not exhibit the narrow interaction features throughout their evolution. An example of a silver-sample object is PTF11iqb, which \citet{Smith_2015} remark is an SN\,IIn that transitions into an SN\,IIP (an SN\,IIn-P) and may have had a red-supergiant progenitor similar to those of SNe\,IIP. PTF\,11iqb shows renewed signs of CSM interaction at later times which \citet{Smith_2015} interpret to be the CSM disc re-emerging as the SN photosphere recedes. We also categorise objects with a single spectrum available to us as silver SNe\,IIn owing to the lack of information regarding the spectral evolution. For a silver SN\,IIn to be promoted to the gold category, more data are required; many of the silver SNe\,IIn are in this category owing to there being only a single available spectrum. If there is a single spectrum where the CSM interaction may be consistent with flash ionisation, we classify these objects as silver SNe\,IIn. They can be demoted to not being an SN\,IIn if the CSM interaction fades and the spectrum resembles a ``standard'' SN\,II. 

\textbf{Not an SN\,IIn}: The spectral data do not show a SN\,IIn, or were noisy or of otherwise poor quality, and a definitive classification cannot be made positively or negatively. In some cases, the available data do not appear to show a SN spectrum at all, or the available data may not exhibit any SN\,IIn features. For example, the spectrum may be contaminated by emission from the underlying \ion{H}{ii} region and the narrow H$\alpha$ feature can be erroneously interpreted as being caused by CSM interaction. 

While we do not use them as part of our classification, in order to highlight the ambiguities in SN\,IIn classification we compare the classifications stated in TNS and OSC to our own reclassifications and also to the results of the template-matching software, SuperNova IDentification (\texttt{SNID}). \texttt{SNID} calculates a quantity, \textit{rlap}, which describes the how well a template is matched to a given spectrum (a full description is given by \citealp{SNID}). The higher the \textit{rlap} value, the better the match. \texttt{SNID} has an \textit{rlap} cutoff of 5, where an \textit{rlap} value $<5$ indicates a poor match.

\section{Results} \label{results}

We obtained usable data for 115 SN\,IIn candidates either from publicly available databases or from the original observers. In \tableref{tab1} we present the breakdown of how many targets are in each category. We present spectra and H$\alpha$ profile Gaussian fits in \figureref{2008jTS} through \figureref{2002kgHa}. In \tableref{reasons} we outline the reasons some of our transients were classified as not being SNe\,IIn. \tableref{changingclass} contains the number of transients which do not have SN\,IIn primary classifications on the online databases, OSC and TNS. Finally, \tableref{snid} shows the number of transients that were classified as SNe\,IIn by \texttt{SNID}. Tables showing the transients we classify into each of our categories can be found in the supplemental online materials.

\begin{table} 
\caption{Breakdown of new categories for our sample of 115 transients.}
\label{tab1}
\begin{center}
\begin{tabular}{cc}
\hline
 Category &Number in sample \\
\hline\hline
 Gold &  37 \\  
 Silver &  50\\    
 Not SN\,IIn & 28 \\
 \hline
\end{tabular}
\end{center}
\end{table}

\subsection{SN\,IIn spectral categories}

In this section we present a selection of spectra from the new SN\,IIn categories described in  \sectionref{methods}. We show SN spectra and the accompanying H$\alpha$ profiles overlaid with the multi-Gaussian fits for each category.

\subsubsection{The gold sample}

From the set of 115 SN\,IIn candidates, we classify 37 as ``gold.'' These exhibit unambiguous CSM interaction over numerous epochs.

An example of a gold SN\,IIn is SN\,2008J in MCG-02-7-33, discovered on 2008 Jan.\ 15 (UT dates are used throughout this paper) with the 0.76\,m Katzman Automatic Imaging Telescope (KAIT) as part of the Lick Observatory Supernova Search \citep[LOSS;][]{Filippenko_2001} at a clear-filter magnitude of 15.4 \citep{Thrasher_2008}. Using the Spitzer Space Telescope, \citet{Fox_2010} detected late-time near-IR emission at 3.6$\,\mu$m and 4.5$\,\mu$m, which they interpret as emission from warm dust. \citet{Fox_2013} did not find late-time CSM interaction in optical spectra obtained with the LRIS and DEIMOS instruments on the Keck 10\,m telescopes.

\citet{Taddia_2012} present evidence that SN\,2008J may be a reddened thermonuclear SN\,IIn (thus, a SN\,Ia-CSM) with $A_V \approx 1.9$\,mag. They find that the near-IR light curves and optical spectra of SN\,2008J are similar to those of the thermonuclear SNe\,IIn (or SNe\,Ia-CSM) SN\,2002ic \citep{Hamuy_2003} and SN\,2005gj \citep{Prieto_2005}. Early-time high-resolution spectra show \ion{Na}{i}\,D $\lambda5891$ absorption features and blueshifted \ion{Si}{ii} $\lambda6355$ absorption.

\begin{figure*} 
\centering
  \includegraphics[width = \textwidth]{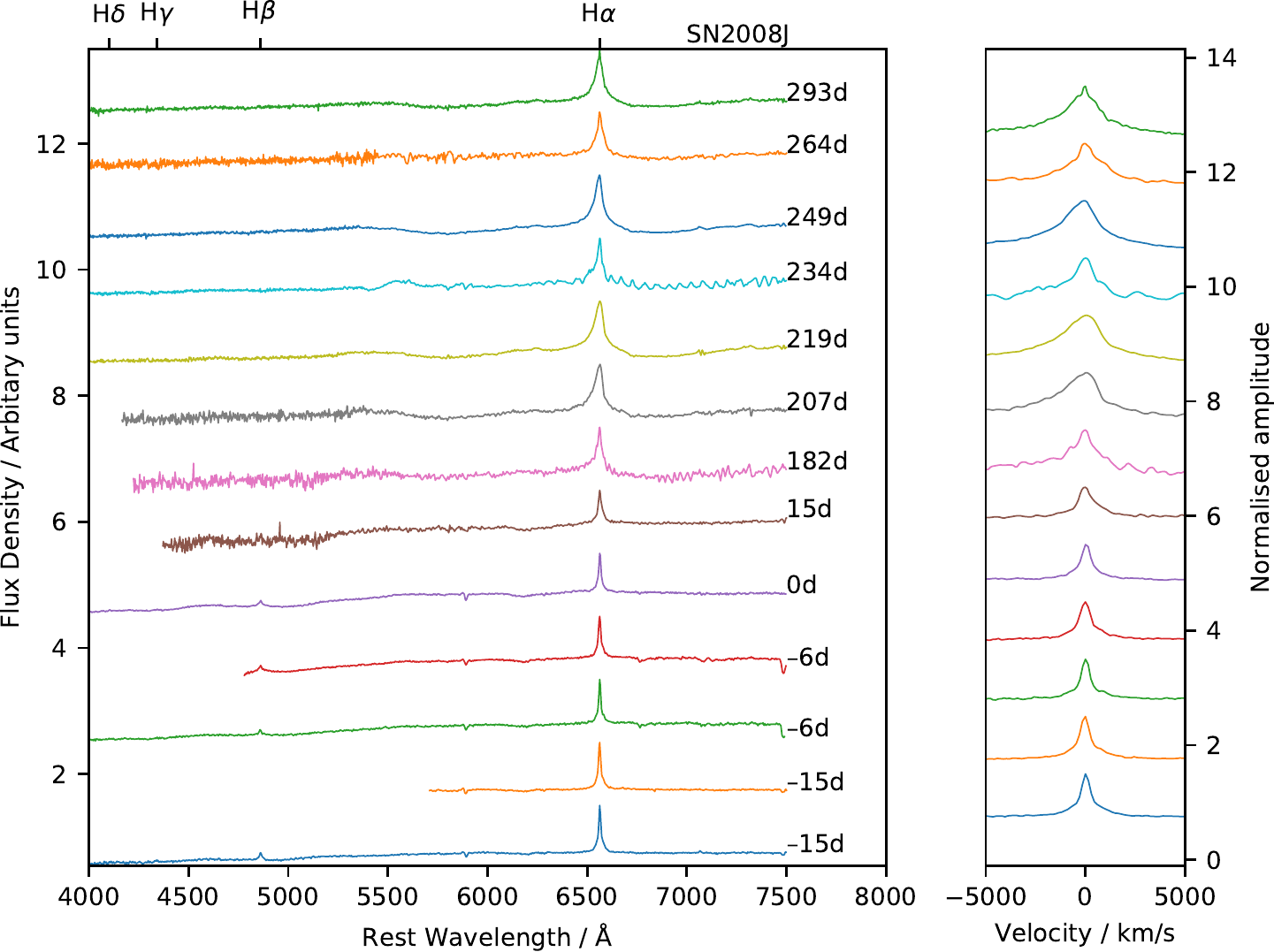}
  \caption{(\textit{Left}) Time-series spectra of an exemplar SN\,IIn -- SN\,2008J. We show optical spectra at 12 epochs over the range 4000--7500\,\AA\ that exhibit strong CSM interaction features. Each spectrum is shifted to the rest frame and plotted with a vertical offset, and the Balmer lines from H$\alpha$ through H$\delta$ are indicated with tick marks. The numbers are the days after maximum brightness (2008 Feb. 02) in the $i$ band. (\textit{Right}) We plot the corresponding H$\alpha$ profiles in velocity space with a vertical offset for clarity.}
  \label{2008jTS}
\end{figure*}

\begin{figure} 
\centering
  \includegraphics[width = 8cm]{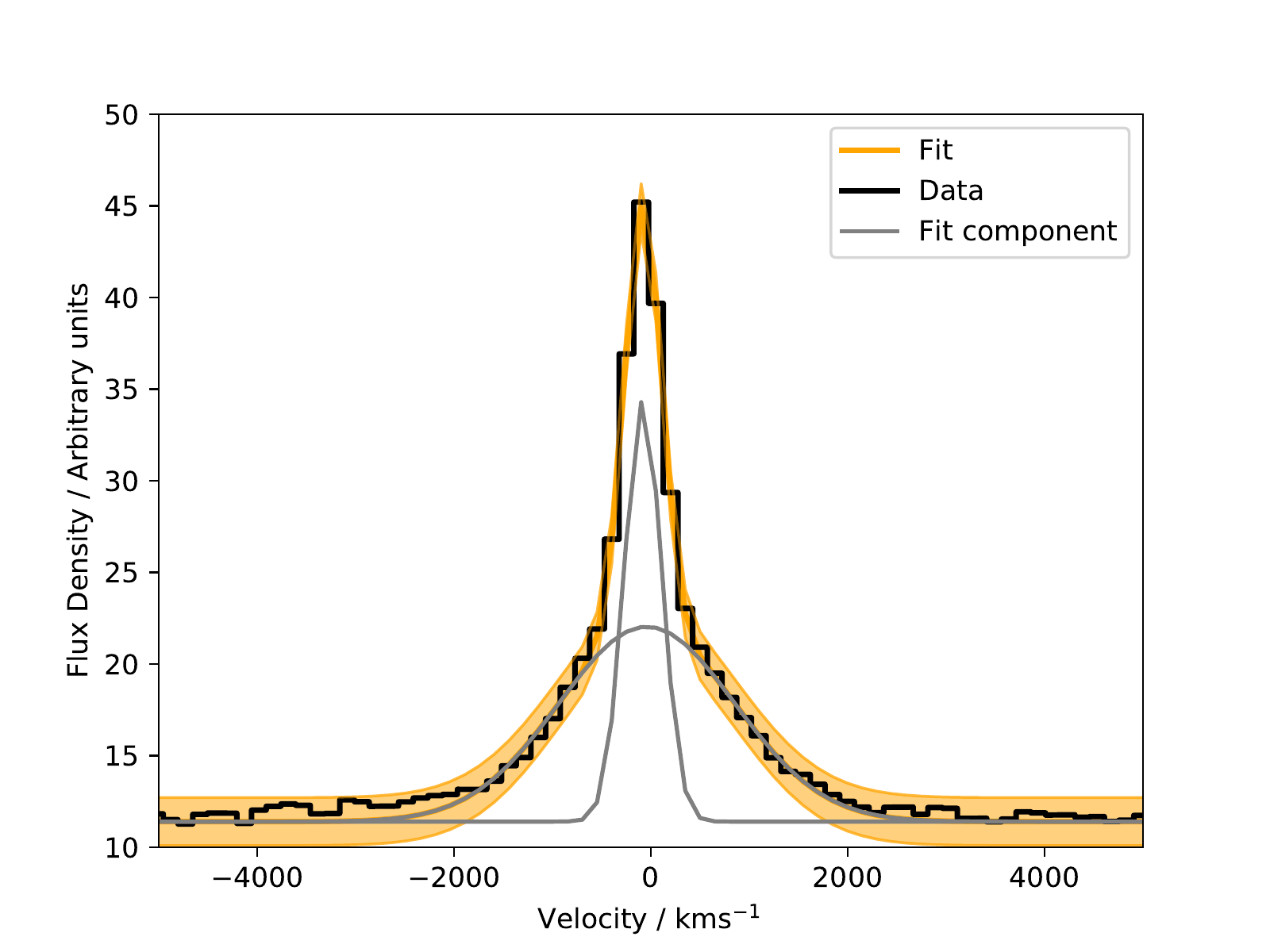}
  \caption{H$\alpha$ profile and the multi-Gaussian fits for the first spectral epoch (15\,days before $i$ band maximum) of SN\,2008J. We fit a broad and a narrow component to the data centred around rest-frame H$\alpha$. The thick black line is the data, the grey lines are the Gaussian components, and the shaded orange area is the total fit along with its 3$\sigma$ uncertainty region.}
  \label{2008jHa}
\end{figure}

We have twelve epochs of spectral data for SN\,2008J shown in  \figureref{2008jTS}. These data cover 311\,days throughout which the H$\alpha$ profile retains strong CSM interaction signatures. The first four spectra show \ion{Na}{i}\,D and some \ion{Si}{ii} absorption \citep{Taddia_2012} and later spectra become noisy. All epochs are fairly featureless or noisy apart from the Balmer lines and the \ion{Na}{i}\,D absorption.  \figureref{2008jHa} displays the H$\alpha$ profile for the first epoch of SN\,2008J. We obtain a good fit to the data from a pair of Gaussian profiles. The broader component has a FWHM of $\sim 2200$\,km\,s$^{-1}$ and the narrow component has FWHM $\approx 400$\,km\,s$^{-1}$. The profile has the classic ``Eiffel Tower'' shape. \texttt{SNID} classifies SN\,2008J as a SN\,IIn.

Our second example of a gold-class SN\,IIn is SN\,2009ip in NGC\,7259, which was discovered on 2009 Aug.\ 26 at mag 17.9 by \citet{Maza_2009} using the 0.41\,m PROMPT-3 telescope at Cerro Tololo Inter-American Observatory. SN\,2009ip is notable for being recognised as a SN ``impostor'' with a peak of about $-13.7$\,mag \citep{Berger_2009, Smith_2010}. Following this initial eruptive phase there were phases of rebrightening before a final eruption in 2012 which has been interpretted as a genuine SN\,IIn \citep[e.g.,][]{Mauerhan_2012, Prieto_2012, Pastorello_2013, Margutti_2014, Ofek_2014, Smith_2014} with a peak around $-18.3$\,mag. However, scenarios where the 2012 eruption was nonterminal are discussed in \sectionref{imps}.

During the eruptions in the initial ``impostor'' phase, it was reported that SN\,2009ip may be the result of LBV outbursts \citep{Miller_2009, Li_2009b}. Those authors and \citet{Nugent_2007} note that archival data from the {\it HST} and the Palomar Oschin Schmidt telescope reveal a transient coincident with SN\,2009ip. This may support the conclusions of \citet{Ofek_2014} that there are precursor eruptions to some SNe\,IIn. \citet{Foley_2011} found that there was a possible progenitor identified in archival {\it HST} data from 1999. \citet{Thoene_2015} reported that SN\,2009ip had become dimmer than its proposed progenitor by 2015 Nov.\ 29, which suggests that the 2012 eruption was terminal and SN\,2009ip had exploded as a genuine SN.

\begin{landscape}

\begin{figure} 
  \includegraphics{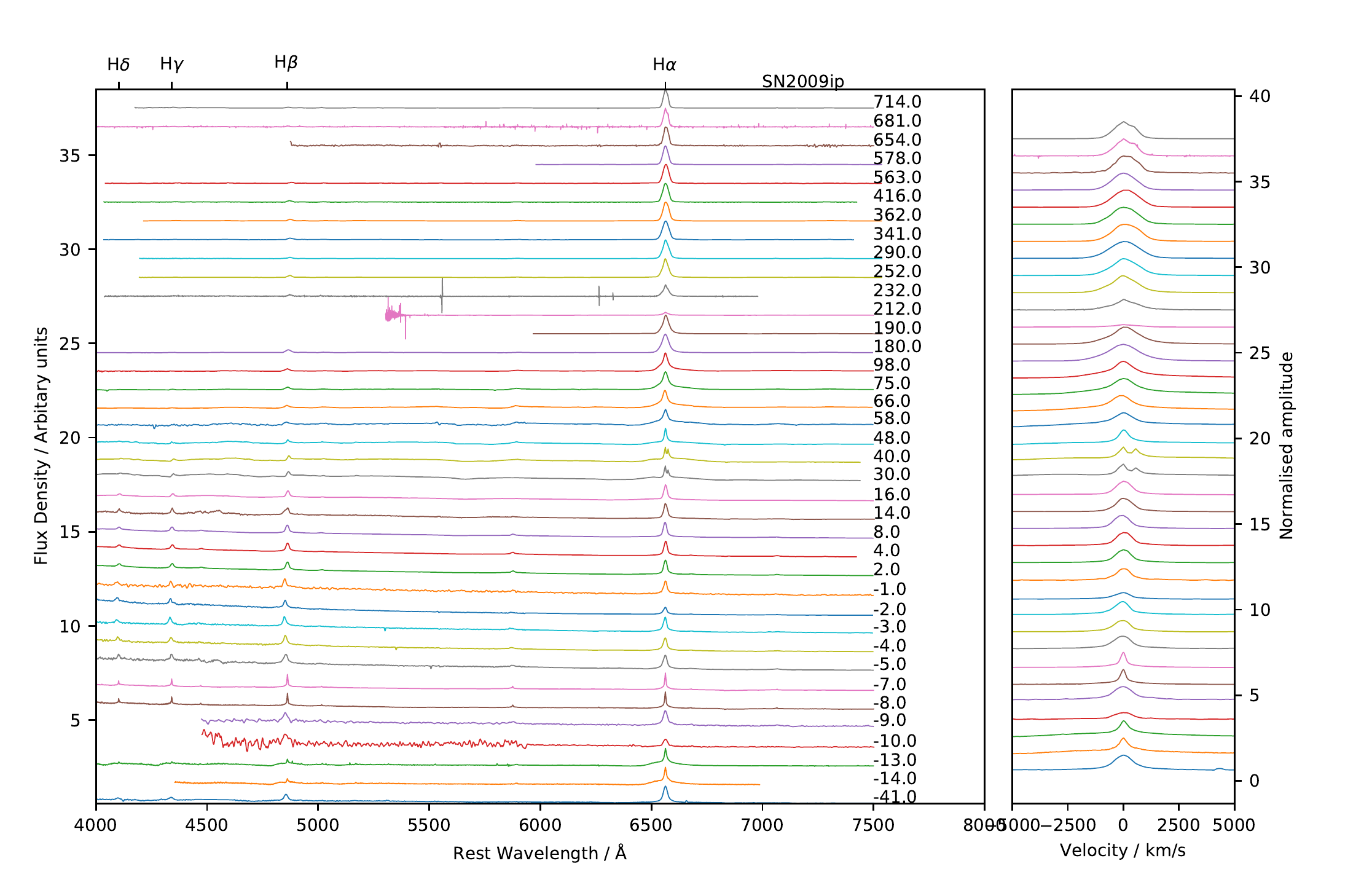} 
  \caption{Time-series spectra of an exemplar SN\,IIn -- SN\,2009ip in NGC\,7259. ($\it{Left}$) Spectra spanning 38 epochs over 755\,days. The dashed lines mark the Balmer series and there is a vertical offset between each spectrum for clarity. The numbers are the days from maximum brightness (2012 Jul. 24) in the {\it UVM2} band. ($\it{Right}$) H$\alpha$ profiles of each epoch in velocity space with an offset.}
  \label{2009ipTS}
\end{figure}

\end{landscape}

\begin{figure} 
\centering
  \includegraphics[width = 8cm]{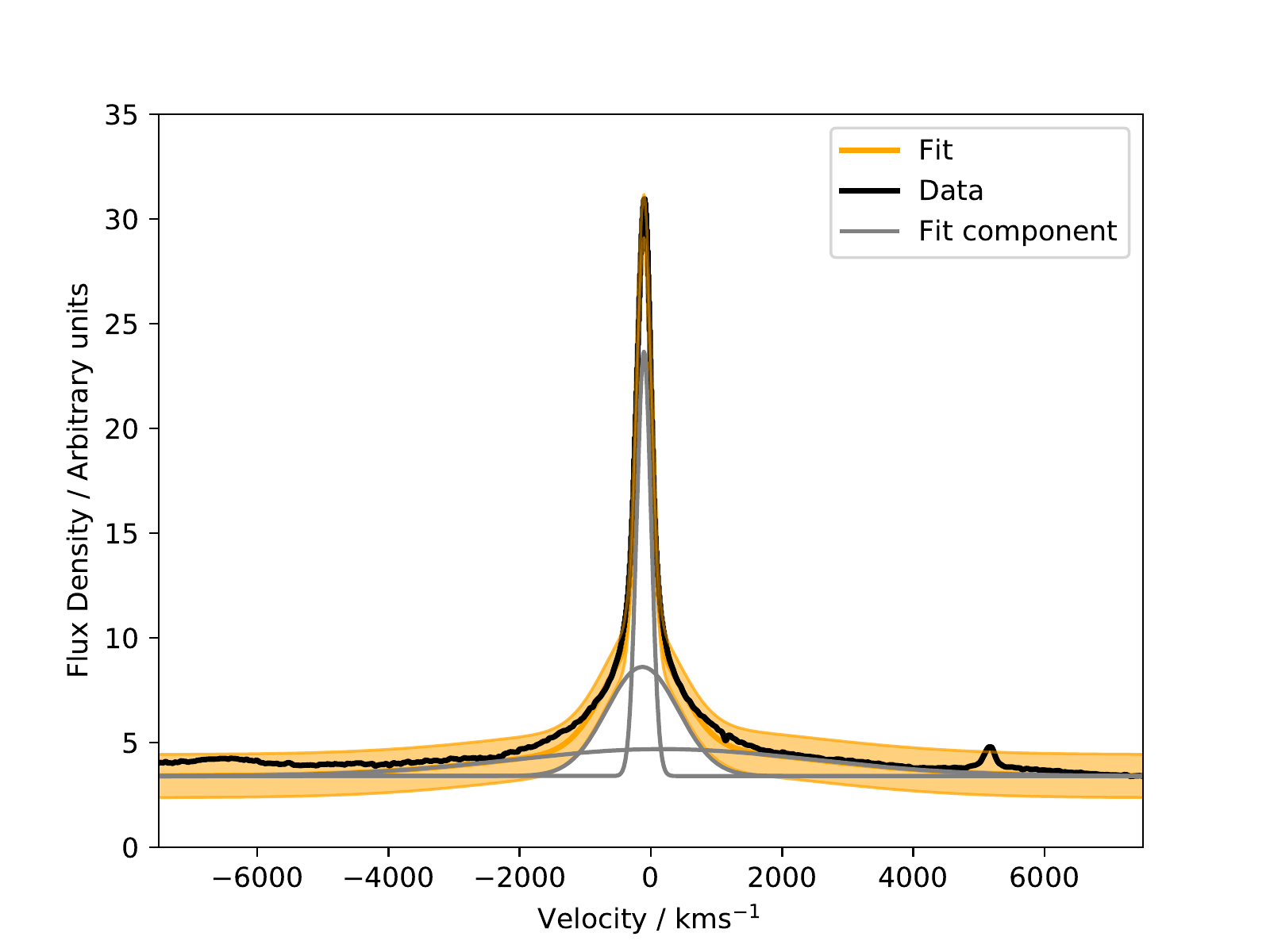}
  \caption{Multicomponent Gaussian fits to epoch seven (7\,days before {\it UVM2}-band maximum) of SN\,2009ip. We were able to fit three components to the data: a broad feature, an intermediate-width component, and a narrower component. The thick black line is the data, the grey lines are the Gaussian components, and the shaded orange area is the total fit along with its 3$\sigma$ uncertainty region.}
  \label{2009ipHa}
\end{figure}

We present the time-series spectra of SN\,2009ip in \figureref{2009ipTS}. 38 spectral epochs spanning 755\,days were collected. All epochs show prominent H$\alpha$ emission with the classic SN\,IIn profile shape. H$\beta$ is also visible at many of the epochs, and H$\gamma$ and H$\delta$ are seen at earlier epochs along with possible \ion{He}{i} $\lambda5876$ emission. Apart from these lines the spectra are featureless. We show the example H$\alpha$ profile at epoch seven of SN\,2009ip in \figureref{2009ipHa}. We fit three Gaussian components to this line, with the broader component having FWHM $\approx  5400$\,km\,s$^{-1}$, an intermediate component with FWHM $\approx  1300$\,km\,s$^{-1}$, and the narrow component having FWHM $\approx 300$\,km\,s$^{-1}$. \texttt{SNID} classifies SN\,2009ip as either an SN\,IIP or an AGN (although it most certainly is {\it not} an AGN).


\subsubsection{The silver sample}

Our dataset contains 50 SNe\,IIn in the silver category. This category is for objects having spectra that are consistent with SNe\,IIn but there is only a single spectrum available, or for those where the CSM interaction is short lived such as in the case of ``weak'' SNe\,IIn.

\begin{figure*} 
\centering
  \includegraphics[width = \textwidth]{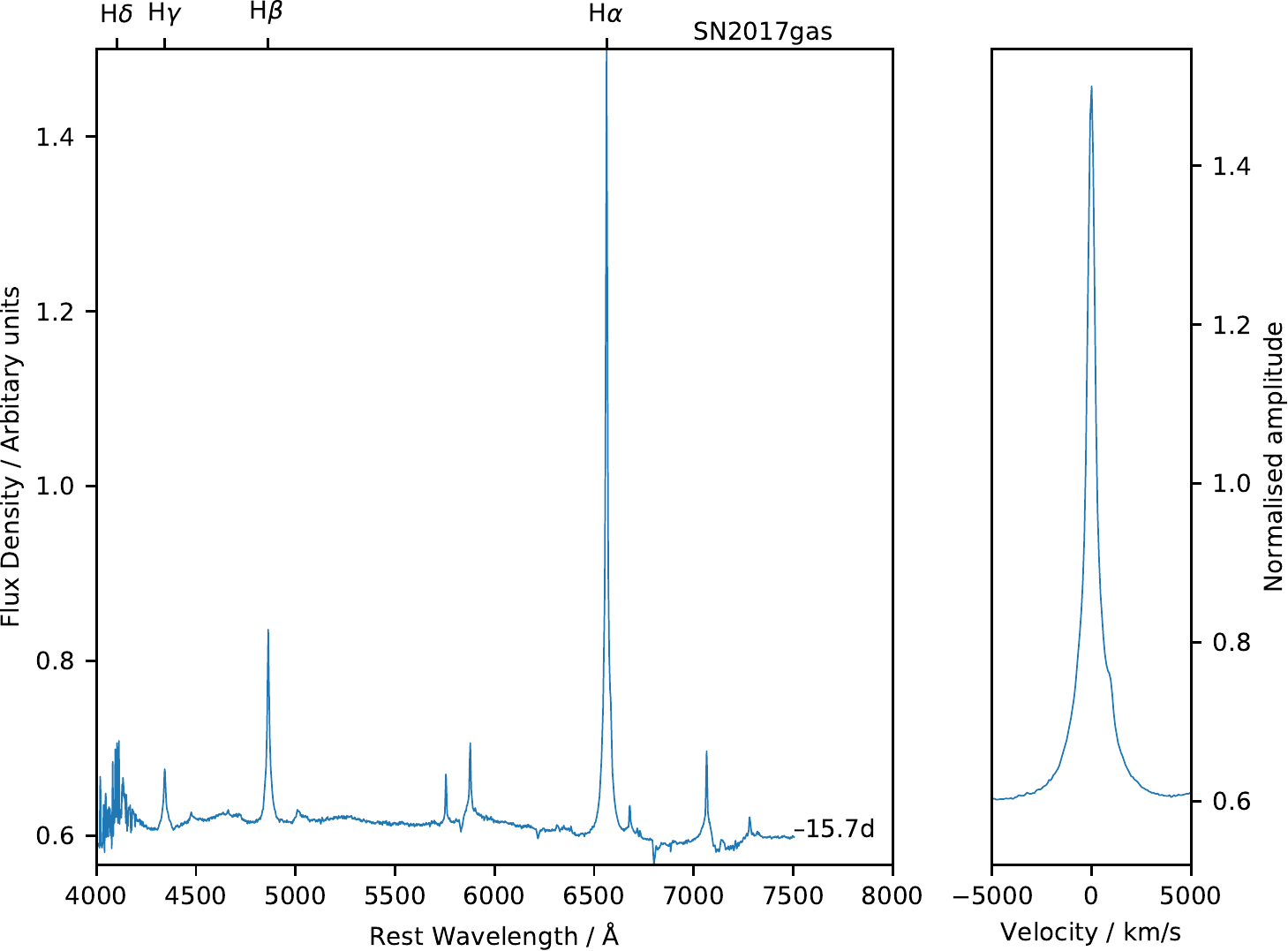}
  \caption{($\it{Left}$) Spectrum of an exemplar silver class SN\,IIn -- SN\,2017gas. We show a single-epoch optical spectrum (15.7\,days before maximum brightness in the $R$ band on 2017 Aug. 10) over the range 4000--7500\,\AA\ that shows strong CSM interaction features. The spectrum is shifted to the rest frame and the Balmer lines are indicated with tick marks.  ($\it{Right}$) The corresponding H$\alpha$ profile in velocity space.}
  \label{2017gasTS}
\end{figure*}

In \figureref{2017gasTS} we display the spectrum of the silver SN\,2017gas. It was discovered in 2MASX\,J20171114+5812094 on 2017 Aug.\ 17 at a $V$-band magnitude of $\sim 16$ as part of the All-Sky Automated Survey for SuperNovae (ASAS-SN) with 14\,cm telescopes in Chile and Hawaii \citep{Stanek_2017}. 
We classify SN\,2017gas in the silver category because there was only a single spectrum available publicly. It shows very strong CSM interaction features with the ``Eiffel Tower'' profile being very apparent in the H$\alpha$, H$\beta$, and H$\gamma$ emission lines. Furthermore, some non-Balmer lines seem to exhibit the narrow features expected from CSM interaction; we may see narrow features in \ion{Fe}{ii} $\lambda5750$ and \ion{Fe}{ii}  $\lambda5872$.

SN\,2017gas appears to be close to the centre of 2MASX\,J20171114+5812094, so we must distinguish this object from an AGN. The host offset of SN\,2017gas is $1.97''$ \citep[according to PS1;][]{PS1}, the host has a semimajor axis of $8.80''$, and \citet{2mass} report an uncertainity of $0.23''$ in the centre of the host right ascension and declination. Discovery images show the transient being off-centre\footnote{\url{https://star.pst.qub.ac.uk/ps1threepi/psdb/candidate/1201711351581208000/}}. The spectrum of SN\,2017gas does not show the strong [\ion{O}{iii}] $\lambda\lambda4959$, 5007 emission lines that may be associated with AGNs. Furthermore, \citet{Ward_2020} found that out of a sample of 5493 AGNs, only nine were offset from the galaxy centre, and the maximum observed offset was $1.649'' \pm 0.004''$. This suggests that the transient is likely an SN rather than an AGN.

\begin{figure} 
\centering
  \includegraphics[width = 8cm]{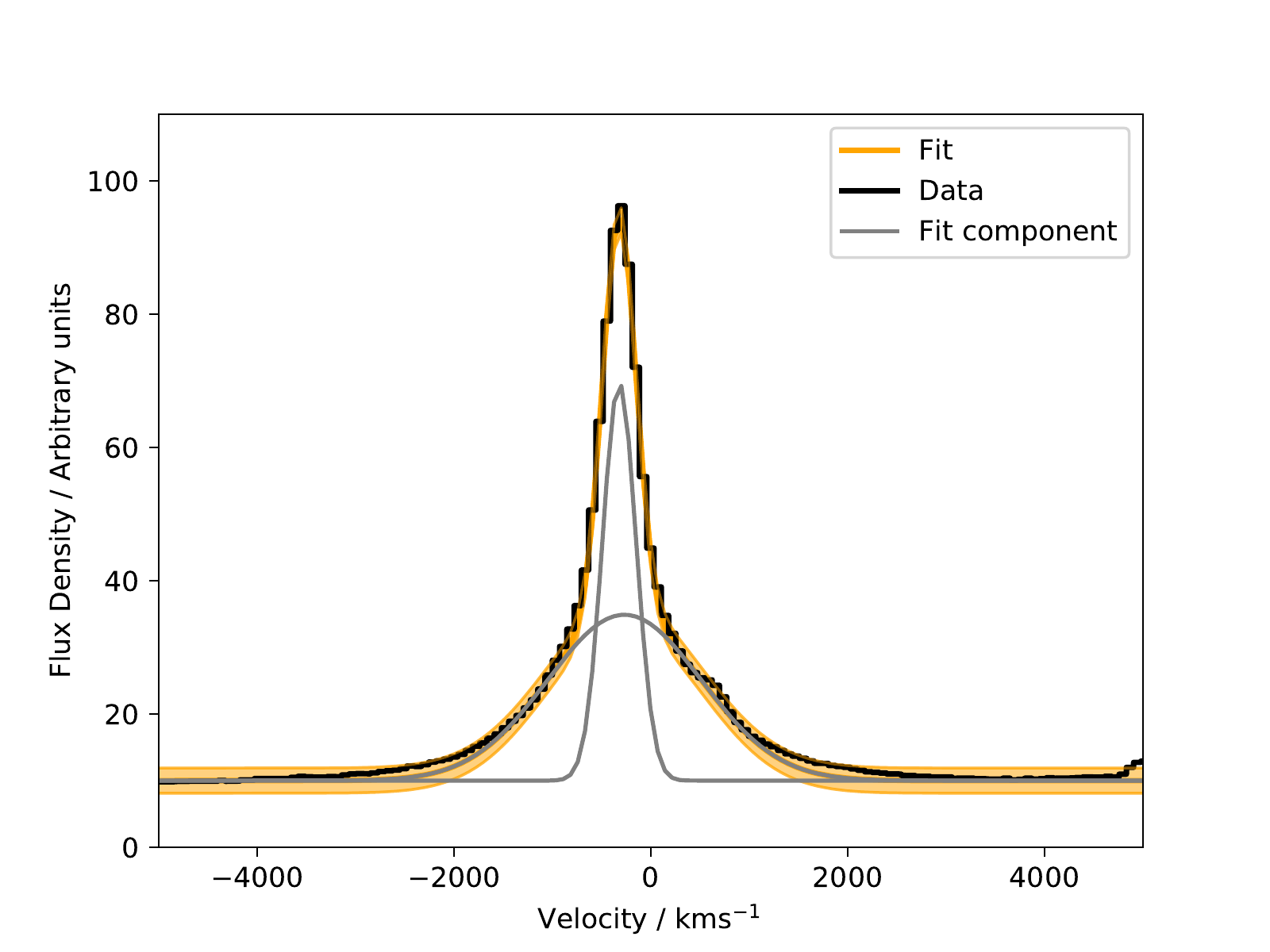}
  \caption{Multicomponent Gaussian fit to the only available spectrum of SN\,2017gas. We were able to fit two Gaussian components, one broad and the other narrower. The thick black line is the data, the grey lines are the Gaussian components, and the shaded orange area is the total fit along with its 3$\sigma$ uncertainty region.}
  \label{2017gasHa}
\end{figure}

The fit to the H$\alpha$ profile of SN\,2017gas is shown in \figureref{2017gasHa}. We were able to fit two Gaussians with the FWHM of the broad component being $\sim 1900$\,km\,s$^{-1}$ and FWHM $\approx400$\,km\,s$^{-1}$ for the narrow component. \texttt{SNID} matches the spectrum of SN\,2017gas to a galaxy template spectrum.

\subsubsection{The non-SN\,IIn spectra}

There are 28 objects that we classify as not belonging to the SN\,IIn class. This group includes SNe with spectra that cannot be positively identified as a SN\,IIn owing to a noisy spectrum, underlying \ion{H}{ii} region pollution, not showing SN-like features or CSM interaction restricted to early times suggesting flash ionisation.

An example of a transient that is not a SN\,IIn is SN\,2009kr in NGC\,1823. It was discovered on 2009 Nov. 6 with a 0.6\,m reflecting telescope at a magnitude of $\sim 16$ \citep{Nakano_2009}. \cite{Tendulkar_2009} obtained a spectrum of SN\,2009kr with the Palomar 5\,m telescope and used \texttt{SNID} to compare with standard templates. They concluded that the apparent prominent narrow feature in the H$\alpha$ emission line indicates that SN\,2009kr is a SN\,IIn. However, \citet{Steele_2009} analysed spectra of SN\,2009kr from the 3\,m Shane telescope at Lick Observatory and found that the narrow H$\alpha$ emission component is likely \ion{H}{ii} region pollution; the prior classification as an SN\,IIn is due to this and blending of the [\ion{N}{ii}] lines on either side of H$\alpha$. They subsequently classify SN\,2009kr as an ordinary SN\,II. 

Analysis of images by \citet{Elias-Rosa_2010} showed that SN\,2009kr may be a SN\,IIL; however, \citet{Fraser_2010} claim that the early-time light curve is more consistent with SN\,2009kr being a SN\,IIP. \citet{Elias-Rosa_2010} also found that the progenitor system of SN\,2009kr may be a yellow supergiant. \citet{Fraser_2010} concluded that the progenitor is ambiguous; if the progenitor was a single star it would be yellow supergiant, but the progenitor could not be ruled out as being in a cluster. The progenitor was quite luminous in {\it HST} images ($M_I \approx -8.5$\,mag). \citet{Li_2009} stated that the progenitor may be in a compact star cluster, and \citet{Maund_2015} suggested that the progenitor may be in a compact cluster with a mass of $\sim 6000$\,M$_\odot$ after analysis of late-time observations of SN\,2009kr.

\begin{figure*} 
\centering
  \includegraphics[width = \textwidth]{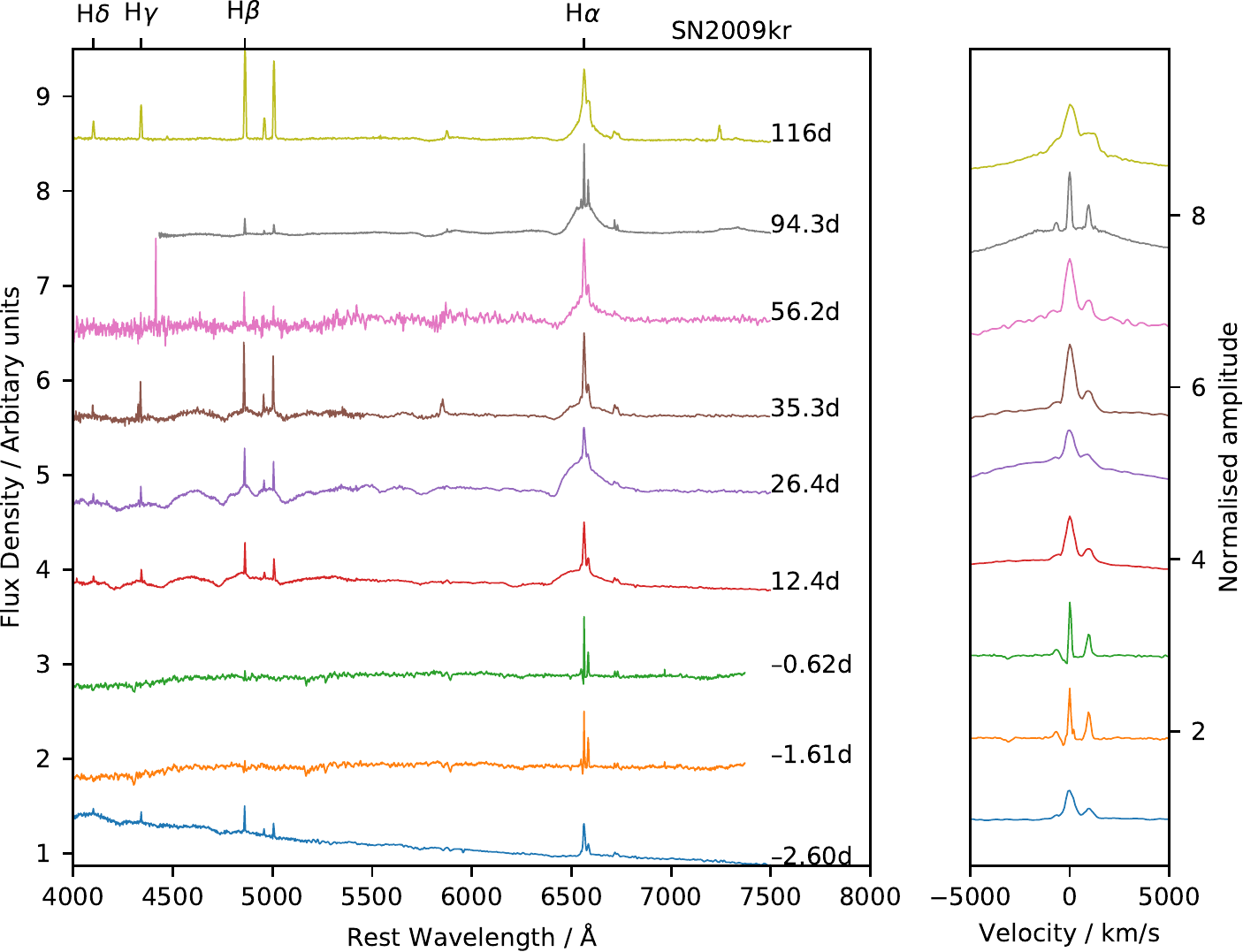}
  \caption{Temporal series of spectra of an example non-SN\,IIn classification -- SN\,2009kr. We display optical spectra over the range 4000--7500\,\AA\ at nine epochs, showing that the previously claimed CSM interaction features are likely due to the spectra being polluted by emission lines from the underlying \ion{H}{ii} region; the [\ion{N}{ii}] lines on either side of H$\alpha$ and the [\ion{S}{ii}] lines redward of H$\alpha$ are clearly visible at all epochs. This object therefore may be a standard SN\,II rather than an SN\,IIn. Each spectrum is shifted to the rest frame, a vertical offset is added for clarity, and the Balmer series from H$\alpha$ through H$\delta$ is indicated with tick marks. The numbers on the right of the spectra denote the days from maximum brightness (2009 Nov. 6) in the $K$ band. ($\it{Right}$) Corresponding H$\alpha$ profiles in velocity space, with vertical offsets.}
  \label{2009krTS}
\end{figure*}

\begin{figure} 
\centering
  \includegraphics[width = 8cm]{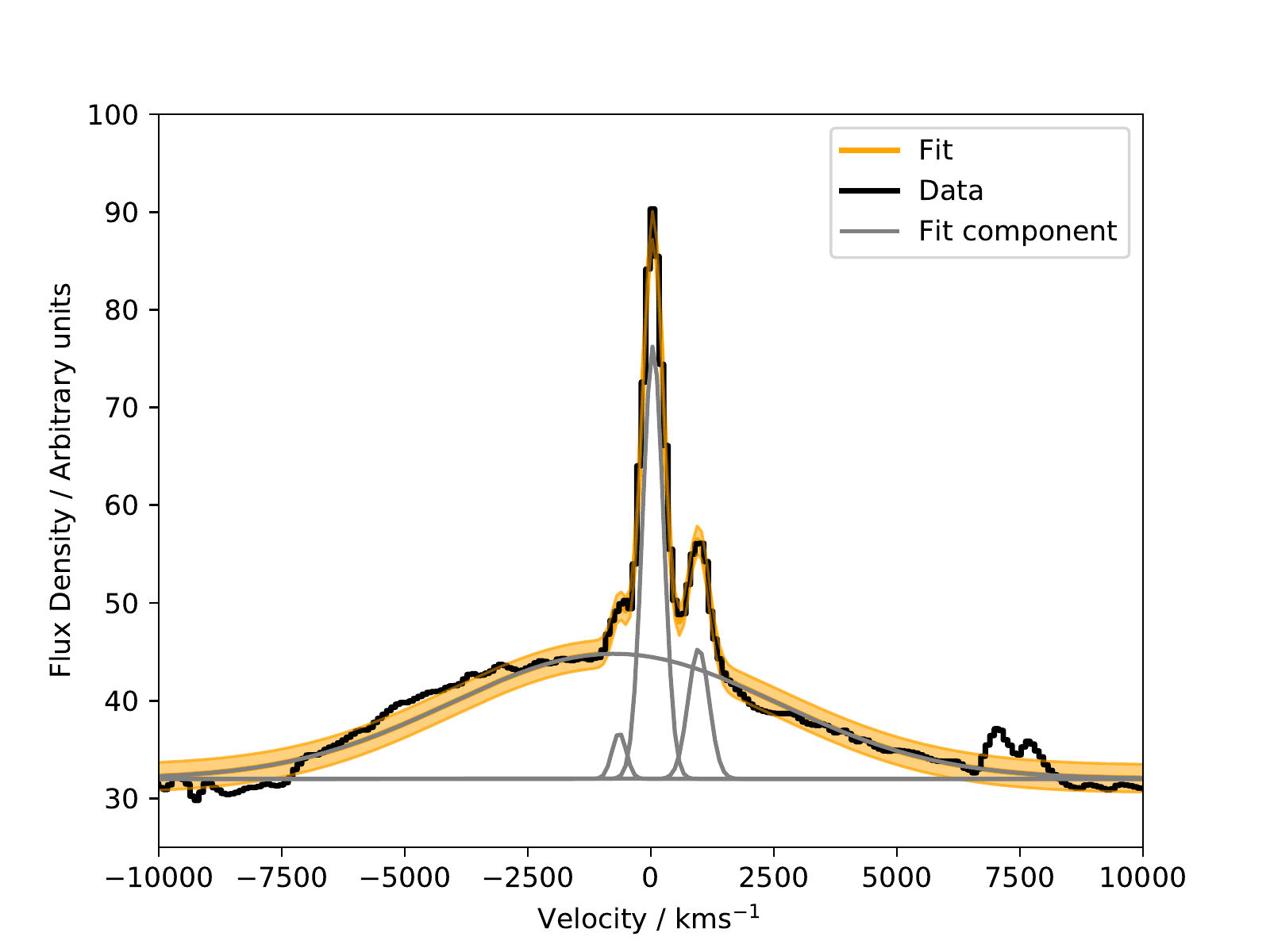}
  \caption{Multicomponent Gaussian fits to the fourth epoch of SN\,2009kr just over 12\,days past maximum brightness in the $K$ band. We were able to fit a broad component to the H$\alpha$ profile along with three narrow components at H$\alpha$ and [\ion{N}{ii}] $\lambda\lambda6548$, 6584. The [\ion{S}{ii}] $\lambda\lambda6717$, 6731 lines are also visible. The presence of these lines indicates that the narrow features seen in this H$\alpha$ profile probably originate from the underlying \ion{H}{ii} region. The thick black line is the data, the grey lines are the Gaussian components, and the shaded orange area is the total fit along with its 3$\sigma$ uncertainty region.}
  \label{2009krHa}
\end{figure}

We display a temporal series of spectra of SN\,2009kr in  \figureref{2009krTS}. At early times (the three epochs before maximum brightness) there are narrow Balmer emission lines. A narrow feature superimposed upon the H$\alpha$ line persists over all observed epochs. The post-maximum spectra show a broadening H$\alpha$ line with a P\,Cygni profile. In  \figureref{2009krHa} we show the fourth epoch of SN\,2009kr (12.4\,days post-maximum); this is when the H$\alpha$ emission line starts to broaden with a FWHM of $\sim 7500$\,km\,s$^{-1}$. We fit narrow features at H$\alpha$ and also [\ion{N}{ii}] $\lambda\lambda6548$, 6584, consistent with the spectrum being contaminated by underlying emission from the local \ion{H}{ii} region. Furthermore, one can see probable emission features of [\ion{S}{ii}] $\lambda\lambda6717$, 6731, supporting the argument that the narrow features previously interpreted as being due to CSM interaction are actually line pollution from the local interstellar medium. Our analysis is therefore consistent with that of \citet{Steele_2009}. Despite the evidence that suggests SN\,2009kr is not a SN\,IIn, the OSC has the primary classification of SN\,2009kr as an SN\,IIn. \texttt{SNID} classifies SN\,2009kr as an SN\,IIP.

\begin{figure*} 
\centering
  \includegraphics[width = \textwidth]{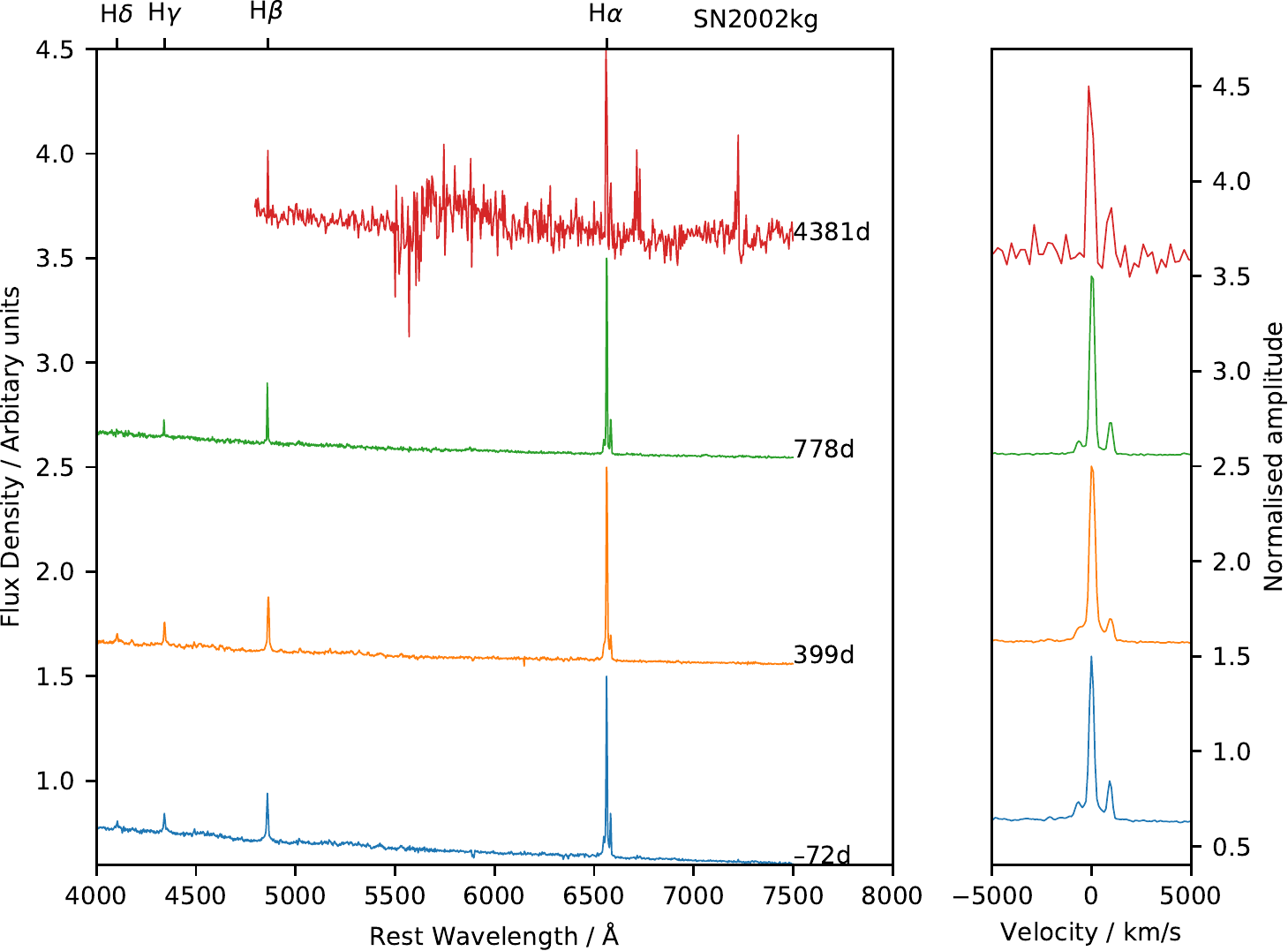}
  \caption{Time-series spectra of an example of a negative SN\,IIn classification - SN\,2002kg. We show optical spectra from 4000-7500\,\AA\, over four epochs that are mostly featureless apart from Balmer emission and [\ion{N}{ii}]. The H$\alpha$ line is strong and narrow and the H$\beta$ line is also apparent in these spectra. Each epoch is shifted to the rest-frame, an offset is added for clarity and the Balmer series from H$\delta$-H$\alpha$ is marked with a tick. The numbers are the days from maximum (2002 Oct. 26) in the Clear band. ($\it{Right}$) Corresponding H$\alpha$ profiles in velocity space with offsets.}
 \label{2002kgTS}
\end{figure*}

\begin{figure} 
\centering
  \includegraphics[width = 8cm]{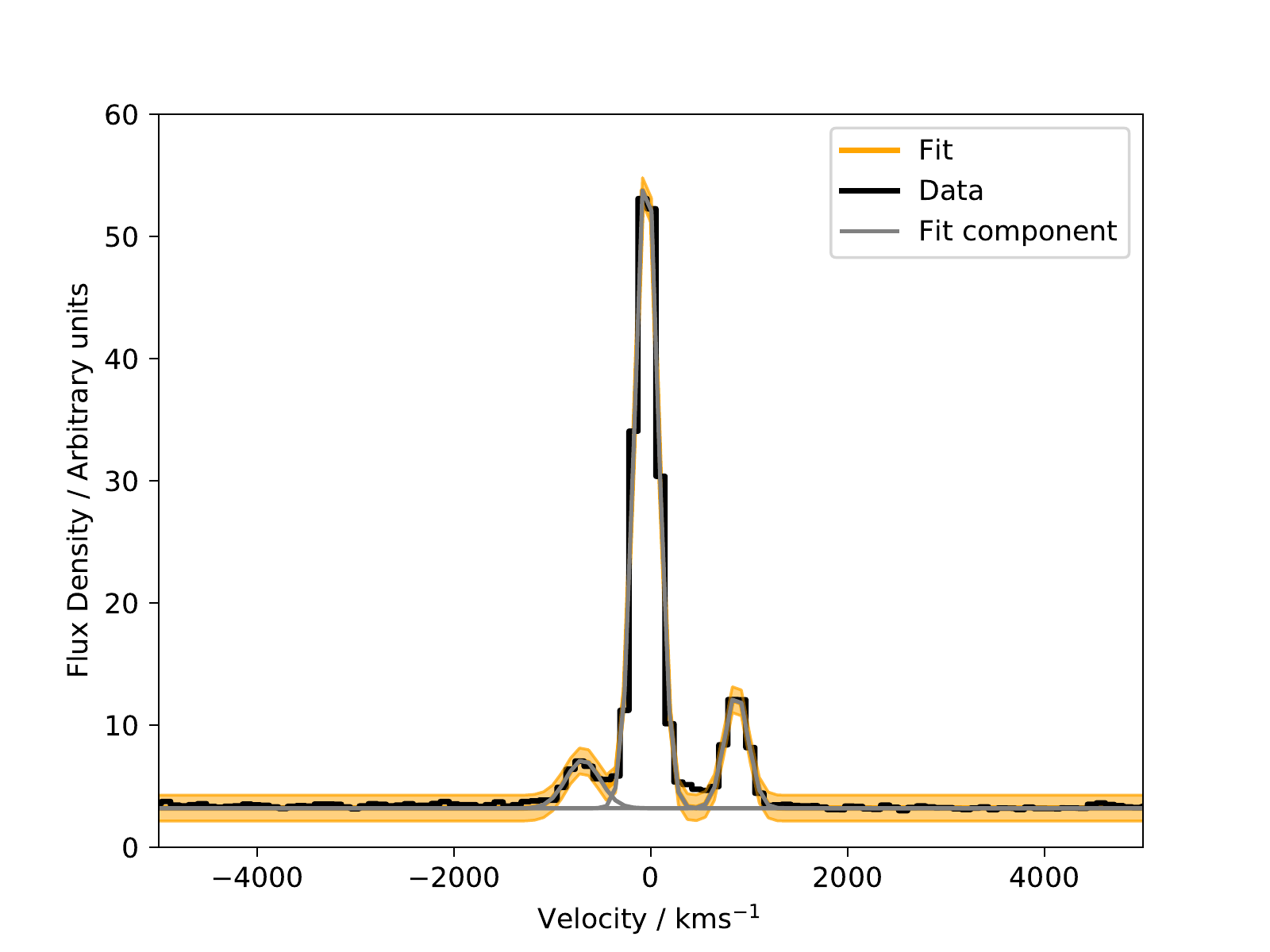}
  \caption{Multicomponent Gaussian fits to the H$\alpha$ profile of the third epoch (778\,days post maximum brightness in the Clear band) of SN\,2002kg. We fit a narrow H$\alpha$ profile with FWHM $\approx 300$\,km\,s$^{-1}$. [\ion{N}{ii}] lines are also visible and may indicate that SN\,2002kg is an SN ``impostor'' where the transient is the brightening of an LBV due to a mass-loss event. The thick black line is the data, the grey lines are the Gaussian components, and the shaded orange area is the total fit along with its 3$\sigma$ uncertainty region.}
  \label{2002kgHa}
\end{figure}

Another example is SN\,2002kg in NGC\,2403. SN\,2002kg was reported on 2006 Oct.\ 26 with KAIT as part of LOTOSS (the Lick Observatory and Tenagra Observatory Supernova Search), with a discovery magnitude of $\sim 19.0$ \citep{Schwartz_2003}. Subsequent spectroscopic observations were carried out by \citet{Filippenko_2003} using the Keck Observatory. We show the temporal series of spectra of SN\,2002kg in  \figureref{2002kgTS}, covering four epochs over 4309\,days. \figureref{2002kgHa} displays the H$\alpha$ profile of SN\,2002kg, overlaid with Gaussian fits to the data. The spectra are mostly featureless apart from some Balmer emission and [\ion{N}{ii}] $\lambda\lambda6548$, 6584 lines. SN\,2002kg was initially classified as an SN\,IIn owing to the narrow H$\alpha$ feature ($\sim 300$\,km\,s$^{-1}$) and a possible broader component. However, \citet{Filippenko_2003} drew comparisons to gap transients such as SN\,2000ch \citep{Wagner_2000}. They noted that the forbidden nitrogen lines, absence of any other forbidden lines, and an absolute magnitude of about $-9$ point toward a nitrogen-enriched wind or CSM consistent with the outburst of a massive star such as an LBV rather than an SN. However, these authors did not rule out an SN explosion until there is further evidence that this transient is an ``impostor.''

KAIT photometry, Keck spectroscopy, archival data, and subsequent observations were used to constrain the properties of SN\,2002kg. It was found that that the outburst is less energetic than most other ``impostor'' events and that the progenitor is the LBV V37 \citep{Tammann_1968}. It has been reported that the outburst may be forming an LBV nebula, with the eruption being consistent with an S~Dor LBV phase and a progenitor mass of 60--80\,M$_\odot$ \citep{Weis_2005, VanDyk_2006, Maund_2006, Humphreys_2017}. There is no multicomponent H$\alpha$ feature present in the spectra, and the spectra are not at an early stage, so we classify SN\,2002kg as not being an SN\,IIn. \texttt{SNID} classifies SN\,2002kg as an AGN (though it certainly was {\it not} an AGN).

\begin{figure*} 
\centering
  \includegraphics[width = \textwidth]{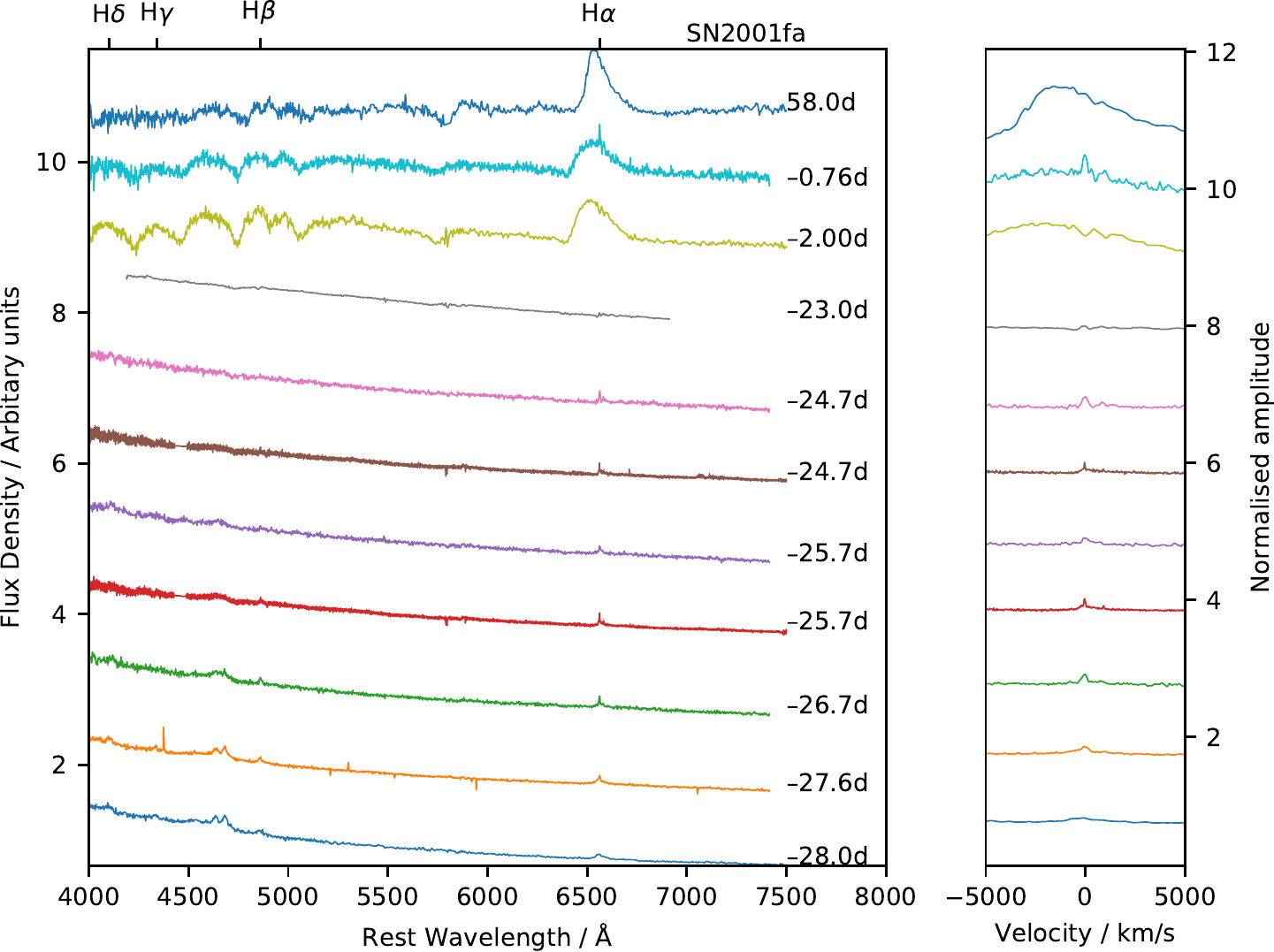}
  \caption{Time series of spectra of an exemplar silver-class SN\,IIn -- SN\,2001fa. We display optical spectra over the range 4000--7500\,\AA\ at nine epochs that show CSM interaction features at earlier times, followed by more-standard SN\,II characteristics with broad H$\alpha$ emission and possible P\,Cygni features in several lines. Each spectrum is shifted to the rest frame, a vertical offset is added for clarity, and the Balmer series from H$\alpha$ through H$\delta$ is indicated with tick marks. The numbers are the days from maximum (2001 Nov. 17) in the Clear band. ($\it{Right}$) Corresponding H$\alpha$ profiles in velocity space, with vertical offsets.}
  \label{2001faTS}
\end{figure*}

A third example of a transient we do not classify as an SN\,IIn is SN\,2001fa in NGC\,673, which was discovered on 2001 Oct.\ 18 with KAIT as part of LOTOSS, with a discovery magnitude of $\sim 16.9$. Later, a spectrum obtained with the Lick 3\,m Shane telescope revealed that SN\,2001fa was a young SN\,IIn that also exhibited WR features including \ion{He}{ii}, \ion{C}{iii}, and \ion{N}{iii} emission lines \citep{Papenkova_2001}. There are spectra at eleven epochs spanning 86\,days that we present in \figureref{2001faTS}. The early-time spectra are mostly featureless with the exception of the Balmer lines, in particular H$\alpha$, as is characteristic of SNe\,IIn. SN\,2001fa exhibited signatures of CSM interaction only visible for $\sim 5$\,days and the spectrum subsequently resembled a more standard SN\,II with a broadened H$\alpha$ profile. There may be a narrow feature on the tenth epoch (0.76\,days before maximum brightness), but it could be due to contamination from an \ion{H}{ii} region. We therefore do not classify SN\,2001fa as an SN\,IIn, as the early-time CSM interaction is representative of flash ionisation and later spectra are consistent with those of a standard SN\,II. Light-curve analysis by \citet{Faran_2014} presents SN\,2001fa as part of a sample of SNe\,IIL. \texttt{SNID} classifies SN\,2001fa as evolving from an SN\,IIn to an SN\,IIP.

\begin{figure} 
\centering
  \includegraphics[width = 8cm]{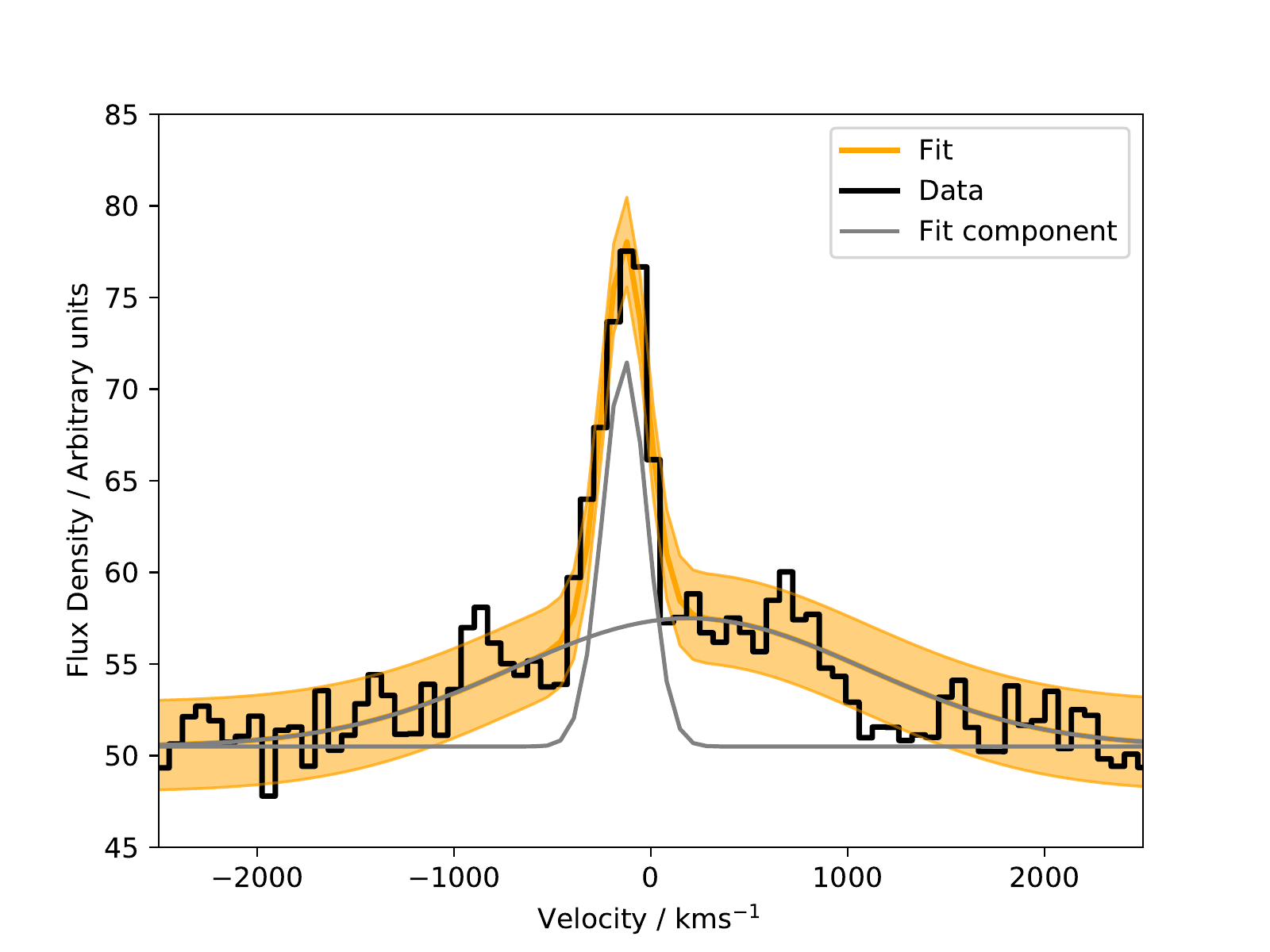}
  \caption{Multicomponent Gaussian fits to the fourth epoch (25.7\,days before maximum brightness in the Clear band) of SN\,2001fa. We were able to fit two Gaussian components to the data, one intermediate-broad and the other narrow. The thick black line is the data, the grey lines are the Gaussian components, and the shaded orange area is the total fit along with its 3$\sigma$ uncertainty region.}
  \label{2001faHa}
\end{figure}

 \figureref{2001faHa} shows the H$\alpha$ profile for the fourth epoch of SN\,2001fa. We fit two components to it, with the broad component having FWHM $\approx 1500$\,km\,s$^{-1}$ and the narrow component having FWHM $\approx 100$\,km\,s$^{-1}$.

\begin{table}\caption{Objects that are not SNe\,IIn and the reasons for classifications.} 
\label{reasons}
\begin{center}
\begin{tabular}{cc}
\hline
 Name & Reason for a not SN\,IIn classification \\
\hline\hline

 SN\,2001dk & Consistent with \ion{H}{ii} region contamination \\
 SN\,2001fa & Flash-ionisation SN \\
 SN\,2004F & Consistent with \ion{H}{ii} region contamination  \\
 SN\,2001I & Consistent with \ion{H}{ii} region contamination\\
 SN\,2002bj & Consistent with \ion{H}{ii} region contamination\\
 SN\,2009kr & Consistent with \ion{H}{ii} region contamination \\
 SN\,2004gd & Consistent with \ion{H}{ii} region contamination\\
 SN\,2002kg & Gap transient\\
 SN\,1999gb &Consistent with \ion{H}{ii} region contamination\\
 SN\,2010al &Consistent with \ion{H}{ii} region contamination\\
 SN\,2002an & No narrow feature \\
 SN\,2003ke & Consistent with \ion{H}{ii} region contamination\\
 SN\,1999bw & Gap transient\\
 SN\,2000ch & Gap transient \\
 SN\,2014G & Flash-ionisation SN\\
 SN\,2001ac & Gap transient\\
 SN\,1997bs & Gap transient\\
 SN\,2002bu & Gap transient \\
 SN\,2017jfs & Gap transient \\
 SN\,2006bv & Gap transient \\
 SN\,2007cm & Flash-ionisation SN \\
 SN\,2001dc & No narrow feature \\
 ASASSN-15hs &Consistent with \ion{H}{ii} region contamination\\
 SN\,2013by &Consistent with \ion{H}{ii} region contamination\\
 SN\,2001ad & No narrow feature \\
 SN\,2018dfy & Consistent with \ion{H}{ii} region contamination\\
 SN\,2006fp & Gap transient \\
 SN\,2008gm &Consistent with \ion{H}{ii} region contamination\\

\hline
\end{tabular}
\end{center}
\end{table}

We find that 28 of our objects are not SNe\,IIn. In \tableref{reasons} we show that out of these, twelve may have been misclassified owing to \ion{H}{ii} region lines being interpreted as CSM interaction. The remaining nine were gap transients that have retained their SN\,IIn classification in OSC and/or TNS.

\subsubsection{Changing classifications}

The motivation for this study was to provide a catalogue of systematically reclassified SNe\,IIn, as the classifications reported in public databases can be incorrect. We compare our classifications to those recorded as the main class stated by the OSC and the TNS, shown in \tableref{changingclass}. Furthermore, we note that of the 115 SNe in our sample, TNS did not give a classification or spectral data for twelve objects.

\begin{table} 
\caption{Breakdown of the number of objects in the OSC and TNS having non-SN\,IIn primary classifications (out of 115) and the number of occurrences where the OSC and TNS disagree on the primary classification.}
\label{changingclass}
\begin{center}
\begin{tabular}{ llll }
\hline
 Class & OSC& TNS& Disagreement \\
\hline\hline
Gold&16&6&16 \\
Silver&18&6&18 \\
Not a SN\,IIn&16&10&15 \\
\hline
\end{tabular}
\end{center}
\end{table}

\subsubsection{\texttt{SNID} results}

In \sectionref{snid}, we present the results of the template matching using \texttt{SNID}.
\texttt{SNID} was able to match 32 of our targets to template SNe\,IIn. In the instances where there was no match to an SN\,IIn spectrum, we note that there are 26 matches to AGN spectra and 39 have matches to other SN classes. We find that 18 match galaxy spectra and five match the spectra of LBV templates. We also find that six of our objects have spectra which evolve into/out of the SN\,IIn subclass according to \texttt{SNID}. 

\begin{table} 
\caption{Results of using \texttt{SNID} to match the spectral data to templates. 
In total, \texttt{SNID} matched 32 of our 115 targets to SN\,IIn templates.}
\label{snid}
\begin{center}
\begin{tabular}{ ll }
\hline
 Class & \texttt{SNID} IIn matches \\
\hline\hline
Gold&15 out of 37 \\
Silver&14 out of 50 \\
Not SN\,IIn&3 out of 28 \\
\hline
\end{tabular} 
\end{center}
\end{table}

\section{Discussion and Summary} \label{discuss}

In this work, we have  explored the spectral diversity in the SN\,IIn subclass. We spectroscopically reclassified 115 objects within $z<0.02$ that have at some point been classified as an SN\,IIn in the OSC. The main result is that 40\% of the SNe in our sample had been misclassified when one takes account of the primary classification in OSC. Furthermore, $\sim 25$\% of the objects in our sample are not SNe\,IIn when considering our classification scheme. Moreover, for objects that are not SNe\,IIn, eleven or 18 (OSC or TNS, respectively) claim a primary SN\,IIn classification. We also find that there is disagreement in the classification for 45\% of our sample between the OSC and TNS. Therefore, any work based on such catalogues of SNe\,IIn could have included SNe which are misclassified or miss some targets.

This disagreement is clearly problematic for work that relies upon samples and catalogues of SNe. Our work was initiated when it was noticed that when an individual SN was searched in databases, the claimed class may differ between databases and also changes in the OSC depending on search parameters. Without our reclassification, the SN\,IIn catalogue would contain at least 28 targets that do not appear to be SNe\,IIn, possibly leading to erroneous conclusions in subsequent studies. Confusion in the classification of SNe\,IIn may come from a number of sources that contribute to the diversity in the SN\,IIn subclass.

As described in \sectionref{sec:intro}, some of this diversity may be explained by considering multiple progenitor paths. The proposed progenitors of SNe\,IIn often quoted in the literature are LBVs. The environmental analysis of \citet{hab14} shows the SF environments of SNe\,IIn are not consistent with high-mass LBVs as the sole progenitor path, suggesting that there may be lower-mass progenitor routes. Further to this, the stellar-evolution models of \citet{Maeder_2008} show that LBVs must become WR stars before a SN eruption, and some examples from our sample (such as SN\,2001fa) show WR lines in their spectra. However, \citet{Groh_2013} find that when a rotational component is added to evolution models, LBVs may terminate in a SN explosion without requiring a WR phase.

The mass-loss mechanism of LBVs is poorly constrained. \citet{Dwarkadas_2011} find that previous assumptions regarding the LBV mass-loss rate being constant may be flawed, as this does not account for the density of CSM required for the SN\,IIn-like features to be apparent. Instead, the mass-loss mechanism may be episodic eruptions preceding the SN, and the progenitor may have transitioned to a WR star by the time the SN occurs. This is consistent with \citet{Ofek_2014}, who find that mass-loss events which cause brightening are common in SN\,IIn progenitors; they are seen in around half of a sample of SNe\,IIn observed by PTF. Furthermore, \citet{Strotjohann_2020} find that month-long precursor events brighter than an absolute magnitude of $-13$ were found in $\sim 25$\% of a sample of 131 SNe\,IIn from ZTF within the three months prior to the SN explosion.

As well as this single-star route to LBVs, there are proposed binary-evolution paths for LBVs or LBV-like objects. \citet{Sana_2012} found that most massive stars evolve in binary systems ($\sim 70$\% of the O-type stars in their sample were in a binary). \citet{Tombleson_2015} showed that LBVs are often isolated, away from regions of ongoing star formation. As discussed in \sectionref{lbv}, this may be due to the stars being kicked from their original cluster through interactions with companions or stellar mergers \cite{Justham_2014, Aghakhanloo_2017}. 

Since the CSM interaction that causes the SN\,IIn phenomenon is an environmental effect, we include SNe\,IIn that may be SNe\,Ia-CSM or flash-ionisation SNe in our analysis; they might be counted in the lower-mass progenitor channel, or in the case of the flash ionisation SNe not be SNe\,IIn at all. In our sample, it is possible that some of the silver SNe\,IIn may show short-lived flash-ionisation behaviour where the progenitor undergoes mass loss shortly before the SN explosion, creating a confined CSM shell \citep{Khazov_2016}. For example, SN\,2007cm may exhibit a flash-ionisation spectrum on day 0, when there is a SN\,IIn-like H$\alpha$ profile, but by day 15 this narrow feature has faded and the spectrum represents a fairly standard SN\,II spectrum. When there is data to show the transition to a standard SN\,II, we do not classify it as a SN\,IIn. \citet{Khazov_2016} have the ``weak'' SN\,IIn PTF\,11iqb in their sample of flash ionisation SNe. While many flash ionisation SNe evolve into standard SNe\,II with no subsequent CSM interaction \citep{Bruch_2020}, others exhibit CSM interaction at later times, such as PTF\,11iqb \citep{Smith_2015}. Moreover, this flash-ionisation phase is similar to the spectra of many young SN\,IIn, with a mostly featureless continuum and narrow CSM interaction signatures on the H$\alpha$ emission line we. We therefore can not rule out young SNe with single spectral epochs, they are classified in the silver category. The CSM around SNe\,IIn is complex, and in order to better understand the nature of these objects, longer-term observations are needed to determine whether CSM interaction re-emerges. These weaker SNe\,IIn may indicate that the mass loss required for the SN\,IIn-like features have a continuum starting from ``normal'' SN progenitors such as RSGs, rather than being exclusive to the massive LBVs. SNe that are classified at early times may be classified as SNe\,IIn owing to flash-ionisation features that endure for $\sim 10$\,days post-explosion. Without follow-up data there may be a number of misclassifications, as some of these SNe might evolve into ``standard'' SNe\,II.

To be able to promote a silver-class object to the gold class, follow-up observations are required. The silver-class SNe\,IIn with a single spectrum and young SNe\,IIn with no immediate follow-up data may show strong CSM interaction over multiple epochs and be promoted to the gold category if subsequent observations indicate ongoing interaction. The SN may have short-lived CSM interaction at later times and be more like the proposed SN\,IIn-L or SN\,IIn-P subclasses. Additional observations of the objects that we did not classify as SNe\,IIn may reduce uncertainty in their classification if the \ion{H}{ii} regions were omitted from the spectra. Observations of the \ion{H}{ii} regions in the host galaxies could be utilised to create templates that can be subtracted from the spectra to remove the contamination, and then CSM interaction features would become clearer if present. Conversely, if the only CSM interaction shown is fleeting and at early times (i.e., flash ionisation), with the spectrum then evolving into a standard SN\,II, we can demote a silver SN\,IIn to not being an SN\,IIn. We demoted three such transients in this study: SN\,2001fa, SN\,2007cm, and SN\,2014G. 

Follow-up observations are the key to add clarity to the SN\,IIn classifications. Current transient surveys such as ZTF find enormous numbers of transients with thousands of SNe discovered each year \citep{Feindt_2019}, and the Vera Rubin Observatory \citep[formerly the Large Synoptic Survey Telescope;][]{LSST} might discover $\sim 10^{5}$ SNe per year. While this may be advantageous for catching the early-time flash spectra, prompt follow-up observations of enormous numbers of transients will be challenging. Such data are particularly important for gap transients. Suspected SN ``impostors'' may be dust-obscured SNe leading to a substantially reduced observed brightness, so a gap transient actually being a fully fledged SN cannot always be ruled out. Late-time observations of gap transients can confirm the survival of the progenitor. An example of a debated gap transient is SN\,2008S, which is often used as a prototypical SN ``impostor''; however, \citet{Adams_2016} suggest that extreme dust behaviour must be invoked to account for a star surviving while SN\,2008S became dimmer than its progenitor. Conversely, as discussed in \sectionref{imps}, some transients such as the final eruption of SN\,2009ip may be interpreted as a terminal CCSN explosion \citep{Pastorello_2013, Smith_2014_2009ip}, but this is disputed \citep{Fraser_2013, Fraser_2015}. Nonterminal eruptions cannot be ruled out until late-time observations can confirm the survival or death of the progenitor.  

In our group of SNe that are not reclassified as SNe\,IIn, the gap transients such as SN\,2002kg have primary classifications as SNe\,IIn in OSC or TNS. As previously discussed, in the case of SN\,2002kg, \citet{Schwartz_2003} recognise that SN\,2002kg may be an ``impostor'' but hold an SN\,IIn classification until further evidence of a nonterminal explosion is revealed. \citet{VanDyk_2006} found that the photometric and spectroscopic properties of SN\,2002kg represent LBV outbursts in the S~Dor phase. \citet{Weis_2005} found that the position of SN\,2002kg in Isaac Newton Telescope (INT) observations was coincident with the LBV NGC 2403-V37; however, the classifications in the public databases were not updated to reflect the new evidence. 

In \sectionref{sec:intro}, we described the classic ``Eiffel Tower'' H$\alpha$ profile of SNe\,IIn. The actual H$\alpha$ profile may deviate from the classic profile, possibly complicating efforts to classify the object. The standard symmetrical SN\,IIn profile is formed from a spherical shell of H-rich CSM, but \citet{Dwarkadas_2011} find that a clump of CSM is sufficient to create SN\,IIn-like features. Indeed, there seems to be a variety of H$\alpha$ profile shapes observed which may result from the CSM having a variety of different geometries, with the inclination to the line of sight also having an effect. \citet{Harvey_2020} shows how different shell morphologies can affect line profiles. An example of a deviation from a symmetrical profile is shown in the middle spectral epochs of SN\,2009ip (\figureref{2009ipTS}), where there may be a double-peaked H$\alpha$ profile, and at later epochs a ``hump'' appears on the red side of the profile. A further example of this profile diversity is the proposed subclass SN\,IId \citep{Benetti_2000}; examples of SNe\,IId within our catalogue may include SN\,2009kn and SN\,2013gc \citep{Reguitti_2019}. There are signs of an aspherical CSM exhibited by SN\,1998S \citep{Leonard_2000}, and the CSM may be contained in a disc around PTF\,11iqb \citep{Smith_2015}.

\subsection{SN 1978K and other caveats}

We have chosen a simple classification scheme that only relies upon the Balmer line profiles. This allows our classifications to be independent of the complexities and heterogeneity of SN\,IIn spectra. However, there are a small number of objects that do not fit into this regime but are an important consideration.

SN\,1978K in NGC\,1313 was discovered in 1990 by \citet{Dopita_1990} and was initially classified as a classical nova system. Subsequent observations of NGC\,1313 found that this ``nova'' was both a strong radio and X-ray source \citep{Schlegel_1999}. Archival photographic plates reveal a dim, possibly dust-enshrouded SN at the ``nova'' location in mid-1978 and the object was designated SN\,1978K \citep{Chapman_1992}. A spectrum was taken over 13,000\,days post-eruption \citep{Chugai_1995}. The narrow H$\alpha$ line has a FWHM of $\sim 500$\,km\,s$^{-1}$. \citet{Ryder_1993} observed SN\,1978K at optical, X-ray, and radio wavelengths, and propose that the ongoing emission from the SN remnant may be explained by SN ejecta interacting with a dense shell of CSM. \citet{Kuncarayakti_2016} made observations of SN\,1978K with the Very Large Telescope and found that the CSM interaction is ongoing over 35 years post-explosion. Furthermore, \texttt{SNID} matches the spectrum of SN\,1978K to the template spectrum of SN\,1996L 322\,days post-explosion.

Very early-time SN\,IIn spectra may exhibit a strong blue continuum but no broad component in the H$\alpha$ profile, yet showing narrow H$\alpha$ emission \citep{Filippenko_1997}. In the case of SN\,1978K, the broad SN ejecta components have faded, leaving only the relatively narrow lines from ongoing CSM interaction. Therefore, it is important to consider the age of the spectrum. This is not always possible, however, as observation dates are not recorded on public databases in every instance.

As discussed previously, caution must be taken to ensure that gap transients and AGNs do not infiltrate the SN\,IIn sample, since their H$\alpha$ profiles can be similar \citep{Filippenko_1989}. We did not identify any AGNs in our sample, but after comparing peak magnitudes and searching the literature we find that nine of our objects are gap transients. One of these transients, SN\,2017jfs (also known as AT\,2017jfs), has a SN\,IIn-like spectrum and was initially classified as a SN\,IIn, but photometric and spectroscopic analysis revealed that this transient was a luminous red nova \citep[LRN;][]{PastorelloLRN}.

Our classification scheme may miss late-time CSM interaction that does not conform with the ``Eiffel Tower'' H$\alpha$ profile shape. We do not classify SN\,2014G as an SN\,IIn, since the apparent CSM interaction features were seen only at early times and thus indicative of flash ionisation. \citet{Terreren_2016} found that SN\,2014G photometrically behaved as an SN\,IIL, but at $\sim 100$\,days post-maximum luminosity, the spectrum started forming a feature on the blue side of H$\alpha$. This feature grew over time along with the [\ion{O}{i}] $\lambda\lambda6300$, 6363 doublet. The authors interpreted this unusual feature as late-time interaction with a highly asymmetric CSM. As our classification scheme is localised to the H$\alpha$ region, we would not consider this possible CSM interaction feature.

\subsection{SNID and \ion{H}{II} region contamination}

Most of the objects that are not SNe\,IIn were initially categorised as SNe\,IIn owing to contamination from the underlying \ion{H}{ii} regions. These \ion{H}{ii} region features include narrow H$\alpha$ and [\ion{N}{ii}] $\lambda\lambda$6548, 6584 lines. If these lines are not considered, it is possible that the narrow features on top of the broad SN H$\alpha$ emission could be misinterpreted as signatures of CSM interaction, hence leading to an erroneous classification of SN\,IIn. An example of a SN with such \ion{H}{ii} contamination that has a primary classifications of SN\,IIn in online databases is SN\,2009kr (see  \figureref{2009krTS}). This SN has multiple spectral epochs with consistent \ion{H}{ii} contamination. Moreover, \citet{Steele_2009} point out the \ion{H}{ii} contamination of SN\,2009kr, but the object still retains a primary SN\,IIn classification in OSC. This highlights the importance of keeping transient entries on public databases up to date, to reflect new data. In cases where there may be \ion{H}{ii} region contamination superimposed upon an H$\alpha$ profile that shows genuine CSM interaction, our classification scheme could be applied to H$\beta$ where the relatively narrow features and intermediate width components may be present \citep{Filippenko_1997}.

Template-matching routines such as \texttt{SNID} are a popular tool for the classification of SN spectra. These routines rely on templates of archetypal members of a spectral class. The heterogeneity found in SNe\,IIn spectra may cause issues with such template-matching methods that have a limited number of templates for less-common SN classes. This is apparent when we find that only 32 of our targets (all from the gold and silver groups) were matched to an SN\,IIn template by \texttt{SNID} (see \tableref{snid}). In the case of SN ``impostors,'' photometric data or late-time observations are required to differentiate an object as a gap transient rather than an SN\,IIn. Similarly, \texttt{SNID} classifies 23 objects in our sample as AGNs, but we do not identify any genuine AGNs in our sample, as none lie near the centre of their host galaxy and also show the characteristic nebular [\ion{O}{iii}] $\lambda\lambda$4959, 5007 lines.

For the SNe with \ion{H}{ii} region pollution, if a spectrum has low resolution ($R \lesssim 500$), or if a noisy spectrum is heavily binned, it is possible that the [\ion{N}{ii}] lines blend with H$\alpha$ to make an illusion of an SN\,IIn-like profile. Spectroscopic templates of the gold SNe\,IIn will be produced in the hope of increasing the sample of reference templates to better reflect the great diversity we see in SNe\,IIn.

This study has relied on the collection of data from public sources such as OSC and from the original classifiers. However, data from the public databases often have little metadata available, with the spectra files in ASCII format. For example, these spectra frequently lack information about the observer, the telescope or instrument, and in some cases even the exact date of observation. While the OSC quotes times in terms of days from maximum brightness, the band in which the maximum is observed may be IR rather than optical, so quoted times may not correspond well to the time of explosion. There are also many unpublished spectra that further complicate any studies wishing to use large SN\,IIn samples. We were unable to obtain data for 28 objects. While TNS is missing classifications for twelve of the objects, it has fewer misclassifications when compared to OSC, with twelve primary classifications in the gold and silver groups not being an SN\,IIn compared to the 34 for OSC.

\subsection{Summary}

 We have outlined a simple classification scheme for SNe\,IIn which takes advantage of their characteristic H$\alpha$ profile and applied this to a sample of 115 SN\,IIn candidates at $z<0.02$ which have publicly available data or data that we were able to collect from the original observers. Our sample is based on a catalogue collected for an environmental survey of SNe\,IIn, and candidates outside of this sample (e.g., at a higher redshift) are beyond the scope of this project. Here we briefly summarise our key findings and recommendations for future work.

\begin{enumerate}
  \item We compiled a catalogue of 115 reclassified transients that have at some point held an SN\,IIn classification.
  \item 87 of the 115 SNe have been reconfirmed to be SNe\,IIn based on their multicomponent H$\alpha$ profile.
  \item There are 37 gold and 50 silver category spectra. 
  \item Gold SNe\,IIn exhibit CSM interaction over multiple epochs. Silver-category transients are consistent with SNe\,IIn, but they may have short-lived CSM interaction in the case of transitional objects (e.g., when the classification evolves from SN\,IIn to SN\,IIP), or they have quite limited data. Silver SNe\,IIn can be promoted to gold SNe\,IIn with more data from follow-up observations. Objects can also be demoted if it becomes apparent that the CSM interaction features are due to flash ionisation.
  \item We do not classify 28 SNe as SNe\,IIn. In most cases, \ion{H}{ii} region lines contaminate the spectrum and were interpreted as CSM interaction, gap transients, or are consistent with flash ionisation.
  \item 34 of the SNe\,IIn have a primary classification from the OSC that is not an SN\,IIn, and TNS has twelve such cases. 
  \item OSC reports that eleven of the targets we classified as not being SNe\,IIn are SNe\,IIn, and TNS reports 18 such cases.
  \item The public databases OSC and TNS disagree on the classification of 51 of our objects; however, TNS primary classifications seem more reliable, despite lacking classifications for twelve objects. 
  \item \texttt{SNID} classifies 15 out of 37 gold SNe\,IIn as SNe\,IIn, 14 out of 50 silver SNe\,IIn as SNe\,IIn. \texttt{SNID} classifies the three SNe with early narrow components which disappear over time, consistent with flash-ionisation signatures, as SNe\,IIn.
  \item Inconsistencies and ambiguities in the classification of SNe\,IIn may be due to the great range in properties that SNe\,IIn exhibit. The heterogeneity in the spectral features is perhaps a result of multiple progenitor paths and different environments.
\end{enumerate}

To solidify classifications of SNe\,IIn, follow-up observations are vital. They provide information on the evolution of the spectra, allowing estimations of the duration of CSM interaction. Furthermore, when an SN\,IIn classification is found to be erroneous, it is important that the new data are reflected in public databases, particularly in cases where specific SNe have been studied in depth in the literature. Our classification scheme should be applied to a larger sample of SN\,IIn candidates, since the current catalogue is limited. We will produce a set of template spectra from the gold category to increase the template selection in \texttt{SNID} to improve classifications in the future. Our SN\,IIn classifications will be integrated into an ongoing survey of SN\,IIn environments.

\section*{Acknowledgements}

We would like to thank the referee for their thorough review which provided constructive comments. C.L.R. acknowledges a PhD studentship from the UK Science and Technology Facilities Council (STFC). S.M.H.-M. and M.J.D. acknowledge partial funding from STFC. A.V.F. is grateful for funding from the TABASGO Foundation, the Christopher J. Redlich Fund, and the U.C. Berkeley Miller Institute for Basic Research in Science (where he is a Senior Miller Fellow). We would like to thank F.~Bauer, N.~Blagorodnova, T.G.~Brink, P.~Challis, N.~Elias-Rosa, A.~Gal-Yam, M.~Hamuy,  A.~Pastorello, the Padova Classification Program\footnote{\url{https://sngroup.oapd.inaf.it/asiago\textunderscore class.html}}$^,$\footnote{Data used for SN\,2011ir \citep{Tomasella_2014}.}, A.~Reguitti, and J.~Zhang for providing spectral data used in this work.

Research at Lick Observatory is partially supported by a generous gift from Google.
Some of the data presented herein were obtained at the W. M. Keck Observatory, which is operated as a scientific partnership among the California Institute of Technology, the University of California, and
NASA; the Observatory was made possible by the generous financial support of the W. M. Keck Foundation.

\section*{Data Availability}

The data used in this work will be shared upon reasonable request to the author.

\bibliographystyle{mnras} 
\bibliography{example} 

\begin{thebibliography}{}
\makeatletter
\relax
\def\mn@urlcharsother{\let\do\@makeother \do\$\do\&\do\#\do\^\do\_\do\%\do\~}
\def\mn@doi{\begingroup\mn@urlcharsother \@ifnextchar [ {\mn@doi@}
  {\mn@doi@[]}}
\def\mn@doi@[#1]#2{\def\@tempa{#1}\ifx\@tempa\@empty \href
  {http://dx.doi.org/#2} {doi:#2}\else \href {http://dx.doi.org/#2} {#1}\fi
  \endgroup}
\def\mn@eprint#1#2{\mn@eprint@#1:#2::\@nil}
\def\mn@eprint@arXiv#1{\href {http://arxiv.org/abs/#1} {{\tt arXiv:#1}}}
\def\mn@eprint@dblp#1{\href {http://dblp.uni-trier.de/rec/bibtex/#1.xml}
  {dblp:#1}}
\def\mn@eprint@#1:#2:#3:#4\@nil{\def\@tempa {#1}\def\@tempb {#2}\def\@tempc
  {#3}\ifx \@tempc \@empty \let \@tempc \@tempb \let \@tempb \@tempa \fi \ifx
  \@tempb \@empty \def\@tempb {arXiv}\fi \@ifundefined
  {mn@eprint@\@tempb}{\@tempb:\@tempc}{\expandafter \expandafter \csname
  mn@eprint@\@tempb\endcsname \expandafter{\@tempc}}}

\bibitem[\protect\citeauthoryear{{Adams}, {Kochanek}, {Prieto}, {Dai},
  {Shappee}  \& {Stanek}}{{Adams} et~al.}{2016}]{Adams_2016}
{Adams} S.~M.,  {Kochanek} C.~S.,  {Prieto} J.~L.,  {Dai} X.,  {Shappee} B.~J.,
    {Stanek} K.~Z.,  2016, \mn@doi [\mnras] {10.1093/mnras/stw1059}, \href
  {https://ui.adsabs.harvard.edu/abs/2016MNRAS.460.1645A} {460, 1645}

\bibitem[\protect\citeauthoryear{{Aghakhanloo}, {Murphy}, {Smith}  \&
  {Hlo{\v{z}}ek}}{{Aghakhanloo} et~al.}{2017}]{Aghakhanloo_2017}
{Aghakhanloo} M.,  {Murphy} J.~W.,  {Smith} N.,   {Hlo{\v{z}}ek} R.,  2017,
  \mn@doi [\mnras] {10.1093/mnras/stx2050}, \href
  {https://ui.adsabs.harvard.edu/abs/2017MNRAS.472..591A} {472, 591}

\bibitem[\protect\citeauthoryear{{Akashi} \& {Kashi}}{{Akashi} \&
  {Kashi}}{2020}]{Akashi_2020}
{Akashi} M.,  {Kashi} A.,  2020, \mn@doi [\mnras] {10.1093/mnras/staa1014},
  \href {https://ui.adsabs.harvard.edu/abs/2020MNRAS.494.3186A} {494, 3186}

\bibitem[\protect\citeauthoryear{{Aldering} et~al.,}{{Aldering}
  et~al.}{2006}]{Aldering_2006}
{Aldering} G.,  et~al., 2006, \mn@doi [\apj] {10.1086/507020}, \href
  {https://ui.adsabs.harvard.edu/abs/2006ApJ...650..510A} {650, 510}

\bibitem[\protect\citeauthoryear{{Anderson} \& {James}}{{Anderson} \&
  {James}}{2008}]{and08}
{Anderson} J.~P.,  {James} P.~A.,  2008, \mn@doi [\mnras]
  {10.1111/j.1365-2966.2008.13843.x}, \href
  {https://ui.adsabs.harvard.edu/abs/2008MNRAS.390.1527A} {390, 1527}

\bibitem[\protect\citeauthoryear{{Anderson}, {Habergham}, {James}  \&
  {Hamuy}}{{Anderson} et~al.}{2012}]{and12}
{Anderson} J.~P.,  {Habergham} S.~M.,  {James} P.~A.,   {Hamuy} M.,  2012,
  \mn@doi [\mnras] {10.1111/j.1365-2966.2012.21324.x}, \href
  {https://ui.adsabs.harvard.edu/abs/2012MNRAS.424.1372A} {424, 1372}

\bibitem[\protect\citeauthoryear{{Anderson} et~al.,}{{Anderson}
  et~al.}{2014}]{Anderson_2014}
{Anderson} J.~P.,  et~al., 2014, \mn@doi [\apj] {10.1088/0004-637X/786/1/67},
  \href {https://ui.adsabs.harvard.edu/abs/2014ApJ...786...67A} {786, 67}

\bibitem[\protect\citeauthoryear{{Andrews} et~al.,}{{Andrews}
  et~al.}{2020}]{Andrews_2020}
{Andrews} J.~E.,  et~al., 2020, arXiv e-prints, \href
  {https://ui.adsabs.harvard.edu/abs/2020arXiv200913541A} {p. arXiv:2009.13541}

\bibitem[\protect\citeauthoryear{{Arbour}}{{Arbour}}{2008}]{Arbour_2008}
{Arbour} R.,  2008, Central Bureau Electronic Telegrams, \href
  {https://ui.adsabs.harvard.edu/abs/2008CBET.1235....2A} {1235, 2}

\bibitem[\protect\citeauthoryear{{Ayani}, {Kawabata}  \& {Yamaoka}}{{Ayani}
  et~al.}{2002}]{2002bu}
{Ayani} K.,  {Kawabata} T.,   {Yamaoka} H.,  2002, \iaucirc, \href
  {https://ui.adsabs.harvard.edu/abs/2002IAUC.7864....4A} {7864, 4}

\bibitem[\protect\citeauthoryear{{Benetti}}{{Benetti}}{2000}]{Benetti_2000}
{Benetti} S.,  2000, \memsai, \href
  {https://ui.adsabs.harvard.edu/abs/2000MmSAI..71..323B} {71, 323}

\bibitem[\protect\citeauthoryear{{Benetti} \& {Zwitter}}{{Benetti} \&
  {Zwitter}}{1996}]{1996ae}
{Benetti} S.,  {Zwitter} T.,  1996, \iaucirc, \href
  {https://ui.adsabs.harvard.edu/abs/1996IAUC.6410....1B} {6410, 1}

\bibitem[\protect\citeauthoryear{{Benetti} et~al.,}{{Benetti}
  et~al.}{2002}]{2002a}
{Benetti} S.,  et~al., 2002, \iaucirc, \href
  {https://ui.adsabs.harvard.edu/abs/2002IAUC.7789....2B} {7789, 2}

\bibitem[\protect\citeauthoryear{{Berger} et~al.,}{{Berger}
  et~al.}{2009a}]{Berger_2009b}
{Berger} E.,  et~al., 2009a, \mn@doi [\apj] {10.1088/0004-637X/699/2/1850},
  \href {https://ui.adsabs.harvard.edu/abs/2009ApJ...699.1850B} {699, 1850}

\bibitem[\protect\citeauthoryear{{Berger}, {Foley}  \& {Ivans}}{{Berger}
  et~al.}{2009b}]{Berger_2009}
{Berger} E.,  {Foley} R.,   {Ivans} I.,  2009b, The Astronomer's Telegram,
  \href {https://ui.adsabs.harvard.edu/abs/2009ATel.2184....1B} {2184, 1}

\bibitem[\protect\citeauthoryear{{Bersier}, {Smartt}  \& {Yaron}}{{Bersier}
  et~al.}{2016}]{2016cvk}
{Bersier} D.,  {Smartt} S.,   {Yaron} O.,  2016, Transient Name Server
  Classification Report, \href
  {https://ui.adsabs.harvard.edu/abs/2016TNSCR.650....1B} {2016-650, 1}

\bibitem[\protect\citeauthoryear{{Bevan} et~al.,}{{Bevan}
  et~al.}{2020}]{Bevan20}
{Bevan} A.~M.,  et~al., 2020, \mn@doi [\apj] {10.3847/1538-4357/ab86a2}, \href
  {https://ui.adsabs.harvard.edu/abs/2020ApJ...894..111B} {894, 111}

\bibitem[\protect\citeauthoryear{{Blanc} et~al.,}{{Blanc}
  et~al.}{2005a}]{2005db}
{Blanc} N.,  et~al., 2005a, The Astronomer's Telegram, \href
  {https://ui.adsabs.harvard.edu/abs/2005ATel..570....1B} {570, 1}

\bibitem[\protect\citeauthoryear{{Blanc} et~al.,}{{Blanc}
  et~al.}{2005b}]{2005gl}
{Blanc} N.,  et~al., 2005b, The Astronomer's Telegram, \href
  {https://ui.adsabs.harvard.edu/abs/2005ATel..630....1B} {630, 1}

\bibitem[\protect\citeauthoryear{{Blondin} \& {Tonry}}{{Blondin} \&
  {Tonry}}{2007}]{SNID}
{Blondin} S.,  {Tonry} J.~L.,  2007, \mn@doi [\apj] {10.1086/520494}, \href
  {https://ui.adsabs.harvard.edu/abs/2007ApJ...666.1024B} {666, 1024}

\bibitem[\protect\citeauthoryear{{Blondin}, {Modjaz}, {Kirshner}, {Challis}  \&
  {Brown}}{{Blondin} et~al.}{2006a}]{2006bv}
{Blondin} S.,  {Modjaz} M.,  {Kirshner} R.,  {Challis} P.,   {Brown} W.,
  2006a, Central Bureau Electronic Telegrams, \href
  {https://ui.adsabs.harvard.edu/abs/2006CBET..494....1B} {494, 1}

\bibitem[\protect\citeauthoryear{{Blondin}, {Masters}, {Modjaz}, {Kirshner},
  {Challis}, {Matheson}  \& {Berlind}}{{Blondin} et~al.}{2006b}]{2006fp}
{Blondin} S.,  {Masters} K.,  {Modjaz} M.,  {Kirshner} R.,  {Challis} P.,
  {Matheson} T.,   {Berlind} P.,  2006b, Central Bureau Electronic Telegrams,
  \href {https://ui.adsabs.harvard.edu/abs/2006CBET..636....1B} {636, 1}

\bibitem[\protect\citeauthoryear{{Blondin}, {Modjaz}, {Kirshner}, {Challis},
  {Matheson}  \& {Mamajek}}{{Blondin} et~al.}{2006c}]{2006jd}
{Blondin} S.,  {Modjaz} M.,  {Kirshner} R.,  {Challis} P.,  {Matheson} T.,
  {Mamajek} E.,  2006c, Central Bureau Electronic Telegrams, \href
  {https://ui.adsabs.harvard.edu/abs/2006CBET..679....1B} {679, 1}

\bibitem[\protect\citeauthoryear{{Blondin}, {Modjaz}, {Kirshner}, {Challis}  \&
  {Berlind}}{{Blondin} et~al.}{2006d}]{2006am}
{Blondin} S.,  {Modjaz} M.,  {Kirshner} R.,  {Challis} P.,   {Berlind} P.,
  2006d, \iaucirc, \href
  {https://ui.adsabs.harvard.edu/abs/2006IAUC.8680....2B} {8680, 2}

\bibitem[\protect\citeauthoryear{{Blondin}, {Calkins}, {Ayani}  \&
  {Yamaoka}}{{Blondin} et~al.}{2008}]{2008b}
{Blondin} S.,  {Calkins} M.,  {Ayani} K.,   {Yamaoka} H.,  2008, Central Bureau
  Electronic Telegrams, \href
  {https://ui.adsabs.harvard.edu/abs/2008CBET.1201....1B} {1201, 1}

\bibitem[\protect\citeauthoryear{Boian \& Groh}{Boian \&
  Groh}{2018}]{Boian_2018}
Boian I.,  Groh J.~H.,  2018, \mn@doi [Astronomy & Astrophysics]
  {10.1051/0004-6361/201731794}, 617, A115

\bibitem[\protect\citeauthoryear{{Bond}, {Bedin}, {Bonanos}, {Humphreys},
  {Monard}, {Prieto}  \& {Walter}}{{Bond} et~al.}{2009}]{Bond_2009}
{Bond} H.~E.,  {Bedin} L.~R.,  {Bonanos} A.~Z.,  {Humphreys} R.~M.,  {Monard}
  L.~A.~G.~B.,  {Prieto} J.~L.,   {Walter} F.~M.,  2009, \mn@doi [\apjl]
  {10.1088/0004-637X/695/2/L154}, \href
  {https://ui.adsabs.harvard.edu/abs/2009ApJ...695L.154B} {695, L154}

\bibitem[\protect\citeauthoryear{{Bonnaud} et~al.,}{{Bonnaud}
  et~al.}{2005}]{2005kj}
{Bonnaud} C.,  et~al., 2005, Central Bureau Electronic Telegrams, \href
  {https://ui.adsabs.harvard.edu/abs/2005CBET..296....1B} {296, 1}

\bibitem[\protect\citeauthoryear{{Bose}, {Dong}, {Sun}, {Prieto}  \&
  {Stanek}}{{Bose} et~al.}{2017}]{2017gas}
{Bose} S.,  {Dong} S.,  {Sun} F.,  {Prieto} J.~L.,   {Stanek} K.~Z.,  2017, The
  Astronomer's Telegram, \href
  {https://ui.adsabs.harvard.edu/abs/2017ATel10669....1B} {10669, 1}

\bibitem[\protect\citeauthoryear{{Bose}, {Dong}, {Chen}, {Rupert}, {Stanek}  \&
  {Stritzinger}}{{Bose} et~al.}{2018}]{2018hpb}
{Bose} S.,  {Dong} S.,  {Chen} P.,  {Rupert} J.,  {Stanek} K.~Z.,
  {Stritzinger} M.,  2018, Transient Name Server Classification Report, \href
  {https://ui.adsabs.harvard.edu/abs/2018TNSCR1770....1B} {2018-1770, 1}

\bibitem[\protect\citeauthoryear{Bostroem et~al.,}{Bostroem
  et~al.}{2019}]{Bostroem_2019}
Bostroem K.~A.,  et~al., 2019, \mn@doi [Monthly Notices of the Royal
  Astronomical Society] {10.1093/mnras/stz570}, 485, 5120–5141

\bibitem[\protect\citeauthoryear{{Botticella} et~al.,}{{Botticella}
  et~al.}{2009}]{Botticella_2009}
{Botticella} M.~T.,  et~al., 2009, \mn@doi [\mnras]
  {10.1111/j.1365-2966.2009.15082.x}, \href
  {https://ui.adsabs.harvard.edu/abs/2009MNRAS.398.1041B} {398, 1041}

\bibitem[\protect\citeauthoryear{{Brennan} et~al.,}{{Brennan}
  et~al.}{2021a}]{BrennanA_2021}
{Brennan} S.~J.,  et~al., 2021a, arXiv e-prints, \href
  {https://ui.adsabs.harvard.edu/abs/2021arXiv210209572B} {p. arXiv:2102.09572}

\bibitem[\protect\citeauthoryear{{Brennan} et~al.,}{{Brennan}
  et~al.}{2021b}]{BrennanB_2021}
{Brennan} S.~J.,  et~al., 2021b, arXiv e-prints, \href
  {https://ui.adsabs.harvard.edu/abs/2021arXiv210209576B} {p. arXiv:2102.09576}

\bibitem[\protect\citeauthoryear{{Bruch} et~al.,}{{Bruch}
  et~al.}{2020}]{Bruch_2020}
{Bruch} R.~J.,  et~al., 2020, arXiv e-prints, \href
  {https://ui.adsabs.harvard.edu/abs/2020arXiv200809986B} {p. arXiv:2008.09986}

\bibitem[\protect\citeauthoryear{{Bufano} et~al.,}{{Bufano}
  et~al.}{2018}]{2017jfs}
{Bufano} F.,  et~al., 2018, The Astronomer's Telegram, \href
  {https://ui.adsabs.harvard.edu/abs/2018ATel11135....1B} {11135, 1}

\bibitem[\protect\citeauthoryear{{Bullivant} et~al.,}{{Bullivant}
  et~al.}{2018}]{Bullivant18}
{Bullivant} C.,  et~al., 2018, \mn@doi [\mnras] {10.1093/mnras/sty045}, \href
  {https://ui.adsabs.harvard.edu/abs/2018MNRAS.476.1497B} {476, 1497}

\bibitem[\protect\citeauthoryear{{Cao}, {Qiu}, {Qiao}, {Hu}, {Li}  \&
  {Filippenko}}{{Cao} et~al.}{1999}]{1999el}
{Cao} L.,  {Qiu} Y.~L.,  {Qiao} Q.~Y.,  {Hu} J.~Y.,  {Li} W.,   {Filippenko}
  A.,  1999, \iaucirc, \href
  {https://ui.adsabs.harvard.edu/abs/1999IAUC.7288....1C} {7288, 1}

\bibitem[\protect\citeauthoryear{{Cartier}, {Terreran}, {Margutti},
  {Blanchard}, {Espinoza}  \& {Ugarte}}{{Cartier} et~al.}{2018}]{2018cvn}
{Cartier} R.,  {Terreran} G.,  {Margutti} R.,  {Blanchard} P.,  {Espinoza} J.,
   {Ugarte} P.,  2018, The Astronomer's Telegram, \href
  {https://ui.adsabs.harvard.edu/abs/2018ATel12080....1C} {12080, 1}

\bibitem[\protect\citeauthoryear{{Challis} \& {Calkins}}{{Challis} \&
  {Calkins}}{2009}]{2008ip}
{Challis} P.,  {Calkins} M.,  2009, Central Bureau Electronic Telegrams, \href
  {https://ui.adsabs.harvard.edu/abs/2009CBET.1649....1C} {1649, 1}

\bibitem[\protect\citeauthoryear{{Challis}, {Kirshner}  \& {Smith}}{{Challis}
  et~al.}{2010}]{2010jp}
{Challis} P.,  {Kirshner} R.,   {Smith} N.,  2010, Central Bureau Electronic
  Telegrams, \href {https://ui.adsabs.harvard.edu/abs/2010CBET.2548....1C}
  {2548, 1}

\bibitem[\protect\citeauthoryear{{Chapman}, {Norris}, {Reynolds}  \& {te Lintel
  Hekkert}}{{Chapman} et~al.}{1992}]{Chapman_1992}
{Chapman} J.~M.,  {Norris} R.~P.,  {Reynolds} J.~E.,   {te Lintel Hekkert} P.,
  1992, \iaucirc, \href {https://ui.adsabs.harvard.edu/abs/1992IAUC.5616....2C}
  {5616, 2}

\bibitem[\protect\citeauthoryear{{Childress}, {Scalzo}, {Yuan}, {Schmidt}  \&
  {Tucker}}{{Childress} et~al.}{2013}]{2013fs}
{Childress} M.,  {Scalzo} R.,  {Yuan} F.,  {Schmidt} B.,   {Tucker} B.,  2013,
  The Astronomer's Telegram, \href
  {https://ui.adsabs.harvard.edu/abs/2013ATel.5455....1C} {5455, 1}

\bibitem[\protect\citeauthoryear{{Childress}, {Tucker}, {Scalzo}, {Yuan},
  {Zhang}, {Ruiter}, {Seitenzahl}  \& {Schmidt}}{{Childress}
  et~al.}{2015}]{asassn15hs}
{Childress} M.,  {Tucker} B.,  {Scalzo} R.,  {Yuan} F.,  {Zhang} B.,  {Ruiter}
  A.,  {Seitenzahl} I.,   {Schmidt} B.,  2015, The Astronomer's Telegram, \href
  {https://ui.adsabs.harvard.edu/abs/2015ATel.7458....1C} {7458, 1}

\bibitem[\protect\citeauthoryear{{Chugai}}{{Chugai}}{1991}]{Chugai_1991}
{Chugai} N.~N.,  1991, \mn@doi [\mnras] {10.1093/mnras/250.3.513}, \href
  {https://ui.adsabs.harvard.edu/abs/1991MNRAS.250..513C} {250, 513}

\bibitem[\protect\citeauthoryear{{Chugai}}{{Chugai}}{2001}]{Chugai_2001}
{Chugai} N.~N.,  2001, \mn@doi [\mnras] {10.1111/j.1365-2966.2001.04717.x},
  \href {https://ui.adsabs.harvard.edu/abs/2001MNRAS.326.1448C} {326, 1448}

\bibitem[\protect\citeauthoryear{{Chugai}, {Danziger}  \& {della
  Valle}}{{Chugai} et~al.}{1995}]{Chugai_1995}
{Chugai} N.~N.,  {Danziger} I.~J.,   {della Valle} M.,  1995, \mn@doi [\mnras]
  {10.1093/mnras/276.2.530}, \href
  {https://ui.adsabs.harvard.edu/abs/1995MNRAS.276..530C} {276, 530}

\bibitem[\protect\citeauthoryear{Chugai et~al.,}{Chugai
  et~al.}{2004}]{Chugai_1994}
Chugai N.~N.,  et~al., 2004, \mn@doi [Monthly Notices of the Royal Astronomical
  Society] {10.1111/j.1365-2966.2004.08011.x}, 352, 1213

\bibitem[\protect\citeauthoryear{{Clocchiatti} et~al.,}{{Clocchiatti}
  et~al.}{1994}]{1994y}
{Clocchiatti} A.,  et~al., 1994, \iaucirc, \href
  {https://ui.adsabs.harvard.edu/abs/1994IAUC.6065....1C} {6065, 1}

\bibitem[\protect\citeauthoryear{{Darnley} et~al.,}{{Darnley}
  et~al.}{2019}]{Darnley_2019}
{Darnley} M.~J.,  et~al., 2019, \mn@doi [\nat] {10.1038/s41586-018-0825-4},
  \href {https://ui.adsabs.harvard.edu/abs/2019Natur.565..460D} {565, 460}

\bibitem[\protect\citeauthoryear{{Deng} et~al.,}{{Deng}
  et~al.}{2004}]{Deng_2004}
{Deng} J.,  et~al., 2004, \mn@doi [The Astrophysical Journal] {10.1086/420698},
  \href {https://ui.adsabs.harvard.edu/abs/2004ApJ...605L..37D} {605, L37}

\bibitem[\protect\citeauthoryear{{Dessart}, {Hillier}, {Gezari}, {Basa}  \&
  {Matheson}}{{Dessart} et~al.}{2009}]{Dessart_2009}
{Dessart} L.,  {Hillier} D.~J.,  {Gezari} S.,  {Basa} S.,   {Matheson} T.,
  2009, \mn@doi [\mnras] {10.1111/j.1365-2966.2008.14042.x}, \href
  {https://ui.adsabs.harvard.edu/abs/2009MNRAS.394...21D} {394, 21}

\bibitem[\protect\citeauthoryear{{Dilday} et~al.,}{{Dilday}
  et~al.}{2012}]{Dilday_2012}
{Dilday} B.,  et~al., 2012, \mn@doi [Science] {10.1126/science.1219164}, \href
  {https://ui.adsabs.harvard.edu/abs/2012Sci...337..942D} {337, 942}

\bibitem[\protect\citeauthoryear{{Dong}, {Qiu}, {Hu}  \& {Li}}{{Dong}
  et~al.}{2001}]{2001ad}
{Dong} X.~Y.,  {Qiu} Y.~L.,  {Hu} J.~Y.,   {Li} W.~D.,  2001, \iaucirc, \href
  {https://ui.adsabs.harvard.edu/abs/2001IAUC.7596....2D} {7596, 2}

\bibitem[\protect\citeauthoryear{{Dopita} \& {Ryder}}{{Dopita} \&
  {Ryder}}{1990a}]{Dopita_1990}
{Dopita} M.~A.,  {Ryder} S.~D.,  1990a, \iaucirc, \href
  {https://ui.adsabs.harvard.edu/abs/1990IAUC.4950....3D} {4950, 3}

\bibitem[\protect\citeauthoryear{{Dopita} \& {Ryder}}{{Dopita} \&
  {Ryder}}{1990b}]{1978k}
{Dopita} M.~A.,  {Ryder} S.~D.,  1990b, \iaucirc, \href
  {https://ui.adsabs.harvard.edu/abs/1990IAUC.4950....3D} {4950, 3}

\bibitem[\protect\citeauthoryear{{Duyvendak}}{{Duyvendak}}{1942}]{Duyvendak_1942}
{Duyvendak} J.~J.~L.,  1942, \mn@doi [\pasp] {10.1086/125409}, \href
  {https://ui.adsabs.harvard.edu/abs/1942PASP...54...91D} {54, 91}

\bibitem[\protect\citeauthoryear{{Dwarkadas}}{{Dwarkadas}}{2011}]{Dwarkadas_2011}
{Dwarkadas} V.~V.,  2011, \mn@doi [\mnras] {10.1111/j.1365-2966.2010.18001.x},
  \href {https://ui.adsabs.harvard.edu/abs/2011MNRAS.412.1639D} {412, 1639}

\bibitem[\protect\citeauthoryear{{Elias-Rosa} et~al.,}{{Elias-Rosa}
  et~al.}{2010}]{Elias-Rosa_2010}
{Elias-Rosa} N.,  et~al., 2010, \mn@doi [\apjl] {10.1088/2041-8205/714/2/L254},
  \href {https://ui.adsabs.harvard.edu/abs/2010ApJ...714L.254E} {714, L254}

\bibitem[\protect\citeauthoryear{{Elias-Rosa} et~al.,}{{Elias-Rosa}
  et~al.}{2016}]{2016aiy}
{Elias-Rosa} N.,  et~al., 2016, The Astronomer's Telegram, \href
  {https://ui.adsabs.harvard.edu/abs/2016ATel.8727....1E} {8727, 1}

\bibitem[\protect\citeauthoryear{{Elias-Rosa} et~al.,}{{Elias-Rosa}
  et~al.}{2018}]{EliasRosa_2018}
{Elias-Rosa} N.,  et~al., 2018, \mn@doi [\mnras] {10.1093/mnras/sty009}, \href
  {https://ui.adsabs.harvard.edu/abs/2018MNRAS.475.2614E} {475, 2614}

\bibitem[\protect\citeauthoryear{{Ergon} et~al.,}{{Ergon} et~al.}{2014}]{2014p}
{Ergon} M.,  et~al., 2014, The Astronomer's Telegram, \href
  {https://ui.adsabs.harvard.edu/abs/2014ATel.5857....1E} {5857, 1}

\bibitem[\protect\citeauthoryear{{Faran} et~al.,}{{Faran}
  et~al.}{2014}]{Faran_2014}
{Faran} T.,  et~al., 2014, \mn@doi [\mnras] {10.1093/mnras/stu1760}, \href
  {https://ui.adsabs.harvard.edu/abs/2014MNRAS.445..554F} {445, 554}

\bibitem[\protect\citeauthoryear{Feindt, Nordin, Rigault, Brinnel, Dhawan,
  Goobar  \& Kowalski}{Feindt et~al.}{2019}]{Feindt_2019}
Feindt U.,  Nordin J.,  Rigault M.,  Brinnel V.,  Dhawan S.,  Goobar A.,
  Kowalski M.,  2019, \mn@doi [Journal of Cosmology and Astroparticle Physics]
  {10.1088/1475-7516/2019/10/005}, 2019, 005–005

\bibitem[\protect\citeauthoryear{{Filippenko}}{{Filippenko}}{1989}]{Filippenko_1989}
{Filippenko} A.~V.,  1989, \mn@doi [\aj] {10.1086/115018}, \href
  {https://ui.adsabs.harvard.edu/abs/1989AJ.....97..726F} {97, 726}

\bibitem[\protect\citeauthoryear{{Filippenko}}{{Filippenko}}{1997}]{Filippenko_1997}
{Filippenko} A.~V.,  1997, \mn@doi [\araa] {10.1146/annurev.astro.35.1.309},
  \href {https://ui.adsabs.harvard.edu/abs/1997ARA%26A..35..309F} {35, 309}

\bibitem[\protect\citeauthoryear{{Filippenko}}{{Filippenko}}{2000}]{2000ch}
{Filippenko} A.~V.,  2000, \iaucirc, \href
  {https://ui.adsabs.harvard.edu/abs/2000IAUC.7421....3F} {7421, 3}

\bibitem[\protect\citeauthoryear{{Filippenko} \& {Barth}}{{Filippenko} \&
  {Barth}}{1997}]{1997eg}
{Filippenko} A.~V.,  {Barth} A.~J.,  1997, \iaucirc, \href
  {https://ui.adsabs.harvard.edu/abs/1997IAUC.6794....1F} {6794, 1}

\bibitem[\protect\citeauthoryear{{Filippenko} \& {Chornock}}{{Filippenko} \&
  {Chornock}}{2001a}]{2001i}
{Filippenko} A.~V.,  {Chornock} R.,  2001a, \iaucirc, \href
  {https://ui.adsabs.harvard.edu/abs/2001IAUC.7571....2F} {7571, 2}

\bibitem[\protect\citeauthoryear{{Filippenko} \& {Chornock}}{{Filippenko} \&
  {Chornock}}{2001b}]{2001fa}
{Filippenko} A.~V.,  {Chornock} R.,  2001b, \iaucirc, \href
  {https://ui.adsabs.harvard.edu/abs/2001IAUC.7737....2F} {7737, 2}

\bibitem[\protect\citeauthoryear{{Filippenko} \& {Chornock}}{{Filippenko} \&
  {Chornock}}{2003}]{Filippenko_2003}
{Filippenko} A.~V.,  {Chornock} R.,  2003, \iaucirc, \href
  {https://ui.adsabs.harvard.edu/abs/2003IAUC.8051....2F} {8051, 2}

\bibitem[\protect\citeauthoryear{{Filippenko} \& {Garnavich}}{{Filippenko} \&
  {Garnavich}}{1999}]{1999gb}
{Filippenko} A.~V.,  {Garnavich} P.,  1999, \iaucirc, \href
  {https://ui.adsabs.harvard.edu/abs/1999IAUC.7328....3F} {7328, 3}

\bibitem[\protect\citeauthoryear{{Filippenko} \& {Schlegel}}{{Filippenko} \&
  {Schlegel}}{1995}]{1995g}
{Filippenko} A.~V.,  {Schlegel} D.,  1995, \iaucirc, \href
  {https://ui.adsabs.harvard.edu/abs/1995IAUC.6139....2F} {6139, 2}

\bibitem[\protect\citeauthoryear{{Filippenko}, {Li}  \& {Modjaz}}{{Filippenko}
  et~al.}{1999}]{1999bw}
{Filippenko} A.~V.,  {Li} W.~D.,   {Modjaz} M.,  1999, \iaucirc, \href
  {https://ui.adsabs.harvard.edu/abs/1999IAUC.7152....2F} {7152, 2}

\bibitem[\protect\citeauthoryear{{Filippenko}, {Li}, {Treffers}  \&
  {Modjaz}}{{Filippenko} et~al.}{2001}]{Filippenko_2001}
{Filippenko} A.~V.,  {Li} W.~D.,  {Treffers} R.~R.,   {Modjaz} M.,  2001, in
  {Paczynski} B.,  {Chen} W.-P.,   {Lemme} C.,  eds,  Astronomical Society of
  the Pacific Conference Series Vol. 246, IAU Colloq. 183: Small Telescope
  Astronomy on Global Scales. p.~121

\bibitem[\protect\citeauthoryear{{Filippenko}, {Silverman}  \&
  {Foley}}{{Filippenko} et~al.}{2007}]{2007pk}
{Filippenko} A.~V.,  {Silverman} J.~M.,   {Foley} R.~J.,  2007, Central Bureau
  Electronic Telegrams, \href
  {https://ui.adsabs.harvard.edu/abs/2007CBET.1129....2F} {1129, 2}

\bibitem[\protect\citeauthoryear{{Flewelling} et~al.,}{{Flewelling}
  et~al.}{2020}]{PS1}
{Flewelling} H.~A.,  et~al., 2020, \mn@doi [\apjs] {10.3847/1538-4365/abb82d},
  \href {https://ui.adsabs.harvard.edu/abs/2020ApJS..251....7F} {251, 7}

\bibitem[\protect\citeauthoryear{{Foley}, {Li}, {Moore}, {Wong}, {Pooley}  \&
  {Filippenko}}{{Foley} et~al.}{2006}]{2006gy}
{Foley} R.~J.,  {Li} W.,  {Moore} M.,  {Wong} D.~S.,  {Pooley} D.,
  {Filippenko} A.~V.,  2006, Central Bureau Electronic Telegrams, \href
  {https://ui.adsabs.harvard.edu/abs/2006CBET..695....1F} {695, 1}

\bibitem[\protect\citeauthoryear{Foley, Berger, Fox, Levesque, Challis, Ivans,
  Rhoads  \& Soderberg}{Foley et~al.}{2011}]{Foley_2011}
Foley R.~J.,  Berger E.,  Fox O.,  Levesque E.~M.,  Challis P.~J.,  Ivans
  I.~I.,  Rhoads J.~E.,   Soderberg A.~M.,  2011, \mn@doi [The Astrophysical
  Journal] {10.1088/0004-637x/732/1/32}, 732, 32

\bibitem[\protect\citeauthoryear{{Fox}, {Chevalier}  \& {Skrutskie}}{{Fox}
  et~al.}{2010}]{Fox_2010}
{Fox} O.~D.,  {Chevalier} R.~A.,   {Skrutskie} M.~F.,  2010, The Astronomer's
  Telegram, \href {https://ui.adsabs.harvard.edu/abs/2010ATel.2665....1F}
  {2665, 1}

\bibitem[\protect\citeauthoryear{{Fox}, {Filippenko}, {Skrutskie}, {Silverman},
  {Ganeshalingam}, {Cenko}  \& {Clubb}}{{Fox} et~al.}{2013}]{Fox_2013}
{Fox} O.~D.,  {Filippenko} A.~V.,  {Skrutskie} M.~F.,  {Silverman} J.~M.,
  {Ganeshalingam} M.,  {Cenko} S.~B.,   {Clubb} K.~I.,  2013, \mn@doi [\aj]
  {10.1088/0004-6256/146/1/2}, \href
  {https://ui.adsabs.harvard.edu/abs/2013AJ....146....2F} {146, 2}

\bibitem[\protect\citeauthoryear{Fox et~al.,}{Fox et~al.}{2015}]{ori15}
Fox O.~D.,  et~al., 2015, \mn@doi [Monthly Notices of the Royal Astronomical
  Society] {10.1093/mnras/stv2270}, 454, 4366

\bibitem[\protect\citeauthoryear{{Fox} et~al.,}{{Fox} et~al.}{2020}]{Fox_2020}
{Fox} O.~D.,  et~al., 2020, \mn@doi [\mnras] {10.1093/mnras/staa2324}, \href
  {https://ui.adsabs.harvard.edu/abs/2020MNRAS.498..517F} {498, 517}

\bibitem[\protect\citeauthoryear{{Fraser} et~al.,}{{Fraser}
  et~al.}{2010}]{Fraser_2010}
{Fraser} M.,  et~al., 2010, \mn@doi [\apjl] {10.1088/2041-8205/714/2/L280},
  \href {https://ui.adsabs.harvard.edu/abs/2010ApJ...714L.280F} {714, L280}

\bibitem[\protect\citeauthoryear{{Fraser} et~al.,}{{Fraser}
  et~al.}{2013}]{Fraser_2013}
{Fraser} M.,  et~al., 2013, \mn@doi [\mnras] {10.1093/mnras/stt813}, \href
  {https://ui.adsabs.harvard.edu/abs/2013MNRAS.433.1312F} {433, 1312}

\bibitem[\protect\citeauthoryear{{Fraser} et~al.,}{{Fraser}
  et~al.}{2015}]{Fraser_2015}
{Fraser} M.,  et~al., 2015, \mn@doi [\mnras] {10.1093/mnras/stv1919}, \href
  {https://ui.adsabs.harvard.edu/abs/2015MNRAS.453.3886F} {453, 3886}

\bibitem[\protect\citeauthoryear{{Gal-Yam} \& {Leonard}}{{Gal-Yam} \&
  {Leonard}}{2009}]{GalYam_2009}
{Gal-Yam} A.,  {Leonard} D.~C.,  2009, \mn@doi [\nat] {10.1038/nature07934},
  \href {https://ui.adsabs.harvard.edu/abs/2009Natur.458..865G} {458, 865}

\bibitem[\protect\citeauthoryear{{Gal-Yam}, {Shemmer}  \& {Dann}}{{Gal-Yam}
  et~al.}{2001}]{2001dk}
{Gal-Yam} A.,  {Shemmer} O.,   {Dann} J.,  2001, \iaucirc, \href
  {https://ui.adsabs.harvard.edu/abs/2001IAUC.7676....2G} {7676, 2}

\bibitem[\protect\citeauthoryear{{Gal-Yam}, {Shemmer}  \& {Dann}}{{Gal-Yam}
  et~al.}{2002}]{2002an}
{Gal-Yam} A.,  {Shemmer} O.,   {Dann} J.,  2002, \iaucirc, \href
  {https://ui.adsabs.harvard.edu/abs/2002IAUC.7818....2G} {7818, 2}

\bibitem[\protect\citeauthoryear{{Gal-Yam} et~al.,}{{Gal-Yam}
  et~al.}{2007}]{Gal-Yam_2007}
{Gal-Yam} A.,  et~al., 2007, \mn@doi [\apj] {10.1086/510523}, \href
  {https://ui.adsabs.harvard.edu/abs/2007ApJ...656..372G} {656, 372}

\bibitem[\protect\citeauthoryear{{Garnavich}, {Challis}, {Riess}, {Kirshner}
  \& {Berlind}}{{Garnavich} et~al.}{1995}]{1994ak}
{Garnavich} P.,  {Challis} P.,  {Riess} A.,  {Kirshner} R.,   {Berlind} P.,
  1995, \iaucirc, \href {https://ui.adsabs.harvard.edu/abs/1995IAUC.6124....3G}
  {6124, 3}

\bibitem[\protect\citeauthoryear{{Germany}, {Reiss}, {Sadler}, {Schmidt}  \&
  {Stubbs}}{{Germany} et~al.}{2000}]{Germany_2000}
{Germany} L.~M.,  {Reiss} D.~J.,  {Sadler} E.~M.,  {Schmidt} B.~P.,   {Stubbs}
  C.~W.,  2000, \mn@doi [The Astrophysical Journal] {10.1086/308639}, \href
  {https://ui.adsabs.harvard.edu/abs/2000ApJ...533..320G} {533, 320}

\bibitem[\protect\citeauthoryear{{Ginsburg} \& {Mirocha}}{{Ginsburg} \&
  {Mirocha}}{2011}]{pyspeckit}
{Ginsburg} A.,  {Mirocha} J.,  2011, {PySpecKit: Python Spectroscopic Toolkit}
  (\mn@eprint {ascl} {1109.001})

\bibitem[\protect\citeauthoryear{{Green}}{{Green}}{2008}]{2008gm}
{Green} D.~W.~E.,  2008, Central Bureau Electronic Telegrams, \href
  {https://ui.adsabs.harvard.edu/abs/2008CBET.1554....2G} {1554, 2}

\bibitem[\protect\citeauthoryear{{Groh}, {Georgy}  \& {Ekstr{\"o}m}}{{Groh}
  et~al.}{2013}]{Groh_2013}
{Groh} J.~H.,  {Georgy} C.,   {Ekstr{\"o}m} S.,  2013, \mn@doi [\aap]
  {10.1051/0004-6361/201322369}, \href
  {https://ui.adsabs.harvard.edu/abs/2013A&A...558L...1G} {558, L1}

\bibitem[\protect\citeauthoryear{{Guillochon}, {Parrent}, {Kelley}  \&
  {Margutti}}{{Guillochon} et~al.}{2017}]{ocs}
{Guillochon} J.,  {Parrent} J.,  {Kelley} L.~Z.,   {Margutti} R.,  2017,
  \mn@doi [\apj] {10.3847/1538-4357/835/1/64}, \href
  {https://ui.adsabs.harvard.edu/abs/2017ApJ...835...64G} {835, 64}

\bibitem[\protect\citeauthoryear{{Gutierrez} et~al.,}{{Gutierrez}
  et~al.}{2017}]{2017cik}
{Gutierrez} C.~P.,  et~al., 2017, The Astronomer's Telegram, \href
  {https://ui.adsabs.harvard.edu/abs/2017ATel10318....1G} {10318, 1}

\bibitem[\protect\citeauthoryear{{Habergham}, {Anderson}, {James}  \&
  {Lyman}}{{Habergham} et~al.}{2014}]{hab14}
{Habergham} S.~M.,  {Anderson} J.~P.,  {James} P.~A.,   {Lyman} J.~D.,  2014,
  \mn@doi [\mnras] {10.1093/mnras/stu684}, \href
  {https://ui.adsabs.harvard.edu/abs/2014MNRAS.441.2230H} {441, 2230}

\bibitem[\protect\citeauthoryear{{Hagen}, {Engels}  \& {Reimers}}{{Hagen}
  et~al.}{1997}]{1997ab}
{Hagen} H.~J.,  {Engels} D.,   {Reimers} D.,  1997, \aap, \href
  {https://ui.adsabs.harvard.edu/abs/1997A&A...324L..29H} {324, L29}

\bibitem[\protect\citeauthoryear{{Hamuy}}{{Hamuy}}{2003}]{Hamuy_2003b}
{Hamuy} M.,  2003, \mn@doi [\apj] {10.1086/344689}, \href
  {https://ui.adsabs.harvard.edu/abs/2003ApJ...582..905H} {582, 905}

\bibitem[\protect\citeauthoryear{{Hamuy} \& {Maza}}{{Hamuy} \&
  {Maza}}{2003}]{2003G}
{Hamuy} M.,  {Maza} J.,  2003, \iaucirc, \href
  {https://ui.adsabs.harvard.edu/abs/2003IAUC.8045....2H} {8045, 2}

\bibitem[\protect\citeauthoryear{{Hamuy}, {Phillips}, {Suntzeff}  \&
  {Maza}}{{Hamuy} et~al.}{2003}]{Hamuy_2003}
{Hamuy} M.,  {Phillips} M.,  {Suntzeff} N.,   {Maza} J.,  2003, International
  Astronomical Union Circular, \href
  {https://ui.adsabs.harvard.edu/abs/2003IAUC.8151....2H} {8151, 2}

\bibitem[\protect\citeauthoryear{{Harvey} et~al.,}{{Harvey}
  et~al.}{2020}]{Harvey_2020}
{Harvey} E.~J.,  et~al., 2020, \mn@doi [\mnras] {10.1093/mnras/staa2896}, \href
  {https://ui.adsabs.harvard.edu/abs/2020MNRAS.499.2959H} {499, 2959}

\bibitem[\protect\citeauthoryear{{Hiramatsu} et~al.,}{{Hiramatsu}
  et~al.}{2020}]{Hiramatsu_2020}
{Hiramatsu} D.,  et~al., 2020, arXiv e-prints, \href
  {https://ui.adsabs.harvard.edu/abs/2020arXiv201102176H} {p. arXiv:2011.02176}

\bibitem[\protect\citeauthoryear{{Huang} \& {Chevalier}}{{Huang} \&
  {Chevalier}}{2018}]{Huang_2018}
{Huang} C.,  {Chevalier} R.~A.,  2018, \mn@doi [\mnras]
  {10.1093/mnras/stx3163}, \href
  {https://ui.adsabs.harvard.edu/abs/2018MNRAS.475.1261H} {475, 1261}

\bibitem[\protect\citeauthoryear{{Humphreys} \& {Davidson}}{{Humphreys} \&
  {Davidson}}{1994}]{HD_1994}
{Humphreys} R.~M.,  {Davidson} K.,  1994, \mn@doi [\pasp] {10.1086/133478},
  \href {https://ui.adsabs.harvard.edu/abs/1994PASP..106.1025H} {106, 1025}

\bibitem[\protect\citeauthoryear{{Humphreys}, {Davidson}, {Jones}, {Pogge},
  {Grammer}, {Prieto}  \& {Pritchard}}{{Humphreys}
  et~al.}{2012}]{Humphreys_2012}
{Humphreys} R.~M.,  {Davidson} K.,  {Jones} T.~J.,  {Pogge} R.~W.,  {Grammer}
  S.~H.,  {Prieto} J.~L.,   {Pritchard} T.~A.,  2012, \mn@doi [\apj]
  {10.1088/0004-637X/760/1/93}, \href
  {https://ui.adsabs.harvard.edu/abs/2012ApJ...760...93H} {760, 93}

\bibitem[\protect\citeauthoryear{{Humphreys}, {Davidson}, {Gordon}, {Weis},
  {Burggraf}, {Bomans}  \& {Martin}}{{Humphreys} et~al.}{2014}]{Humphreys_2014}
{Humphreys} R.~M.,  {Davidson} K.,  {Gordon} M.~S.,  {Weis} K.,  {Burggraf} B.,
   {Bomans} D.~J.,   {Martin} J.~C.,  2014, \mn@doi [\apjl]
  {10.1088/2041-8205/782/2/L21}, \href
  {https://ui.adsabs.harvard.edu/abs/2014ApJ...782L..21H} {782, L21}

\bibitem[\protect\citeauthoryear{{Humphreys}, {Davidson}, {Van Dyk}  \&
  {Gordon}}{{Humphreys} et~al.}{2017}]{Humphreys_2017}
{Humphreys} R.~M.,  {Davidson} K.,  {Van Dyk} S.~D.,   {Gordon} M.~S.,  2017,
  \mn@doi [\apj] {10.3847/1538-4357/aa8a71}, \href
  {https://ui.adsabs.harvard.edu/abs/2017ApJ...848...86H} {848, 86}

\bibitem[\protect\citeauthoryear{Hunter}{Hunter}{2007}]{Hunter_2007}
Hunter J.~D.,  2007, \mn@doi [Computing in Science \& Engineering]
  {10.1109/MCSE.2007.55}, 9, 90

\bibitem[\protect\citeauthoryear{{Hurst} et~al.,}{{Hurst}
  et~al.}{2001}]{2001dc}
{Hurst} G.~M.,  et~al., 2001, \iaucirc, \href
  {https://ui.adsabs.harvard.edu/abs/2001IAUC.7662....1H} {7662, 1}

\bibitem[\protect\citeauthoryear{{Inserra} et~al.,}{{Inserra}
  et~al.}{2012}]{2012ca}
{Inserra} C.,  et~al., 2012, Central Bureau Electronic Telegrams, \href
  {https://ui.adsabs.harvard.edu/abs/2012CBET.3101....2I} {3101, 2}

\bibitem[\protect\citeauthoryear{{Inserra} et~al.,}{{Inserra}
  et~al.}{2016}]{Inserra_2016}
{Inserra} C.,  et~al., 2016, \mn@doi [\mnras] {10.1093/mnras/stw825}, \href
  {https://ui.adsabs.harvard.edu/abs/2016MNRAS.459.2721I} {459, 2721}

\bibitem[\protect\citeauthoryear{{James} \& {Anderson}}{{James} \&
  {Anderson}}{2006}]{ja06}
{James} P.~A.,  {Anderson} J.~P.,  2006, \mn@doi [\aap]
  {10.1051/0004-6361:20054509}, \href
  {https://ui.adsabs.harvard.edu/abs/2006A&A...453...57J} {453, 57}

\bibitem[\protect\citeauthoryear{{Jha}, {Challis}, {Kirshner}  \&
  {Berlind}}{{Jha} et~al.}{2000}]{2000p}
{Jha} S.,  {Challis} P.,  {Kirshner} R.,   {Berlind} P.,  2000, \iaucirc, \href
  {https://ui.adsabs.harvard.edu/abs/2000IAUC.7381....2J} {7381, 2}

\bibitem[\protect\citeauthoryear{Justham, Podsiadlowski  \& Vink}{Justham
  et~al.}{2014}]{Justham_2014}
Justham S.,  Podsiadlowski P.,   Vink J.~S.,  2014, \mn@doi [The Astrophysical
  Journal] {10.1088/0004-637x/796/2/121}, 796, 121

\bibitem[\protect\citeauthoryear{{Kankare} et~al.,}{{Kankare}
  et~al.}{2013}]{2013fc}
{Kankare} E.,  et~al., 2013, The Astronomer's Telegram, \href
  {https://ui.adsabs.harvard.edu/abs/2013ATel.5338....1K} {5338, 1}

\bibitem[\protect\citeauthoryear{{Kankare} et~al.,}{{Kankare}
  et~al.}{2017}]{2017dwq}
{Kankare} E.,  et~al., 2017, The Astronomer's Telegram, \href
  {https://ui.adsabs.harvard.edu/abs/2017ATel10391....1K} {10391, 1}

\bibitem[\protect\citeauthoryear{{Kasliwal}}{{Kasliwal}}{2011}]{kasphd}
{Kasliwal} M.~M.,  2011, PhD thesis, California Institute of Technology

\bibitem[\protect\citeauthoryear{{Khazov} et~al.,}{{Khazov}
  et~al.}{2016}]{Khazov_2016}
{Khazov} D.,  et~al., 2016, \mn@doi [\apj] {10.3847/0004-637X/818/1/3}, \href
  {https://ui.adsabs.harvard.edu/abs/2016ApJ...818....3K} {818, 3}

\bibitem[\protect\citeauthoryear{{Kiewe} et~al.,}{{Kiewe}
  et~al.}{2012}]{Kiewe_2012}
{Kiewe} M.,  et~al., 2012, \mn@doi [\apj] {10.1088/0004-637X/744/1/10}, \href
  {https://ui.adsabs.harvard.edu/abs/2012ApJ...744...10K} {744, 10}

\bibitem[\protect\citeauthoryear{{Kochanek}}{{Kochanek}}{2011}]{Kochanek_2011}
{Kochanek} C.~S.,  2011, \mn@doi [\apj] {10.1088/0004-637X/741/1/37}, \href
  {https://ui.adsabs.harvard.edu/abs/2011ApJ...741...37K} {741, 37}

\bibitem[\protect\citeauthoryear{{Kochanek}, {Szczygie{\l}}  \&
  {Stanek}}{{Kochanek} et~al.}{2012}]{Kochanek_2012}
{Kochanek} C.~S.,  {Szczygie{\l}} D.~M.,   {Stanek} K.~Z.,  2012, \mn@doi
  [\apj] {10.1088/0004-637X/758/2/142}, \href
  {https://ui.adsabs.harvard.edu/abs/2012ApJ...758..142K} {758, 142}

\bibitem[\protect\citeauthoryear{{Kotak}, {Meikle}, {European Supernova
  Collaboration}  \& {Rodriguez-Gil}}{{Kotak} et~al.}{2003}]{2003dv}
{Kotak} R.,  {Meikle} W.~P.~S.,  {European Supernova Collaboration}
  {Rodriguez-Gil} P.,  2003, \iaucirc, \href
  {https://ui.adsabs.harvard.edu/abs/2003IAUC.8124....1K} {8124, 1}

\bibitem[\protect\citeauthoryear{{Kuncarayakti}, {Maeda}, {Anderson}, {Hamuy},
  {Nomoto}, {Galbany}  \& {Doi}}{{Kuncarayakti}
  et~al.}{2016}]{Kuncarayakti_2016}
{Kuncarayakti} H.,  {Maeda} K.,  {Anderson} J.~P.,  {Hamuy} M.,  {Nomoto} K.,
  {Galbany} L.,   {Doi} M.,  2016, \mn@doi [\mnras] {10.1093/mnras/stw430},
  \href {https://ui.adsabs.harvard.edu/abs/2016MNRAS.458.2063K} {458, 2063}

\bibitem[\protect\citeauthoryear{{Kurf{\"u}rst}, {Pejcha}  \&
  {Krti{\v{c}}ka}}{{Kurf{\"u}rst} et~al.}{2020}]{Kurfurst_2020}
{Kurf{\"u}rst} P.,  {Pejcha} O.,   {Krti{\v{c}}ka} J.,  2020, arXiv e-prints,
  \href {https://ui.adsabs.harvard.edu/abs/2020arXiv200813169K} {p.
  arXiv:2008.13169}

\bibitem[\protect\citeauthoryear{{LSST Science Collaboration} et~al.,}{{LSST
  Science Collaboration} et~al.}{2009}]{LSST}
{LSST Science Collaboration} et~al., 2009, arXiv e-prints, \href
  {https://ui.adsabs.harvard.edu/abs/2009arXiv0912.0201L} {p. arXiv:0912.0201}

\bibitem[\protect\citeauthoryear{{Leonard}, {Filippenko}, {Barth}  \&
  {Matheson}}{{Leonard} et~al.}{2000}]{Leonard_2000}
{Leonard} D.~C.,  {Filippenko} A.~V.,  {Barth} A.~J.,   {Matheson} T.,  2000,
  \mn@doi [\apj] {10.1086/308910}, \href
  {https://ui.adsabs.harvard.edu/abs/2000ApJ...536..239L} {536, 239}

\bibitem[\protect\citeauthoryear{{Li}, {Smith}, {Miller}  \& {Filippenko}}{{Li}
  et~al.}{2009a}]{Li_2009b}
{Li} W.,  {Smith} N.,  {Miller} A.~A.,   {Filippenko} A.~V.,  2009a, The
  Astronomer's Telegram, \href
  {https://ui.adsabs.harvard.edu/abs/2009ATel.2212....1L} {2212, 1}

\bibitem[\protect\citeauthoryear{{Li}, {Smith}, {Miller}  \& {Filippenko}}{{Li}
  et~al.}{2009b}]{2009ip}
{Li} W.,  {Smith} N.,  {Miller} A.~A.,   {Filippenko} A.~V.,  2009b, The
  Astronomer's Telegram, \href
  {https://ui.adsabs.harvard.edu/abs/2009ATel.2212....1L} {2212, 1}

\bibitem[\protect\citeauthoryear{{Li}, {Filippenko}, {Miller}, {Cuilland re},
  {Elias-Rosa}  \& {van Dyk}}{{Li} et~al.}{2009c}]{Li_2009}
{Li} W.,  {Filippenko} A.~V.,  {Miller} A.~A.,  {Cuilland re} J.~C.,
  {Elias-Rosa} N.,   {van Dyk} S.~D.,  2009c, The Astronomer's Telegram, \href
  {https://ui.adsabs.harvard.edu/abs/2009ATel.2312....1L} {2312, 1}

\bibitem[\protect\citeauthoryear{{Li} et~al.,}{{Li} et~al.}{2011}]{Li_2011}
{Li} W.,  et~al., 2011, \mn@doi [\mnras] {10.1111/j.1365-2966.2011.18160.x},
  \href {https://ui.adsabs.harvard.edu/abs/2011MNRAS.412.1441L} {412, 1441}

\bibitem[\protect\citeauthoryear{{Li}, {Wang}  \& {Zhang}}{{Li}
  et~al.}{2014}]{2014es}
{Li} W.,  {Wang} X.,   {Zhang} T.,  2014, The Astronomer's Telegram, \href
  {https://ui.adsabs.harvard.edu/abs/2014ATel.6734....1L} {6734, 1}

\bibitem[\protect\citeauthoryear{{Maeder} \& {Meynet}}{{Maeder} \&
  {Meynet}}{2008}]{Maeder_2008}
{Maeder} A.,  {Meynet} G.,  2008, in {de Koter} A.,  {Smith} L.~J.,   {Waters}
  L. B.~F.~M.,  eds,  Astronomical Society of the Pacific Conference Series
  Vol. 388, Mass Loss from Stars and the Evolution of Stellar Clusters. p.~3

\bibitem[\protect\citeauthoryear{{Margutti}, {Soderberg}  \&
  {Milisavljevic}}{{Margutti} et~al.}{2013}]{2013by}
{Margutti} R.,  {Soderberg} A.,   {Milisavljevic} D.,  2013, The Astronomer's
  Telegram, \href {https://ui.adsabs.harvard.edu/abs/2013ATel.5106....1M}
  {5106, 1}

\bibitem[\protect\citeauthoryear{{Margutti} et~al.,}{{Margutti}
  et~al.}{2014}]{Margutti_2014}
{Margutti} R.,  et~al., 2014, \mn@doi [\apj] {10.1088/0004-637X/780/1/21},
  \href {https://ui.adsabs.harvard.edu/abs/2014ApJ...780...21M} {780, 21}

\bibitem[\protect\citeauthoryear{{Marion} \& {Calkins}}{{Marion} \&
  {Calkins}}{2011}]{2011an}
{Marion} G.~H.,  {Calkins} M.,  2011, Central Bureau Electronic Telegrams,
  \href {https://ui.adsabs.harvard.edu/abs/2011CBET.2668....2M} {2668, 2}

\bibitem[\protect\citeauthoryear{{Matheson}, {Jha}, {Challis}, {Kirshner}  \&
  {Calkins}}{{Matheson} et~al.}{2001}]{2001ac}
{Matheson} T.,  {Jha} S.,  {Challis} P.,  {Kirshner} R.,   {Calkins} M.,  2001,
  \iaucirc, \href {https://ui.adsabs.harvard.edu/abs/2001IAUC.7597....3M}
  {7597, 3}

\bibitem[\protect\citeauthoryear{{Matheson}, {Jha}, {Challis}, {Kirshner}  \&
  {Calkins}}{{Matheson} et~al.}{2002a}]{2001ir}
{Matheson} T.,  {Jha} S.,  {Challis} P.,  {Kirshner} R.,   {Calkins} M.,
  2002a, \iaucirc, \href
  {https://ui.adsabs.harvard.edu/abs/2002IAUC.7784....3M} {7784, 3}

\bibitem[\protect\citeauthoryear{{Matheson}, {Jha}, {Challis}, {Kirshner}  \&
  {Berlind}}{{Matheson} et~al.}{2002b}]{2002bj}
{Matheson} T.,  {Jha} S.,  {Challis} P.,  {Kirshner} R.,   {Berlind} P.,
  2002b, \iaucirc, \href
  {https://ui.adsabs.harvard.edu/abs/2002IAUC.7844....5M} {7844, 5}

\bibitem[\protect\citeauthoryear{{Matheson}, {Challis}, {Kirshner}  \&
  {Calkins}}{{Matheson} et~al.}{2003}]{2003ke}
{Matheson} T.,  {Challis} P.,  {Kirshner} R.,   {Calkins} M.,  2003, \iaucirc,
  \href {https://ui.adsabs.harvard.edu/abs/2003IAUC.8246....1M} {8246, 1}

\bibitem[\protect\citeauthoryear{{Matheson}, {Challis}, {Kirshner}  \&
  {Calkins}}{{Matheson} et~al.}{2004}]{2003lo}
{Matheson} T.,  {Challis} P.,  {Kirshner} R.,   {Calkins} M.,  2004, \iaucirc,
  \href {https://ui.adsabs.harvard.edu/abs/2004IAUC.8268....2M} {8268, 2}

\bibitem[\protect\citeauthoryear{{Mauerhan} \& {Smith}}{{Mauerhan} \&
  {Smith}}{2012}]{Mauerhan_2012}
{Mauerhan} J.,  {Smith} N.,  2012, \mn@doi [\mnras]
  {10.1111/j.1365-2966.2012.21325.x}, \href
  {https://ui.adsabs.harvard.edu/abs/2012MNRAS.424.2659M} {424, 2659}

\bibitem[\protect\citeauthoryear{Mauerhan et~al.,}{Mauerhan
  et~al.}{2013}]{Mauerhan13a}
Mauerhan J.~C.,  et~al., 2013, \mn@doi [Monthly Notices of the Royal
  Astronomical Society] {10.1093/mnras/stt009}, 430, 1801

\bibitem[\protect\citeauthoryear{{Mauerhan}, {Filippenko}, {Brink}  \&
  {Zheng}}{{Mauerhan} et~al.}{2017}]{2017hcc}
{Mauerhan} J.~C.,  {Filippenko} A.~V.,  {Brink} T.~G.,   {Zheng} W.,  2017, The
  Astronomer's Telegram, \href
  {https://ui.adsabs.harvard.edu/abs/2017ATel10911....1M} {10911, 1}

\bibitem[\protect\citeauthoryear{{Maund} et~al.,}{{Maund}
  et~al.}{2006}]{Maund_2006}
{Maund} J.~R.,  et~al., 2006, \mn@doi [\mnras]
  {10.1111/j.1365-2966.2006.10308.x}, \href
  {https://ui.adsabs.harvard.edu/abs/2006MNRAS.369..390M} {369, 390}

\bibitem[\protect\citeauthoryear{{Maund}, {Fraser}, {Reilly}, {Ergon}  \&
  {Mattila}}{{Maund} et~al.}{2015}]{Maund_2015}
{Maund} J.~R.,  {Fraser} M.,  {Reilly} E.,  {Ergon} M.,   {Mattila} S.,  2015,
  \mn@doi [\mnras] {10.1093/mnras/stu2658}, \href
  {https://ui.adsabs.harvard.edu/abs/2015MNRAS.447.3207M} {447, 3207}

\bibitem[\protect\citeauthoryear{{Mayall} \& {Oort}}{{Mayall} \&
  {Oort}}{1942}]{Mayall_1942}
{Mayall} N.~U.,  {Oort} J.~H.,  1942, \mn@doi [\pasp] {10.1086/125410}, \href
  {https://ui.adsabs.harvard.edu/abs/1942PASP...54...95M} {54, 95}

\bibitem[\protect\citeauthoryear{{Maza} et~al.,}{{Maza}
  et~al.}{2009}]{Maza_2009}
{Maza} J.,  et~al., 2009, Central Bureau Electronic Telegrams, \href
  {https://ui.adsabs.harvard.edu/abs/2009CBET.1928....1M} {1928, 1}

\bibitem[\protect\citeauthoryear{{Miller}, {Li}, {Nugent}, {Bloom},
  {Filippenko}  \& {Merritt}}{{Miller} et~al.}{2009}]{Miller_2009}
{Miller} A.~A.,  {Li} W.,  {Nugent} P.~E.,  {Bloom} J.~S.,  {Filippenko} A.~V.,
    {Merritt} A.~T.,  2009, The Astronomer's Telegram, \href
  {https://ui.adsabs.harvard.edu/abs/2009ATel.2183....1M} {2183, 1}

\bibitem[\protect\citeauthoryear{{Millour} et~al.,}{{Millour}
  et~al.}{2020}]{Millour_2020}
{Millour} F.,  et~al., 2020, arXiv e-prints, \href
  {https://ui.adsabs.harvard.edu/abs/2020arXiv200615660M} {p. arXiv:2006.15660}

\bibitem[\protect\citeauthoryear{{Miyaji}, {Nomoto}, {Yokoi}  \&
  {Sugimoto}}{{Miyaji} et~al.}{1980}]{Miyaji_1980}
{Miyaji} S.,  {Nomoto} K.,  {Yokoi} K.,   {Sugimoto} D.,  1980, \pasj, \href
  {https://ui.adsabs.harvard.edu/abs/1980PASJ...32..303M} {32, 303}

\bibitem[\protect\citeauthoryear{{Modjaz}, {Li}, {Garnavich}, {Jha}, {Challis},
  {Kirshner}  \& {Berlind}}{{Modjaz} et~al.}{1999}]{1999eb}
{Modjaz} M.,  {Li} W.~D.,  {Garnavich} P.,  {Jha} S.,  {Challis} P.,
  {Kirshner} R.,   {Berlind} P.,  1999, \iaucirc, \href
  {https://ui.adsabs.harvard.edu/abs/1999IAUC.7268....1M} {7268, 1}

\bibitem[\protect\citeauthoryear{{Modjaz}, {Kirshner}, {Challis}  \&
  {Hao}}{{Modjaz} et~al.}{2005a}]{2005aq}
{Modjaz} M.,  {Kirshner} R.,  {Challis} P.,   {Hao} H.,  2005a, \iaucirc, \href
  {https://ui.adsabs.harvard.edu/abs/2005IAUC.8492....3M} {8492, 3}

\bibitem[\protect\citeauthoryear{{Modjaz}, {Kirshner}, {Challis}  \&
  {Calkins}}{{Modjaz} et~al.}{2005b}]{2005ip}
{Modjaz} M.,  {Kirshner} R.,  {Challis} P.,   {Calkins} M.,  2005b, \iaucirc,
  \href {https://ui.adsabs.harvard.edu/abs/2005IAUC.8628....2M} {8628, 2}

\bibitem[\protect\citeauthoryear{{Morales Garoffolo} et~al.,}{{Morales
  Garoffolo} et~al.}{2013}]{2013l}
{Morales Garoffolo} A.,  et~al., 2013, The Astronomer's Telegram, \href
  {https://ui.adsabs.harvard.edu/abs/2013ATel.4767....1M} {4767, 1}

\bibitem[\protect\citeauthoryear{Moriya, Blinnikov, Tominaga, Yoshida, Tanaka,
  Maeda  \& Nomoto}{Moriya et~al.}{2012}]{moriya13}
Moriya T.~J.,  Blinnikov S.~I.,  Tominaga N.,  Yoshida N.,  Tanaka M.,  Maeda
  K.,   Nomoto K.,  2012, \mn@doi [Monthly Notices of the Royal Astronomical
  Society] {10.1093/mnras/sts075}, 428, 1020

\bibitem[\protect\citeauthoryear{Moriya, Maeda, Taddia, Sollerman, Blinnikov
  \& Sorokina}{Moriya et~al.}{2014}]{Moriya_2014}
Moriya T.~J.,  Maeda K.,  Taddia F.,  Sollerman J.,  Blinnikov S.~I.,
  Sorokina E.~I.,  2014, \mn@doi [Monthly Notices of the Royal Astronomical
  Society] {10.1093/mnras/stu163}, 439, 2917–2926

\bibitem[\protect\citeauthoryear{{Moriya} et~al.,}{{Moriya}
  et~al.}{2019}]{Moriya_2019}
{Moriya} T.~J.,  et~al., 2019, arXiv e-prints, \href
  {https://ui.adsabs.harvard.edu/abs/2019arXiv190701633M} {p. arXiv:1907.01633}

\bibitem[\protect\citeauthoryear{{Moriya} et~al.,}{{Moriya}
  et~al.}{2020}]{Moriya_2020}
{Moriya} T.~J.,  et~al., 2020, \mn@doi [\aap] {10.1051/0004-6361/202038118},
  \href {https://ui.adsabs.harvard.edu/abs/2020A&A...641A.148M} {641, A148}

\bibitem[\protect\citeauthoryear{{Nakano} et~al.,}{{Nakano}
  et~al.}{1996}]{1996bu}
{Nakano} S.,  et~al., 1996, \iaucirc, \href
  {https://ui.adsabs.harvard.edu/abs/1996IAUC.6505....1N} {6505, 1}

\bibitem[\protect\citeauthoryear{{Nakano}, {Yusa}  \& {Kadota}}{{Nakano}
  et~al.}{2009}]{Nakano_2009}
{Nakano} S.,  {Yusa} T.,   {Kadota} K.,  2009, Central Bureau Electronic
  Telegrams, \href {https://ui.adsabs.harvard.edu/abs/2009CBET.2006....1N}
  {2006, 1}

\bibitem[\protect\citeauthoryear{{Nomoto}}{{Nomoto}}{1984}]{Nomoto_1984}
{Nomoto} K.,  1984, \mn@doi [\apj] {10.1086/161749}, \href
  {https://ui.adsabs.harvard.edu/abs/1984ApJ...277..791N} {277, 791}

\bibitem[\protect\citeauthoryear{{Nomoto}}{{Nomoto}}{1987}]{Nomoto_1987}
{Nomoto} K.,  1987, \mn@doi [\apj] {10.1086/165716}, \href
  {https://ui.adsabs.harvard.edu/abs/1987ApJ...322..206N} {322, 206}

\bibitem[\protect\citeauthoryear{{Nugent}}{{Nugent}}{2007}]{Nugent_2007}
{Nugent} P.,  2007, The Astronomer's Telegram, \href
  {https://ui.adsabs.harvard.edu/abs/2007ATel.1213....1N} {1213, 1}

\bibitem[\protect\citeauthoryear{{Ochner} et~al.,}{{Ochner}
  et~al.}{2014a}]{2014g}
{Ochner} P.,  et~al., 2014a, The Astronomer's Telegram, \href
  {https://ui.adsabs.harvard.edu/abs/2014ATel.5767....1O} {5767, 1}

\bibitem[\protect\citeauthoryear{{Ochner}, {Benetti}, {Tomasella}, {Terreran},
  {Pastorello}, {Cappellaro}, {Elias-Rosa}  \& {Turatto}}{{Ochner}
  et~al.}{2014b}]{gaia14ahl}
{Ochner} P.,  {Benetti} S.,  {Tomasella} L.,  {Terreran} G.,  {Pastorello} A.,
  {Cappellaro} E.,  {Elias-Rosa} N.,   {Turatto} M.,  2014b, The Astronomer's
  Telegram, \href {https://ui.adsabs.harvard.edu/abs/2014ATel.6488....1O}
  {6488, 1}

\bibitem[\protect\citeauthoryear{{Ochner} et~al.,}{{Ochner}
  et~al.}{2014c}]{2014ee}
{Ochner} P.,  et~al., 2014c, The Astronomer's Telegram, \href
  {https://ui.adsabs.harvard.edu/abs/2014ATel.6750....1O} {6750, 1}

\bibitem[\protect\citeauthoryear{{Ofek} et~al.,}{{Ofek}
  et~al.}{2014}]{Ofek_2014}
{Ofek} E.~O.,  et~al., 2014, \mn@doi [\apj] {10.1088/0004-637X/789/2/104},
  \href {https://ui.adsabs.harvard.edu/abs/2014ApJ...789..104O} {789, 104}

\bibitem[\protect\citeauthoryear{{Osterbrock} \& {Ferland}}{{Osterbrock} \&
  {Ferland}}{2006}]{Osterbrock_AGN}
{Osterbrock} D.~E.,  {Ferland} G.~J.,  2006, {Astrophysics of gaseous nebulae
  and active galactic nuclei}.
University Science Books; Sausalito, CA

\bibitem[\protect\citeauthoryear{{Papenkova} \& {Li}}{{Papenkova} \&
  {Li}}{2001}]{Papenkova_2001}
{Papenkova} M.,  {Li} W.~D.,  2001, \iaucirc, \href
  {https://ui.adsabs.harvard.edu/abs/2001IAUC.7737....1P} {7737, 1}

\bibitem[\protect\citeauthoryear{{Parrent} et~al.,}{{Parrent}
  et~al.}{2011}]{ptf11iqb}
{Parrent} J.,  et~al., 2011, The Astronomer's Telegram, \href
  {https://ui.adsabs.harvard.edu/abs/2011ATel.3510....1P} {3510, 1}

\bibitem[\protect\citeauthoryear{{Pastorello} et~al.,}{{Pastorello}
  et~al.}{2007}]{Pastorello_2007}
{Pastorello} A.,  et~al., 2007, \mn@doi [\nat] {10.1038/nature05825}, \href
  {https://ui.adsabs.harvard.edu/abs/2007Natur.447..829P} {447, 829}

\bibitem[\protect\citeauthoryear{{Pastorello}, {Stanishev}, {Smartt}, {Fraser}
  \& {Lindborg}}{{Pastorello} et~al.}{2011}]{2011ht}
{Pastorello} A.,  {Stanishev} V.,  {Smartt} S.~J.,  {Fraser} M.,   {Lindborg}
  M.,  2011, Central Bureau Electronic Telegrams, \href
  {https://ui.adsabs.harvard.edu/abs/2011CBET.2851....2P} {2851, 2}

\bibitem[\protect\citeauthoryear{Pastorello et~al.,}{Pastorello
  et~al.}{2013}]{Pastorello_2013}
Pastorello A.,  et~al., 2013, \mn@doi [The Astrophysical Journal]
  {10.1088/0004-637x/767/1/1}, 767, 1

\bibitem[\protect\citeauthoryear{Pastorello et~al.,}{Pastorello
  et~al.}{2017}]{Pastorello_2017}
Pastorello A.,  et~al., 2017, \mn@doi [Monthly Notices of the Royal
  Astronomical Society] {10.1093/mnras/stx2668}, 474, 197–218

\bibitem[\protect\citeauthoryear{{Pastorello} et~al.,}{{Pastorello}
  et~al.}{2019a}]{PastorelloLRN}
{Pastorello} A.,  et~al., 2019a, \mn@doi [\aap] {10.1051/0004-6361/201935511},
  \href {https://ui.adsabs.harvard.edu/abs/2019A&A...625L...8P} {625, L8}

\bibitem[\protect\citeauthoryear{{Pastorello} et~al.,}{{Pastorello}
  et~al.}{2019b}]{Pastorello_2019}
{Pastorello} A.,  et~al., 2019b, \mn@doi [\aap] {10.1051/0004-6361/201935420},
  \href {https://ui.adsabs.harvard.edu/abs/2019A&A...628A..93P} {628, A93}

\bibitem[\protect\citeauthoryear{{Pollas}, {Albanese}, {Benetti}, {Bouchet}  \&
  {Schwarz}}{{Pollas} et~al.}{1995}]{1995n}
{Pollas} C.,  {Albanese} D.,  {Benetti} S.,  {Bouchet} P.,   {Schwarz} H.,
  1995, \iaucirc, \href {https://ui.adsabs.harvard.edu/abs/1995IAUC.6170....1P}
  {6170, 1}

\bibitem[\protect\citeauthoryear{{Prieto}, {Garnavich}, {Depoy}, {Marshall},
  {Eastman}  \& {Frank}}{{Prieto} et~al.}{2005}]{Prieto_2005}
{Prieto} J.,  {Garnavich} P.,  {Depoy} D.,  {Marshall} J.,  {Eastman} J.,
  {Frank} S.,  2005, Central Bureau Electronic Telegrams, \href
  {https://ui.adsabs.harvard.edu/abs/2005CBET..302....1P} {302, 1}

\bibitem[\protect\citeauthoryear{{Prieto} et~al.,}{{Prieto}
  et~al.}{2008}]{Prieto_2008}
{Prieto} J.~L.,  et~al., 2008, \mn@doi [\apjl] {10.1086/589922}, \href
  {https://ui.adsabs.harvard.edu/abs/2008ApJ...681L...9P} {681, L9}

\bibitem[\protect\citeauthoryear{{Prieto}, {Brimacombe}  \& {Drake}}{{Prieto}
  et~al.}{2012}]{Prieto_2012}
{Prieto} J.~L.,  {Brimacombe} J.,   {Drake} A.~J.,  2012, The Astronomer's
  Telegram, \href {https://ui.adsabs.harvard.edu/abs/2012ATel.4439....1P}
  {4439, 1}

\bibitem[\protect\citeauthoryear{{Puckett} et~al.,}{{Puckett}
  et~al.}{2005}]{2005ma}
{Puckett} T.,  et~al., 2005, \iaucirc, \href
  {https://ui.adsabs.harvard.edu/abs/2005IAUC.8647....1P} {8647, 1}

\bibitem[\protect\citeauthoryear{{Reguitti} et~al.,}{{Reguitti}
  et~al.}{2019}]{Reguitti_2019}
{Reguitti} A.,  et~al., 2019, \mn@doi [\mnras] {10.1093/mnras/sty2870}, \href
  {https://ui.adsabs.harvard.edu/abs/2019MNRAS.482.2750R} {482, 2750}

\bibitem[\protect\citeauthoryear{{Reynolds} et~al.,}{{Reynolds}
  et~al.}{2016}]{2016eso}
{Reynolds} T.,  et~al., 2016, The Astronomer's Telegram, \href
  {https://ui.adsabs.harvard.edu/abs/2016ATel.9357....1R} {9357, 1}

\bibitem[\protect\citeauthoryear{{Rui}, {Wang}, {Zhang}, {Jia}  \& {Lu}}{{Rui}
  et~al.}{2016}]{2016hgf}
{Rui} L.,  {Wang} X.,  {Zhang} T.,  {Jia} J.,   {Lu} Y.,  2016, The
  Astronomer's Telegram, \href
  {https://ui.adsabs.harvard.edu/abs/2016ATel.9667....1R} {9667, 1}

\bibitem[\protect\citeauthoryear{{Ryder}, {Staveley-Smith}, {Dopita}, {Petre},
  {Colbert}, {Malin}  \& {Schlegel}}{{Ryder} et~al.}{1993}]{Ryder_1993}
{Ryder} S.,  {Staveley-Smith} L.,  {Dopita} M.,  {Petre} R.,  {Colbert} E.,
  {Malin} D.,   {Schlegel} E.,  1993, \mn@doi [\apj] {10.1086/173223}, \href
  {https://ui.adsabs.harvard.edu/abs/1993ApJ...416..167R} {416, 167}

\bibitem[\protect\citeauthoryear{{Sana} et~al.,}{{Sana}
  et~al.}{2012}]{Sana_2012}
{Sana} H.,  et~al., 2012, \mn@doi [Science] {10.1126/science.1223344}, \href
  {https://ui.adsabs.harvard.edu/abs/2012Sci...337..444S} {337, 444}

\bibitem[\protect\citeauthoryear{{Sanders} et~al.,}{{Sanders}
  et~al.}{2015}]{Sanders_2015}
{Sanders} N.~E.,  et~al., 2015, \mn@doi [\apj] {10.1088/0004-637X/799/2/208},
  \href {https://ui.adsabs.harvard.edu/abs/2015ApJ...799..208S} {799, 208}

\bibitem[\protect\citeauthoryear{{Schaefer} \& {Roscherr}}{{Schaefer} \&
  {Roscherr}}{1998}]{1998s}
{Schaefer} B.~E.,  {Roscherr} B.,  1998, \iaucirc, \href
  {https://ui.adsabs.harvard.edu/abs/1998IAUC.7058....2S} {7058, 2}

\bibitem[\protect\citeauthoryear{{Schlegel}}{{Schlegel}}{1990}]{Schlegel_1990}
{Schlegel} E.~M.,  1990, \mnras, \href
  {https://ui.adsabs.harvard.edu/abs/1990MNRAS.244..269S} {244, 269}

\bibitem[\protect\citeauthoryear{{Schlegel}, {Kirshner}, {Huchra}  \&
  {Schild}}{{Schlegel} et~al.}{1996}]{Schlegel_1996}
{Schlegel} E.~M.,  {Kirshner} R.~P.,  {Huchra} J.~P.,   {Schild} R.~E.,  1996,
  \mn@doi [\aj] {10.1086/117939}, \href
  {https://ui.adsabs.harvard.edu/abs/1996AJ....111.2038S} {111, 2038}

\bibitem[\protect\citeauthoryear{{Schlegel}, {Ryder}, {Staveley-Smith},
  {Petre}, {Colbert}, {Dopita}  \& {Campbell-Wilson}}{{Schlegel}
  et~al.}{1999}]{Schlegel_1999}
{Schlegel} E.~M.,  {Ryder} S.,  {Staveley-Smith} L.,  {Petre} R.,  {Colbert}
  E.,  {Dopita} M.,   {Campbell-Wilson} D.,  1999, \mn@doi [\aj]
  {10.1086/301145}, \href
  {https://ui.adsabs.harvard.edu/abs/1999AJ....118.2689S} {118, 2689}

\bibitem[\protect\citeauthoryear{{Schwartz}, {Li}, {Filippenko}  \&
  {Chornock}}{{Schwartz} et~al.}{2003}]{Schwartz_2003}
{Schwartz} M.,  {Li} W.,  {Filippenko} A.~V.,   {Chornock} R.,  2003, \iaucirc,
  \href {https://ui.adsabs.harvard.edu/abs/2003IAUC.8051....1S} {8051, 1}

\bibitem[\protect\citeauthoryear{Shivvers, Groh, Mauerhan, Fox, Leonard  \&
  Filippenko}{Shivvers et~al.}{2015}]{Shivvers_2015}
Shivvers I.,  Groh J.~H.,  Mauerhan J.~C.,  Fox O.~D.,  Leonard D.~C.,
  Filippenko A.~V.,  2015, \mn@doi [The Astrophysical Journal]
  {10.1088/0004-637x/806/2/213}, 806, 213

\bibitem[\protect\citeauthoryear{{Siebert}, {Pan}  \& {Foley}}{{Siebert}
  et~al.}{2018}]{2018dfy}
{Siebert} M.~R.,  {Pan} Y.~C.,   {Foley} R.~J.,  2018, Transient Name Server
  Classification Report, \href
  {https://ui.adsabs.harvard.edu/abs/2018TNSCR1003....1S} {2018-1003, 1}

\bibitem[\protect\citeauthoryear{{Silverman}, {Kleiser}, {Morton}  \&
  {Filippenko}}{{Silverman} et~al.}{2010}]{2010al}
{Silverman} J.~M.,  {Kleiser} I.~K.~W.,  {Morton} A.~J.~L.,   {Filippenko}
  A.~V.,  2010, Central Bureau Electronic Telegrams, \href
  {https://ui.adsabs.harvard.edu/abs/2010CBET.2223....1S} {2223, 1}

\bibitem[\protect\citeauthoryear{{Silverman} et~al.,}{{Silverman}
  et~al.}{2013}]{Silverman_2013}
{Silverman} J.~M.,  et~al., 2013, \mn@doi [\apjs] {10.1088/0067-0049/207/1/3},
  \href {https://ui.adsabs.harvard.edu/abs/2013ApJS..207....3S} {207, 3}

\bibitem[\protect\citeauthoryear{{Skrutskie} et~al.,}{{Skrutskie}
  et~al.}{2006}]{2mass}
{Skrutskie} M.~F.,  et~al., 2006, \mn@doi [\aj] {10.1086/498708}, \href
  {https://ui.adsabs.harvard.edu/abs/2006AJ....131.1163S} {131, 1163}

\bibitem[\protect\citeauthoryear{{Smith}}{{Smith}}{2006}]{Smith_2006}
{Smith} N.,  2006, \mn@doi [\apj] {10.1086/503766}, \href
  {https://ui.adsabs.harvard.edu/abs/2006ApJ...644.1151S} {644, 1151}

\bibitem[\protect\citeauthoryear{{Smith}}{{Smith}}{2013}]{Smith_Crab}
{Smith} N.,  2013, \mn@doi [\mnras] {10.1093/mnras/stt1004}, \href
  {https://ui.adsabs.harvard.edu/abs/2013MNRAS.434..102S} {434, 102}

\bibitem[\protect\citeauthoryear{Smith}{Smith}{2014}]{Smith_2014}
Smith N.,  2014, \mn@doi [Annual Review of Astronomy and Astrophysics]
  {10.1146/annurev-astro-081913-040025}, 52, 487

\bibitem[\protect\citeauthoryear{{Smith} \& {Andrews}}{{Smith} \&
  {Andrews}}{2020}]{Smith20}
{Smith} N.,  {Andrews} J.~E.,  2020, arXiv e-prints, \href
  {https://ui.adsabs.harvard.edu/abs/2020arXiv200914215S} {p. arXiv:2009.14215}

\bibitem[\protect\citeauthoryear{{Smith} \& {Hartigan}}{{Smith} \&
  {Hartigan}}{2006}]{Smith_Hartigan_2006}
{Smith} N.,  {Hartigan} P.,  2006, \mn@doi [\apj] {10.1086/498860}, \href
  {https://ui.adsabs.harvard.edu/abs/2006ApJ...638.1045S} {638, 1045}

\bibitem[\protect\citeauthoryear{Smith \& Tombleson}{Smith \&
  Tombleson}{2014}]{snt15}
Smith N.,  Tombleson R.,  2014, \mn@doi [Monthly Notices of the Royal
  Astronomical Society] {10.1093/mnras/stu2430}, 447, 598

\bibitem[\protect\citeauthoryear{{Smith} \& {Tombleson}}{{Smith} \&
  {Tombleson}}{2015}]{Tombleson_2015}
{Smith} N.,  {Tombleson} R.,  2015, \mn@doi [\mnras] {10.1093/mnras/stu2430},
  \href {https://ui.adsabs.harvard.edu/abs/2015MNRAS.447..598S} {447, 598}

\bibitem[\protect\citeauthoryear{{Smith}, {Gehrz}, {Hinz}, {Hoffmann}, {Hora},
  {Mamajek}  \& {Meyer}}{{Smith} et~al.}{2003}]{Smith_2003}
{Smith} N.,  {Gehrz} R.~D.,  {Hinz} P.~M.,  {Hoffmann} W.~F.,  {Hora} J.~L.,
  {Mamajek} E.~E.,   {Meyer} M.~R.,  2003, \mn@doi [\aj] {10.1086/346278},
  \href {https://ui.adsabs.harvard.edu/abs/2003AJ....125.1458S} {125, 1458}

\bibitem[\protect\citeauthoryear{Smith et~al.,}{Smith
  et~al.}{2007}]{Smith_2007}
Smith N.,  et~al., 2007, \mn@doi [The Astrophysical Journal] {10.1086/519949},
  666, 1116

\bibitem[\protect\citeauthoryear{{Smith} et~al.,}{{Smith}
  et~al.}{2009}]{Smith_2009}
{Smith} N.,  et~al., 2009, \mn@doi [\apjl] {10.1088/0004-637X/697/1/L49}, \href
  {https://ui.adsabs.harvard.edu/abs/2009ApJ...697L..49S} {697, L49}

\bibitem[\protect\citeauthoryear{{Smith} et~al.,}{{Smith}
  et~al.}{2010}]{Smith_2010}
{Smith} N.,  et~al., 2010, \mn@doi [\aj] {10.1088/0004-6256/139/4/1451}, \href
  {https://ui.adsabs.harvard.edu/abs/2010AJ....139.1451S} {139, 1451}

\bibitem[\protect\citeauthoryear{Smith, Li, Silverman, Ganeshalingam  \&
  Filippenko}{Smith et~al.}{2011}]{Smith_2011}
Smith N.,  Li W.,  Silverman J.~M.,  Ganeshalingam M.,   Filippenko A.~V.,
  2011, \mn@doi [Monthly Notices of the Royal Astronomical Society]
  {10.1111/j.1365-2966.2011.18763.x}, 415, 773

\bibitem[\protect\citeauthoryear{{Smith}, {Mauerhan}, {Kasliwal}  \&
  {Burgasser}}{{Smith} et~al.}{2013}]{Smith_2013}
{Smith} N.,  {Mauerhan} J.~C.,  {Kasliwal} M.~M.,   {Burgasser} A.~J.,  2013,
  \mn@doi [\mnras] {10.1093/mnras/stt944}, \href
  {https://ui.adsabs.harvard.edu/abs/2013MNRAS.434.2721S} {434, 2721}

\bibitem[\protect\citeauthoryear{{Smith}, {Mauerhan}  \& {Prieto}}{{Smith}
  et~al.}{2014}]{Smith_2014_2009ip}
{Smith} N.,  {Mauerhan} J.~C.,   {Prieto} J.~L.,  2014, \mn@doi [\mnras]
  {10.1093/mnras/stt2269}, \href
  {https://ui.adsabs.harvard.edu/abs/2014MNRAS.438.1191S} {438, 1191}

\bibitem[\protect\citeauthoryear{Smith et~al.,}{Smith
  et~al.}{2015}]{Smith_2015}
Smith N.,  et~al., 2015, \mn@doi [Monthly Notices of the Royal Astronomical
  Society] {10.1093/mnras/stv354}, 449, 1876–1896

\bibitem[\protect\citeauthoryear{{Smith} et~al.,}{{Smith}
  et~al.}{2016a}]{Smith_2016b}
{Smith} N.,  et~al., 2016a, \mn@doi [\mnras] {10.1093/mnras/stw219}, \href
  {https://ui.adsabs.harvard.edu/abs/2016MNRAS.458..950S} {458, 950}

\bibitem[\protect\citeauthoryear{Smith et~al.,}{Smith
  et~al.}{2016b}]{Smith_2016}
Smith N.,  et~al., 2016b, \mn@doi [Monthly Notices of the Royal Astronomical
  Society] {10.1093/mnras/stw3204}, 466, 3021

\bibitem[\protect\citeauthoryear{{Smith} et~al.,}{{Smith}
  et~al.}{2017}]{Smith17}
{Smith} N.,  et~al., 2017, \mn@doi [\mnras] {10.1093/mnras/stw3204}, \href
  {https://ui.adsabs.harvard.edu/abs/2017MNRAS.466.3021S} {466, 3021}

\bibitem[\protect\citeauthoryear{{Smith} et~al.,}{{Smith}
  et~al.}{2018}]{Smith_2018}
{Smith} N.,  et~al., 2018, \mn@doi [\mnras] {10.1093/mnras/sty1500}, \href
  {https://ui.adsabs.harvard.edu/abs/2018MNRAS.480.1466S} {480, 1466}

\bibitem[\protect\citeauthoryear{{Stanek}}{{Stanek}}{2017}]{Stanek_2017}
{Stanek} K.~Z.,  2017, Transient Name Server Discovery Report, \href
  {https://ui.adsabs.harvard.edu/abs/2017TNSTR.862....1S} {2017-862, 1}

\bibitem[\protect\citeauthoryear{{Steele}, {Silverman}, {Ganeshalingam}, {Lee},
  {Li}  \& {Filippenko}}{{Steele} et~al.}{2008}]{2008s}
{Steele} T.~N.,  {Silverman} J.~M.,  {Ganeshalingam} M.,  {Lee} N.,  {Li} W.,
  {Filippenko} A.~V.,  2008, Central Bureau Electronic Telegrams, \href
  {https://ui.adsabs.harvard.edu/abs/2008CBET.1275....1S} {1275, 1}

\bibitem[\protect\citeauthoryear{{Steele}, {Cobb}  \& {Filippenko}}{{Steele}
  et~al.}{2009a}]{Steele_2009}
{Steele} T.~N.,  {Cobb} B.,   {Filippenko} A.~V.,  2009a, Central Bureau
  Electronic Telegrams, \href
  {https://ui.adsabs.harvard.edu/abs/2009CBET.2011....1S} {2011, 1}

\bibitem[\protect\citeauthoryear{{Steele}, {Cobb}  \& {Filippenko}}{{Steele}
  et~al.}{2009b}]{2009kn}
{Steele} T.~N.,  {Cobb} B.,   {Filippenko} A.~V.,  2009b, Central Bureau
  Electronic Telegrams, \href
  {https://ui.adsabs.harvard.edu/abs/2009CBET.2011....1S} {2011, 1}

\bibitem[\protect\citeauthoryear{{Stritzinger}, {Folatelli}  \&
  {Morrell}}{{Stritzinger} et~al.}{2008}]{2008J}
{Stritzinger} M.,  {Folatelli} G.,   {Morrell} N.,  2008, Central Bureau
  Electronic Telegrams, \href
  {https://ui.adsabs.harvard.edu/abs/2008CBET.1218....1S} {1218, 1}

\bibitem[\protect\citeauthoryear{{Stritzinger}, {Morrell}, {Folatelli},
  {Covarrubias}  \& {Phillips}}{{Stritzinger} et~al.}{2009}]{2009au}
{Stritzinger} M.,  {Morrell} N.,  {Folatelli} G.,  {Covarrubias} R.,
  {Phillips} M.~M.,  2009, Central Bureau Electronic Telegrams, \href
  {https://ui.adsabs.harvard.edu/abs/2009CBET.1725....1S} {1725, 1}

\bibitem[\protect\citeauthoryear{{Stritzinger}, {Prieto}, {Morrell}  \&
  {Pignata}}{{Stritzinger} et~al.}{2011}]{2011a}
{Stritzinger} M.,  {Prieto} J.~L.,  {Morrell} N.,   {Pignata} G.,  2011,
  Central Bureau Electronic Telegrams, \href
  {https://ui.adsabs.harvard.edu/abs/2011CBET.2623....2S} {2623, 2}

\bibitem[\protect\citeauthoryear{{Stritzinger} et~al.,}{{Stritzinger}
  et~al.}{2012}]{Stritzinger_2012}
{Stritzinger} M.,  et~al., 2012, \mn@doi [\apj] {10.1088/0004-637X/756/2/173},
  \href {https://ui.adsabs.harvard.edu/abs/2012ApJ...756..173S} {756, 173}

\bibitem[\protect\citeauthoryear{{Strolger}, {Seguel}, {Krick}, {Block},
  {Candia}, {Smith}, {Suntzeff}  \& {Phillips}}{{Strolger}
  et~al.}{2000}]{2000eo}
{Strolger} L.~G.,  {Seguel} J.~C.,  {Krick} J.,  {Block} A.,  {Candia} P.,
  {Smith} R.~C.,  {Suntzeff} N.~B.,   {Phillips} M.~M.,  2000, \iaucirc, \href
  {https://ui.adsabs.harvard.edu/abs/2000IAUC.7524....2S} {7524, 2}

\bibitem[\protect\citeauthoryear{{Strotjohann} et~al.,}{{Strotjohann}
  et~al.}{2021}]{Strotjohann_2020}
{Strotjohann} N.~L.,  et~al., 2021, \mn@doi [\apj] {10.3847/1538-4357/abd032},
  \href {https://ui.adsabs.harvard.edu/abs/2021ApJ...907...99S} {907, 99}

\bibitem[\protect\citeauthoryear{{Taddia} et~al.,}{{Taddia}
  et~al.}{2012}]{Taddia_2012}
{Taddia} F.,  et~al., 2012, \mn@doi [\aap] {10.1051/0004-6361/201220105}, \href
  {https://ui.adsabs.harvard.edu/abs/2012A&A...545L...7T} {545, L7}

\bibitem[\protect\citeauthoryear{{Taddia} et~al.,}{{Taddia}
  et~al.}{2013}]{tadd13}
{Taddia} F.,  et~al., 2013, \mn@doi [\aap] {10.1051/0004-6361/201321180}, \href
  {https://ui.adsabs.harvard.edu/abs/2013A&A...555A..10T} {555, A10}

\bibitem[\protect\citeauthoryear{{Tammann} \& {Sandage}}{{Tammann} \&
  {Sandage}}{1968}]{Tammann_1968}
{Tammann} G.~A.,  {Sandage} A.,  1968, \mn@doi [\apj] {10.1086/149487}, \href
  {https://ui.adsabs.harvard.edu/abs/1968ApJ...151..825T} {151, 825}

\bibitem[\protect\citeauthoryear{{Tendulkar}, {Kasliwal}, {Quimby}  \&
  {Kulkarni}}{{Tendulkar} et~al.}{2009a}]{Tendulkar_2009}
{Tendulkar} S.,  {Kasliwal} M.,  {Quimby} R.,   {Kulkarni} S.,  2009a, The
  Astronomer's Telegram, \href {http://www.astronomerstelegram.org/?read=2291}
  {2291, 1}

\bibitem[\protect\citeauthoryear{{Tendulkar}, {Kasliwal}, {Quimby}  \&
  {Kulkarni}}{{Tendulkar} et~al.}{2009b}]{2009kr}
{Tendulkar} S.~P.,  {Kasliwal} M.~M.,  {Quimby} R.,   {Kulkarni} S.~R.,  2009b,
  The Astronomer's Telegram, \href
  {https://ui.adsabs.harvard.edu/abs/2009ATel.2291....1T} {2291, 1}

\bibitem[\protect\citeauthoryear{{Terreran} et~al.,}{{Terreran}
  et~al.}{2016}]{Terreren_2016}
{Terreran} G.,  et~al., 2016, \mn@doi [\mnras] {10.1093/mnras/stw1591}, \href
  {https://ui.adsabs.harvard.edu/abs/2016MNRAS.462..137T} {462, 137}

\bibitem[\protect\citeauthoryear{{Thoene}, {de Ugarte Postigo}, {Leloudas},
  {Cano}  \& {Maeda}}{{Thoene} et~al.}{2015}]{Thoene_2015}
{Thoene} C.,  {de Ugarte Postigo} A.,  {Leloudas} G.,  {Cano} Z.,   {Maeda} K.,
   2015, The Astronomer's Telegram, \href
  {https://ui.adsabs.harvard.edu/abs/2015ATel.8417....1T} {8417, 1}

\bibitem[\protect\citeauthoryear{{Thompson}, {Prieto}, {Stanek}, {Kistler},
  {Beacom}  \& {Kochanek}}{{Thompson} et~al.}{2009}]{Thompson_2009}
{Thompson} T.~A.,  {Prieto} J.~L.,  {Stanek} K.~Z.,  {Kistler} M.~D.,  {Beacom}
  J.~F.,   {Kochanek} C.~S.,  2009, \mn@doi [\apj]
  {10.1088/0004-637X/705/2/1364}, \href
  {https://ui.adsabs.harvard.edu/abs/2009ApJ...705.1364T} {705, 1364}

\bibitem[\protect\citeauthoryear{{Th{\"o}ne} et~al.,}{{Th{\"o}ne}
  et~al.}{2017}]{Thone_2017}
{Th{\"o}ne} C.~C.,  et~al., 2017, \mn@doi [\aap] {10.1051/0004-6361/201629968},
  \href {https://ui.adsabs.harvard.edu/abs/2017A&A...599A.129T} {599, A129}

\bibitem[\protect\citeauthoryear{{Thrasher}, {Li}  \& {Filippenko}}{{Thrasher}
  et~al.}{2008}]{Thrasher_2008}
{Thrasher} P.,  {Li} W.,   {Filippenko} A.~V.,  2008, Central Bureau Electronic
  Telegrams, \href {https://ui.adsabs.harvard.edu/abs/2008CBET.1211....1T}
  {1211, 1}

\bibitem[\protect\citeauthoryear{{Tomasella } et~al.,}{{Tomasella }
  et~al.}{2014}]{Tomasella_2014}
{Tomasella } L.,  et~al., 2014, \mn@doi [Astronomische Nachrichten]
  {10.1002/asna.201412068}, \href
  {https://ui.adsabs.harvard.edu/abs/2014AN....335..841T} {335, 841}

\bibitem[\protect\citeauthoryear{{Tomasella}, {Valenti}, {Ochner}, {Benetti},
  {Cappellaro}  \& {Pastorello}}{{Tomasella} et~al.}{2011}]{2011fx}
{Tomasella} L.,  {Valenti} S.,  {Ochner} P.,  {Benetti} S.,  {Cappellaro} E.,
  {Pastorello} A.,  2011, Central Bureau Electronic Telegrams, \href
  {https://ui.adsabs.harvard.edu/abs/2011CBET.2830....2T} {2830, 2}

\bibitem[\protect\citeauthoryear{{Tomasella}, {Cappellaro}, {Benetti}  \&
  {Turatto}}{{Tomasella} et~al.}{2019}]{2019rz}
{Tomasella} L.,  {Cappellaro} E.,  {Benetti} S.,   {Turatto} M.,  2019, The
  Astronomer's Telegram, \href
  {https://ui.adsabs.harvard.edu/abs/2019ATel12397....1T} {12397, 1}

\bibitem[\protect\citeauthoryear{{Uomoto}}{{Uomoto}}{1989}]{1989l}
{Uomoto} A.,  1989, \iaucirc, \href
  {https://ui.adsabs.harvard.edu/abs/1989IAUC.4792....3U} {4792, 3}

\bibitem[\protect\citeauthoryear{{Valenti} et~al.,}{{Valenti}
  et~al.}{2016}]{Valenti_2016}
{Valenti} S.,  et~al., 2016, \mn@doi [\mnras] {10.1093/mnras/stw870}, \href
  {https://ui.adsabs.harvard.edu/abs/2016MNRAS.459.3939V} {459, 3939}

\bibitem[\protect\citeauthoryear{{Van Dyk}, {Li}, {Filippenko}, {Humphreys},
  {Chornock}, {Foley}  \& {Challis}}{{Van Dyk} et~al.}{2006}]{VanDyk_2006}
{Van Dyk} S.~D.,  {Li} W.,  {Filippenko} A.~V.,  {Humphreys} R.~M.,  {Chornock}
  R.,  {Foley} R.,   {Challis} P.~M.,  2006, arXiv e-prints, \href
  {https://ui.adsabs.harvard.edu/abs/2006astro.ph..3025V} {pp
  astro--ph/0603025}

\bibitem[\protect\citeauthoryear{{Vink}}{{Vink}}{2018}]{Vink_2018}
{Vink} J.~S.,  2018, \mn@doi [\aap] {10.1051/0004-6361/201833352}, \href
  {https://ui.adsabs.harvard.edu/abs/2018A&A...619A..54V} {619, A54}

\bibitem[\protect\citeauthoryear{{Wagner} et~al.,}{{Wagner}
  et~al.}{2000}]{Wagner_2000}
{Wagner} R.~M.,  et~al., 2000, in American Astronomical Society Meeting
  Abstracts. p. 44.13

\bibitem[\protect\citeauthoryear{{Wang}, {Chen}, {Wang}, {Hu}, {Xi}, {Yang},
  {Zhao}  \& {Li}}{{Wang} et~al.}{2018}]{Wang_2018}
{Wang} X.,  {Chen} J.,  {Wang} L.,  {Hu} M.,  {Xi} G.,  {Yang} Y.,  {Zhao} X.,
   {Li} W.,  2018, arXiv e-prints, \href
  {https://ui.adsabs.harvard.edu/abs/2018arXiv181011936W} {p. arXiv:1810.11936}

\bibitem[\protect\citeauthoryear{{Ward} et~al.,}{{Ward}
  et~al.}{2020}]{Ward_2020}
{Ward} C.,  et~al., 2020, arXiv e-prints, \href
  {https://ui.adsabs.harvard.edu/abs/2020arXiv201111656W} {p. arXiv:2011.11656}

\bibitem[\protect\citeauthoryear{{Weis} \& {Bomans}}{{Weis} \&
  {Bomans}}{2005}]{Weis_2005}
{Weis} K.,  {Bomans} D.~J.,  2005, \mn@doi [\aap]
  {10.1051/0004-6361:200400105}, \href
  {https://ui.adsabs.harvard.edu/abs/2005A&A...429L..13W} {429, L13}

\bibitem[\protect\citeauthoryear{{Weis} \& {Bomans}}{{Weis} \&
  {Bomans}}{2020}]{Weis_2020}
{Weis} K.,  {Bomans} D.~J.,  2020, \mn@doi [Galaxies]
  {10.3390/galaxies8010020}, \href
  {https://ui.adsabs.harvard.edu/abs/2020Galax...8...20W} {8, 20}

\bibitem[\protect\citeauthoryear{{Yamanaka}, {Okushima}, {Arai}, {Sasada}  \&
  {Sato}}{{Yamanaka} et~al.}{2010}]{2010jl}
{Yamanaka} M.,  {Okushima} T.,  {Arai} A.,  {Sasada} M.,   {Sato} H.,  2010,
  Central Bureau Electronic Telegrams, \href
  {https://ui.adsabs.harvard.edu/abs/2010CBET.2539....1Y} {2539, 1}

\bibitem[\protect\citeauthoryear{{Zhang}, {Wang}, {Li}, {Dong}  \&
  {Wu}}{{Zhang} et~al.}{2014}]{2014eu}
{Zhang} J.,  {Wang} X.,  {Li} W.,  {Dong} X.,   {Wu} X.,  2014, The
  Astronomer's Telegram, \href
  {https://ui.adsabs.harvard.edu/abs/2014ATel.6723....1Z} {6723, 1}

\bibitem[\protect\citeauthoryear{{Zhang} et~al.,}{{Zhang}
  et~al.}{2016}]{2016bly}
{Zhang} J.,  et~al., 2016, The Astronomer's Telegram, \href
  {https://ui.adsabs.harvard.edu/abs/2016ATel.8933....1Z} {8933, 1}

\bibitem[\protect\citeauthoryear{{Zhang}, {Xu}  \& {Wang}}{{Zhang}
  et~al.}{2018}]{2018zd}
{Zhang} J.,  {Xu} L.,   {Wang} X.,  2018, The Astronomer's Telegram, \href
  {https://ui.adsabs.harvard.edu/abs/2018ATel11379....1Z} {11379, 1}

\bibitem[\protect\citeauthoryear{{Zhang}, {Xin}  \& {Wang}}{{Zhang}
  et~al.}{2019}]{2019el}
{Zhang} J.,  {Xin} Y.,   {Wang} X.,  2019, Transient Name Server Classification
  Report, \href {https://ui.adsabs.harvard.edu/abs/2019TNSCR..70....1Z}
  {2019-70, 1}

\bibitem[\protect\citeauthoryear{{Zhang} et~al.,}{{Zhang}
  et~al.}{2020a}]{Zhang_2020}
{Zhang} J.,  et~al., 2020a, \mn@doi [\mnras] {10.1093/mnras/staa2273}, \href
  {https://ui.adsabs.harvard.edu/abs/2020MNRAS.498...84Z} {498, 84}

\bibitem[\protect\citeauthoryear{{Zhang}, {Xiang}, {Lin}, {Wang}, {Rui},
  {Zhang}  \& {Zhang}}{{Zhang} et~al.}{2020b}]{2018lkg}
{Zhang} X.,  {Xiang} D.,  {Lin} H.,  {Wang} X.,  {Rui} L.,  {Zhang} T.,
  {Zhang} J.,  2020b, Transient Name Server Classification Report, \href
  {https://ui.adsabs.harvard.edu/abs/2020TNSCR2552....1Z} {2020-2552, 1}

\bibitem[\protect\citeauthoryear{{de Jager}}{{de Jager}}{1998}]{deJager_1998}
{de Jager} C.,  1998, \mn@doi [\aapr] {10.1007/s001590050009}, \href
  {https://ui.adsabs.harvard.edu/abs/1998A&ARv...8..145D} {8, 145}

\bibitem[\protect\citeauthoryear{{de Ugarte Postigo}, {Thoene}, {Leloudas}  \&
  {Aceituno}}{{de Ugarte Postigo} et~al.}{2015}]{2015bh}
{de Ugarte Postigo} A.,  {Thoene} C.~C.,  {Leloudas} G.,   {Aceituno} F.,
  2015, The Astronomer's Telegram, \href
  {https://ui.adsabs.harvard.edu/abs/2015ATel.7514....1D} {7514, 1}

\makeatother
\end{thebibliography}

\appendix

\section{Classification tables}


Here we present tables containing our sample of SNe organised into their new spectral classifications. \tableref{gold} lists the SNe\,IIn in the gold category. Tables showing the full lists showing transients in each new classification are shown in the supplementary online materials.

\onecolumn
\begin{table}
\caption{The sample of SNe\,IIn that were reclassified into the ``gold'' spectral category. From our sample of 115 transients within $z<0.02$ with data, 37 SNe are classified as ``gold'' SNe\,IIn where the CSM interaction is visible throughout multiple epochs. The table gives the common SN name, the discovery date, the common host name, the J2000 coordinates, and the redshift.}
\begin{center}
\label{gold}
\begin{tabular}{lllllllllll}
\hline
Name & Disc. Date & Host Name & R.A. & Dec. & $z$ & OSC & TNS &SNID& Class & Source\\
\hline
SN2005db & 2005/07/19 & NGC 214 & 00:41:26.79 & 25:29:51.60 & 0.0151 &IIn & IIn &Gal& Gold IIn & \citet{2005db} \\
SN2005gl & 2005/10/05 & NGC 266 & 00:49:50.02 & 32:16:56.80 & 0.0155&IIn &IIn &IIn& Gold IIn & \citet{2005gl}\\
SN1999eb & 1999/10/02 & NGC 664 & 01:43:45.45 & 04:13:25.90 & 0.0180 & II & IIn &AGN& Gold IIn& \citet{1999eb}\\
SN2003G & 2003/01/08 & IC 208 & 02:08:28.13 & 06:23:51.90 & 0.0120 & IIn & IIn &AGN& Gold IIn & \citet{2003G}\\
SN2008J & 2008/01/15 & MCG -02-07-33 & 02:34:24.20 & -10:50:38.50 & 0.0159 &Ia & IIn &IIn& Gold IIn& \citet{2008J} \\
SN2013fc & 2013/08/20 & ESO 154-G10 & 02:45:08.95 & -55:44:27.30 & 0.0186&IIn&IIn&Ia&Gold IIn& \citet{2013fc}\\
SN2000eo & 2000/11/16 & MCG -02-09-03 & 03:09:08.17 & -10:17:55.30 & 0.0100 & IIn & IIn &IIP& Gold IIn & \citet{2000eo}\\
SN2006gy & 2006/09/18 & NGC 1260 & 03:17:27.06 & 41:24:19.51 & 0.0192&II &IIn &IIP &Gold IIn & \citet{2006gy}\\
SN1989R & 1989/10/04 & UGC 2912 & 03:59:32.56 & 42:37:09.20 & 0.0180 & II & IIn &IIP& Gold IIn&P.~Challis\\
SN1995G & 1995/02/23 & NGC 1643 & 04:43:44.26 & -05:18:53.70 & 0.0160&II &IIP &Gal/IIn& Gold IIn& \citet{1995g}\\
SN2010jp & 2010/11/11 & A061630-2124 & 06:16:30.63 & -21:24:36.30 & 0.0092 &IIn &IIn &AGN&Gold IIn &\citet{2010jp}\\
SN2006jd & 2006/10/12 & UGC 4179 & 08:02:07.43 & 00:48:31.50 & 0.0186&IIn &IIb &II-pec& Gold IIn& \citet{2006jd}\\
SN2013gc & 2013/11/07 & ESO 430-G20 & 08:07:11.88 & -28:03:26.30 & 0.0034&IIn &IIn &IIn/IIP& Gold IIn &A.~Reguitti \\
SN2009kn & 2009/10/26 & MCG-03-21-06 & 08:09:43.04 & -17:44:51.30 & 0.0143&IIn &IIn &AGN& Gold IIn&\citet{2009kn}\\
SN2005kj & 2005/11/17 & A084009-0536 & 08:40:09.18 & -05:36:02.20 & 0.0160&IIn &IIn &LBV/IIn/IIP& Gold IIn& \citet{2005kj}\\
SN1994ak & 1994/12/24 & NGC 2782 & 09:14:01.47 & 40:06:21.50 & 0.0085& II & IIn &AGN& Gold IIn& \citet{1994ak}\\
SN2005ip & 2005/11/05 & NGC 2906 & 09:32:06.42 & 08:26:44.40 & 0.00718&IIn & II &IIn &Gold IIn&\citet{2005ip}\\
SN2010jl & 2010/11/03 & UGC 5189A & 09:42:53.33 & 09:29:41.80 & 0.0107 &IIn &IIn &AGN &Gold IIn &\citet{2010jl}\\
SN1989C & 1989/02/03 & MCG+01-25-25 & 09:47:45.49 & 02:37:36.10 & 0.0063&II &IIP &IIP& Gold IIn&P.~Challis\\
SN2011ht & 2011/09/29 & UGC 5460 & 10:08:10.56 & 51:50:57.12 & 0.0036 &IIn&IIn &LBV/IIP& Gold IIn& \citet{2011ht}\\
SN1993N & 1993/04/15 & UGC 5695 & 10:29:46.33 & 13:01:14.00 & 0.0098 &II &IIP &LBV &Gold IIn & This work \\
SN1998S & 1998/03/02 & NGC 3877 & 11:46:06.13 & 47:28:55.40 & 0.0030&IIn&IIn &IIn& Gold IIn&\citet{1998s} \\
SN1994W & 1994/07/29 & NGC 4041 & 12:02:10.92 & 62:08:32.70 & 0.00403&II &IIP &LBV/IIn &Gold IIn & This work \\
SN2011A & 2011/01/02 & NGC 4902 & 13:01:01.19 & -14:31:34.80 & 0.0089&IIn&IIn &AGN& Gold IIn & \citet{2011a}\\
SN2000P & 2000/03/08 & NGC 4965 & 13:07:10.53 & -28:14:02.50 & 0.0075&II&IIn &IIP/IIn& Gold IIn &\citet{2000p} \\
SN2016bdu & 2016/02/28 & - & 13:10:13.95 & 32:31:14.07 & 0.0170&IIn&IIn &AGN &Gold IIn&N.~Elias-Rosa\\
SN1997eg & 1997/12/04 & NGC 5012 & 13:11:36.73 & 22:55:29.40 & 0.0087&II&-&II-pec$\rightarrow$\,IIn&Gold IIn& \citet{1997eg}\\
SN2015da & 2015/01/09 & NGC 5337 & 13:52:24.11 & 39:41:28.60 & 0.0072&IIn&-&AGN& Gold IIn & J.~Zhang\\
SN1994Y & 1994/08/19 & NGC 5371 & 13:55:36.90 & 40:27:53.40 & 0.0085&II&IIn &AGN& Gold IIn&\citet{1994y}\\
SN1995N & 1995/05/05 & MCG-02-38-17 & 14:49:28.29 & -10:10:14.40 & 0.0062&II &IIn &AGN/II-pec &Gold IIn & \citet{1995n}\\
SN2008B & 2008/01/02 & NGC 5829 & 15:02:43.65 & 23:20:07.80 & 0.0188&II&IIn &IIn& Gold IIn & \citet{2008b}\\
SN1987B & 1987/02/24 & NGC 5850 & 15:07:02.92 & 01:30:13.20 & 0.0085&II&-&IIL/IIn&Gold IIn&\citet{Schlegel_1996}\\
SN2012ca & 2012/04/25 & ESO 336-G9 & 18:41:07.25 & -41:47:38.40 & 0.0190&Ia-CSM&IIn &IIn &Gold IIn & \citet{2012ca}\\
SN2008S & 2008/02/01 & NGC 6946 & 20:34:45.35 & 60:05:57.80 & 0.0002&IIn&IIn &IIP/IIn& Gold IIn & \citet{2008s}\\
SN1999el & 1999/10/20 & NGC 6951 & 20:37:18.03 & 66:06:11.90 & 0.0047&IIn&IIn &Gal/IIn& Gold IIn& \citet{1999el}\\
SN2010bt & 2010/04/17 & NGC 7130 & 21:48:20.22 & -34:57:16.50 & 0.0162&II&IIn &Gal& Gold IIn&N.~Elias-Rosa\\
SN2009ip & 2012/07/24 & NGC 7259 & 22:23:08.30 & -28:56:52.40 & 0.0059&IIn&- &IIP/AGN& Gold IIn& \citet{2009ip}\\
\hline
\end{tabular}
\end{center}
\end{table}

\onecolumn
\begin{landscape}
\begin{longtable}{lllllllllll}
\caption{The sample of SNe\,IIn that were reclassified into the ``silver'' spectral category. From our sample of 115 transients within $z<0.02$ with data, \textbf{50} SNe are classified as ``silver'' SNe\,IIn where the CSM interaction is apparent in some epochs but may be short-lived, or we have only one spectrum for the object. The table gives the common SN name, the discovery date, the common host name, the J2000 coordinates, and the redshift.\label{silver}}\\
\hline
Name & Disc. Date & Host Name & R.A. & Dec. & $z$ &OSC &TNS& SNID &Class&Source\\
\hline
\endfirsthead
\multicolumn{11}{l}{\tablename\ \thetable\ --- Continued from previous page\ldots}\\
\hline
Name & Disc. Date & Host Name & R.A. & Dec. & $z$ &OSC &TNS& SNID &Class&Source\\
\hline
\endhead
\hline
\endlastfoot
\hline
\multicolumn{11}{r}{Continued on next page\ldots}\\
\endfoot
SN2019el & 2019/01/02 & - & 00:02:56.70 & 32:32:52.30 & 0.0005&IIn&IIn&-&Silver IIn&\citet{2019el}\\
SN2017hcc & 2017/10/02 & GALEX 2.67E+18 & 00:03:50.58 & -11:28:28.78 & 0.0173&IIn&IIn&Gal&Silver IIn&\citet{2017hcc} \\
SN2011fx & 2011/08/30 & MCG+04-01-48 & 00:17:59.56 & 24:33:46.00 & 0.0193&IIn&IIn&Ia&Silver IIn&\citet{2011fx}\\
PTF11iqb & 2011/07/23 & NGC 151 & 00:34:04.84 & -09:42:17.90 & 0.0125&IIn&-&Ic &Silver IIn& \citet{ptf11iqb} \\
SN2018cvn & 2018/06/26 & ESO -476G & 01:30:27.12 & -26:47:06.79 & 0.0198&IIn&II&Ia-91bg&Silver IIn&\citet{2018cvn} \\
SN2007pk & 2007/11/10 & NGC 579 & 01:31:47.07 & 33:36:54.10 & 0.0167&II&IIn&IIP&Silver IIn&\citet{2007pk} \\
SN2016eem & 2016/07/08 & - & 02:05:59.80 & 47:44:14.00 & 0.0200&IIn&-&Gal&Silver IIn& N.~Blagorodnova\\
SN2002ea & 2002/07/21 & NGC 820 & 02:08:25.08 & 14:20:52.80 & 0.0148 &II &IIn &IIP &Silver IIn & This work \\
SN1978K & 1978/07/31 & NGC 1313 & 03:17:38.60 & -66:33:04.60 & 0.0016&II&-&IIn&Silver IIn& \citet{1978k}\\
SN2003lo & 2003/12/31 & NGC 1376 & 03:37:05.12 & -05:02:17.30 & 0.0140&IIn&IIn&II-pec&Silver IIn&\citet{2003lo} \\
SN2005aq & 2005/03/07 & NGC 1599 & 04:31:38.82 & -04:35:06.80 & 0.0130 &IIn &IIn &AGN& Silver IIn&\citet{2005aq}\\
Gaia14ahl & 2014/09/20 & PGC 1681539 & 04:42:12.09 & 23:06:15.00 & 0.0170&IIn &-  &IIn& Silver IIn& \citet{gaia14ahl}\\
SN2005ma & 2005/12/24 & MCG -02-13-13 & 04:49:53.91 & -10:45:23.40 & 0.0150&II&IIn&AGN&Silver IIn& \citet{2005ma} \\
SN2016hgf & 2016/10/16 & WEIN 69 & 04:51:45.97 & 44:36:03.06 & 0.0172&IIn&IIn&AGN&Silver IIn& \citet{2016hgf} \\
SN2016eso & 2016/08/08 & ESO -422G &04:59:30.04  & -28:51:39.17 &0.01606 &IIn&IIn&II-pec&Silver IIn&\citet{2016eso}\\
SN2018zd & 2018/03/02 & NGC 2146 & 06:18:03.18 & 78:22:00.90 & 0.0030&IIn&II&IIn&Silver IIn& \citet{2018zd}\\
SN2019rz & 2019/01/14 & UGC 3554 & 06:50:25.80 & 43:03:11.70 & 0.0189&IIn &IIn &Gal &Silver IIn & \citet{2019rz}\\
AT2018lkg & 2018/12/30 & UGC 3660 & 07:06:34.76 & 63:50:56.90 & 0.0142 &IIn &IIn &IIn &Silver IIn & \citet{2018lkg}\\
SN2002A & 2002/01/01 & UGC 3804 & 07:22:36.14 & 71:35:41.50 & 0.0096&II&IIn&Gal&Silver IIn& \citet{2002a} \\
AT2014eu & 2014/11/17 & MCG+09-13-02 & 07:28:55.97 & 56:11:46.20 & 0.0179&IIn&-&IIP&Silver IIn& \citet{2014eu}\\
SN2017cik & 2017/03/17 & SDSS J075412.97+214729.9 & 07:54:13.07 & 21:47:35.80 & 0.0160&IIn&IIn&AGN&Silver IIn& \citet{2017cik} \\
SN2014ee & 2014/11/12 & UGC 4132 & 07:59:11.68 & 32:54:39.60 & 0.0174&IIn&IIn&IIP&Silver IIn & \citet{2014ee} \\
SN2011an & 2011/03/01 & UGC 4139 & 07:59:24.42 & 16:25:08.20 & 0.0163&IIn&IIn&-&Silver IIn&\citet{2011an}\\
SN2001ir & 2001/12/19 & MCG-02-22-22 & 08:36:28.12 & -11:50:03.50 & 0.0200&II&IIn&AGN&Silver IIn& \citet{2001ir} \\
SN2002fj & 2002/09/12 & NGC 2642 & 08:40:45.10 & -04:07:38.50 & 0.0140&II&IIn&IIn&Silver IIn& M.~Hamuy \\
SN2017dwq & 2017/05/04 & 2MASX J09033237-2120017 & 09:03:32.49 & -21:20:02.99 & 0.0182&IIn&IIn&IIn&Silver IIn& \citet{2017dwq}\\
SN2015bh & 2015/02/07 & NGC 2770 & 09:09:34.96 & 33:07:20.40 & 0.0064&IIn&IIn&LBV&Silver IIn&\citet{2015bh}  \\
SN2014es & 2014/11/20 & MCG -01-24-12 & 09:20:46.91 & -08:03:34.00 & 0.0196&IIn&IIn&IIn&Silver IIn& \citet{2014es} \\
SN1997ab & 1997/02/28 & A095100+2004 & 09:51:00.40 & 20:04:24.00 & 0.0130&II &IIP & AGN&Silver IIn& \citet{1997ab}\\
SN2014P & 2014/02/02 & ESO 264-G49 & 10:54:04.00 & -45:48:43.60 & 0.0190& IIn&IIn & AGN& Silver IIn& \citet{2014p}\\
SN1996bu & 1996/11/14 & NGC 3631 & 11:20:59.18 & 53:12:08.00 & 0.0039&II&IIn&IIP&Silver IIn&  \citet{1996bu}\\
SN2013L & 2013/01/22 & ESO 216-G39 & 11:45:29.55 & -50:35:53.10 & 0.0170&IIn&IIn&Gal&Silver IIn& \citet{2013l}\\
SN2011ir & 2011/11/21 & UGC 6771 & 11:48:00.32 & 04:29:47.10 & 0.0199&IIn&II&IIn$\rightarrow$IIL&Silver IIn&A.~Pastorello\\
SN1987F & 1987/03/22 & NGC 4615 & 12:41:38.99 & 26:04:22.40 & 0.0160&II &IIP &Gal/AGN &Silver IIn &This work\\ 
SN2008ip & 2008/12/31 & NGC 4846 & 12:57:50.20 & 36:22:33.50 & 0.0151&II&IIn&IIP$\rightarrow$\,IIn&Silver IIn&\citet{2008ip} \\
SN2009au & 2009/03/11 & ESO 443-G21 & 12:59:46.00 & -29:36:07.50 & 0.0094&II&IIn&IIn$\rightarrow$\,IIP&Silver IIn& \citet{2009au} \\
SN2016aiy & 2016/02/17 & ESO 323-G084 & 13:08:25.40 & -41:58:50.10 & 0.0100&IIn&IIn&IIn&Silver IIn& \citet{2016aiy}\\
SN1996cr & 1996/03/16 & ESO 97-G13 & 14:13:10.05 & -65:20:44.40 & 0.0014&II &IIn &Gal &Silver IIn &F.~Bauer\\
SN2006am & 2006/02/22 & NGC 5630 & 14:27:37.24 & 41:15:35.40 & 0.0089&II&IIn&Gal&Silver IIn&\citet{2006am}\\
SN1996ae & 1996/05/21 & NGC 5775 & 14:53:59.81 & 03:31:46.30 & 0.0056&II &IIn &II-pec &Silver IIn & \citet{1996ae}\\
SN2003dv & 2003/04/22 & UGC 9638 & 14:58:04.92 & 58:52:49.90 & 0.0076&IIn&IIn&IIP&Silver IIn&\citet{2003dv}\\
SN2016bly & 2016/04/09 & 2MASX J17224883+1400584 & 17:22:48.90 & 14:00:59.88 & 0.0194&IIn &IIn &Gal &Silver IIn & \citet{2016bly}\\
SN2018hpb & 2018/10/25 & - & 22:01:34.52 & -17:27:45.22 & 0.0177&IIn &IIn &IIn &Silver IIn & \citet{2018hpb}\\
SN2017gas & 2017/08/10 & 2MASX J20171114+5812094 & 20:17:11.32 & 58:12:08.00 & 0.0100&IIn&IIn&Gal&Silver IIn&\citet{2017gas} \\
SN2006bo & 2006/04/05 & UGC 11578 & 20:30:41.90 & 09:11:40.80 & 0.0153 &IIn&IIn&LBV&Silver IIn& F.~Taddia\\
ASASSN-15lx & 2015/06/26 & ESO 47-G4 & 20:36:05.24 & -73:06:32.41 & 0.0126&II&-&IIn&Silver IIn&G.~Hosseinzadeh  \\
SN2016cvk & 2016/06/12 & ESO -344-G21 & 22:19:49.39 & -40:40:03.20 & 0.0107&II-09ip&IIn-pec&AGN&Silver IIn&\citet{2016cvk}\\
SN1989L & 1989/05/04 & NGC 7339 & 22:37:49.71 & 23:47:15.70 & 0.0044&II&II&IIP&Silver IIn&\citet{1989l}\\
SN2015bf & 2015/12/12 & NGC 7653 & 23:24:49.03 & 15:16:52.00 & 0.014227&IIn&IIn&IIn&Silver IIn&J.~Zhang\\
SN2013fs & 2013/10/07 & NGC 7610 & 23:19:44.70 & 10:11:05.00 & 0.0119 &IIP &IIn &IIP &Silver IIn & \citet{2013fs}\\
\end{longtable}
\end{landscape}

\begin{landscape}
\begin{table}
\caption{The sample of SNe\,IIn that were reclassified as not being SNe\,IIn. From our sample of 115 transients within $z<0.02$ with data, \textbf{28} SNe are not SNe\,IIn. The CSM interaction may be confused with pollution from \ion{H}{ii} regions, or the data are noisy, or the object is a gap transient with a spectrum that resembles that of SNe\,IIn. The table gives the common SN name, the discovery date, the common host name, the J2000 coordinates, and the redshift.}\label{notiin}
\begin{center}
\begin{tabular}{lllllllllll}
\hline
Name & Disc. Date & Host Name & R.A. & Dec. & $z$ &OSC&TNS&SNID&Class& Source\\
\hline
SN2001dk & 2001/07/29 & UGC 913 & 01:22:14.59 & 34:40:11.30 & 0.0171&IIP&IIP&II&\ion{H}{ii} region&\citet{2001dk}\\
SN2001fa & 2001/10/18 & NGC 673 & 01:48:22.22 & 11:31:34.40 & 0.0170&IIn&IIn&IIn$\rightarrow$\, IIP&Flash ionisation & \citet{2001fa}\\
SN2004F & 2004/01/16 & NGC 1285 & 03:17:53.80 & -07:17:43.00 & 0.0175&II&-&Gal&\ion{H}{ii} region& This work \\ 
SN2001I & 2001/01/15 & UGC 2836 & 03:43:57.28 & 39:17:39.40 & 0.0170&II&IIn&Gal&\ion{H}{ii} region&\citet{2001i}\\
SN2002bj & 2002/02/28 & NGC 1821 & 05:11:46.41 & -15:08:10.80 & 0.0120&II&IIn&Ic&\ion{H}{ii} region&\citet{2002bj}\\
SN2009kr & 2009/11/06 & NGC 1832 & 05:12:03.30 & -15:41:52.20 & 0.0065&IIn&II&IIP&\ion{H}{ii} region& \citet{2009kr} \\
SN2004gd & 2004/11/06 & NGC 2341 & 07:09:11.71 & 20:36:10.60 & 0.0170&II&IIn&Gal&\ion{H}{ii} region& This work\\ 
SN2002kg & 2002/10/26 & NGC 2403 & 07:37:01.83 & 65:34:29.32 & 0.0008&IIn&IIn&AGN&Gap transient&\citet{Schwartz_2003}\\
SN1999gb & 1999/11/22 & NGC 2532 & 08:10:13.70 & 33:57:29.80 & 0.01751&II&IIn&II&\ion{H}{ii} region&\citet{1999gb}\\
SN2010al & 2010/03/13 & UGC 4286 & 08:14:15.91 & 18:26:18.20 & 0.0172&Ibn&IIn&Ic&\ion{H}{ii} region&\citet{2010al}\\
SN2002an & 2002/01/22 & NGC 2575 & 08:22:47.76 & 24:17:41.70 & 0.0129& II&II &IIP &No narrow features &\citet{2002an}\\
SN2003ke & 2003/11/23 & MCG+06-22-09 & 09:45:52.96 & 34:41:01.40 & 0.0204&II&IIn&Gal&\ion{H}{ii} region&\citet{2003ke} \\
SN1999bw & 1999/04/20 & NGC 3198 & 10:19:46.81 & 45:31:35.00 & 0.0022&IIn?&IIn&LBV&Gap transient& \citet{1999bw} \\
SN2000ch & 2000/05/03 & NGC 3432 & 10:52:41.40 & 36:40:09.50 & 0.0017 &LBV & IIn&AGN &Gap transient & \citet{2000ch}\\
SN2014G & 2014/01/14 & NGC 3448 & 10:54:34.13 & 54:17:56.90 & 0.00450&IIn&IIn&IIn$\rightarrow$\,IIP&Flash ionisation& \citet{2014g}\\
SN2001ac & 2001/03/12 & NGC 3504 & 11:03:15.37 & 27:58:29.50 & 0.0051&IIn&IIn&AGN&Gap transient& \citet{2001ac} \\
SN1997bs & 1997/04/15 & NGC 3627 & 11:20:14.16 & 12:58:19.56 & 0.0019 &IIn &IIn? &AGN &Gap transient& This work\\
SN2002bu & 2002/03/28 & NGC 4242 & 12:17:37.18 & 45:38:47.40 & 0.0020&IIn &IIn &AGN/LBV &Gap transient & \citet{2002bu}\\
SN2017jfs & 2017/12/26 & NGC 4470 & 12:29:37.79 & 07:49:35.18 & 0.0080&IIn&IIn&AGN&Gap transient& \citet{2017jfs} \\
SN2006bv & 2006/04/28 & UGC 7848 & 12:41:01.55 & 63:31:11.60 & 0.0084&IIn? &IIn &LBV &Gap transient & \citet{2006bv}\\
SN2007cm & 2007/05/24 & NGC 4644 & 12:42:45.18 & 55:08:57.10 & 0.0160&II&IIn&IIn$\rightarrow$\,IIP&Flash ionisation& This work \\ 
SN2001dc & 2001/05/30 & NGC 5777 & 14:51:16.15 & 58:59:02.80 & 0.0071&II &IIP & IIP&No narrow feature & \citet{2001dc}\\
ASASSN-15hs & 2015/04/24 & 2MASX J15333488-7807258 & 15:33:34.31 & -78:07:23.46 & 0.0091&IIn&-&Gal&\ion{H}{ii} region&\citet{asassn15hs}\\
SN2013by & 2013/04/23 & ESO 138-G10 & 16:59:02.43 & -60:11:41.80 & 0.0038&II&II&IIP&\ion{H}{ii} region&\citet{2013by}\\
SN2001ad & 2001/03/11 & NGC 6373 & 17:24:02.45 & 58:59:52.20 & 0.0110&II &IIb &IIb &No narrow features & \citet{2001ad}\\
SN2018dfy & 2018/07/10 & UGC 11801 & 21:43:32.91 & 43:32:58.20 & 0.013519&IIn&IIP&IIP&\ion{H}{ii} region&\citet{2018dfy}\\
SN2006fp & 2006/09/17 & UGC 12182 & 22:45:41.13 & 73:09:47.80 & 0.0050 &LBV &IIn &Gal &Gap transient & \cite{2006fp} \\
SN2008gm & 2008/10/22 & NGC 7530 & 23:14:12.39 & -02:46:52.40 & 0.0117&IIn&IIn&Gal&\ion{H}{ii} region&\citet{2008gm}\\

\hline
\end{tabular}
\end{center}
\end{table}
\end{landscape}

\begin{table*}
\caption{The sample of SNe\,IIn for which we were unable to obtain any data. From our full sample of 144 SNe\,IIn within $z<0.02$, 28 SNe have no data. These may have data elsewhere, or in some cases there may have been data loss. The table gives the common SN name, the discovery date, the common host name, the J2000 coordinates, and the redshift.}
\begin{center}
\label{nodata}
\begin{tabular}{|l|l|l|l|l|l|}
\hline
Name & Disc. Date & Host Name & R.A. & Dec. & $z$  \\
\hline

SN2010jj & 2010/11/03 & NGC 812 & 02:06:52.23 & 44:34:17.50 & 0.0172\\
SN1986J & 1986/08/21 & NGC 891 & 02:22:31.33 & 42:19:57.40 & 0.0018\\
PS15cwt & 2015/08/20 & - & 02:33:16.24 & 19:15:25.20 & 0.0135\\
SN2011js & 2011/12/31 & NGC 1103 & 02:48:04.96 & -13:57:51.10 & 0.0138\\
SN2006qt & 2006/10/11 & A034002-0434 & 03:40:02.72 & -04:34:18.70 & 0.0100\\
SN2005kd & 2005/11/12 & PGC 14370 & 04:03:16.88 & 71:43:18.90 & 0.0150\\
SN2007ak & 2007/03/10 & UGC 3293 & 05:20:40.75 & 08:48:16.00 & 0.0156\\
CSS140111:060437-123740 & 2013/12/24 & MCG -02-16-002 & 06:04:36.70 & -12:37:41.00 & 0.0074\\
SN2013ha & 2013/11/06 & MCG +11-08-25 & 06:15:49.85 & 66:50:19.40 & 0.0131\\
SN2000ev & 2000/11/27 & UGC 3500 & 06:47:52.00 & 84:10:02.20 & 0.0150\\
SN2015J & 2015/04/27 & A073505-6907 & 07:35:05.18 & -69:07:53.10 & 0.0054\\
SN1987C & 1987/03/21 & MCG+09-14-47 & 08:30:01.30 & 52:41:33.00 & 0.0142\\
SN2016ehw & 2016/07/20 & MCG+12-08-47 & 08:36:37.60 & 73:35:03.70 & 0.0120\\
SN2000cl & 2000/05/26 & NGC 3318 & 10:37:16.07 & -41:37:47.80 & 0.0092 \\
ASASSN-15lf & 2015/06/15 & NGC 4108 & 12:06:45.56 & 67:09:24.00 & 0.0084\\
SN2012ab & 2012/01/31 & A122248+0536 & 12:22:47.60 & 05:36:25.00 & 0.0180\\
ASASSN-14fd & 2014/08/07 & PGC 43070 & 12:46:14.77 & -40:48:52.53 & 0.0154\\
SN2011fh & 2011/02/24 & NGC 4806 & 12:56:14.01 & -29:29:54.80 & 0.0080\\
SN1986B & 1986/02/13 & NGC 5101 & 13:21:42.40 & -27:27:43.00 & 0.0062\\
PS15aip & 2015/05/02 & KUG 1319+356 & 13:21:55.23 & 35:21:32.00 & 0.0195\\
SN2006M & 2006/01/17 & PGC 47137 & 13:27:19.76 & 31:47:14.50 & 0.0150\\
SN1998E & 1998/01/29 & NGC 5161 & 13:29:14.30 & -33:10:38.00 & 0.0080\\
ASASSN-15ab & 2015/01/02 & ESO 325-G45 & 14:03:06.24 & -38:28:29.90 & 0.0178\\
PSNJ14041297-0938168 & 2013/12/20 & IC 4363 & 14:04:12.97 & -09:38:16.80 & 0.0028\\
SNhunt248 & 2014/05/21 & NGC 5806 & 14:59:59.50 & 01:54:26 & 0.0045\\
SN2005av & 2005/03/24 & NGC 6943 & 20:44:37.58 & -68:45:10.60 & 0.0100\\
SN1978G & 1978/11/24 & IC 5201 & 22:20:48.30 & -46:01:22.00 & 0.0031\\
SN2006dn & 2006/07/05 & UGC 12188 & 22:47:37.84 & 39:52:50.16 & 0.0171\\
\hline
\end{tabular}
\end{center}
\end{table*}

\bsp	
\label{lastpage}
\end{document}